\documentclass[aps,preprint,floats,epsf,epsfig,nofootinbib,letter]{revtex4}

\usepackage{graphicx}
\usepackage{dcolumn}
\usepackage{bm}
\usepackage{subfigure}

\def\be{\begin{eqnarray}}
\def\en{\end{eqnarray}}
\def\non{\nonumber}
\def\la{\langle}
\def\ra{\rangle}

\def\B{{\cal B}}

\begin{document}

\renewcommand{\baselinestretch}{1.10}

\font\el=cmbx10 scaled \magstep2{\obeylines\hfill Sep., 2020}

\vskip 1.5 cm

\centerline{\Large\bf Data-driven study of }
\centerline{\Large\bf the implications of anomalous magnetic moments }
\centerline{\Large\bf and lepton flavor violating processes of $e$, $\mu$ and $\tau$}

\bigskip
\centerline{\bf Chun-Khiang Chua}
\medskip
\centerline{Department of Physics and Center for High Energy Physics}
\centerline{Chung Yuan Christian University}
\centerline{Chung-Li, Taiwan 320, Republic of China}
\medskip

\centerline{\bf Abstract}
We study anomalous magnetic moments and flavor violating processes of $e$, $\mu$ and $\tau$ leptons. We use a data driven approach to investigate the implications of the present data on the parameters of a class of models, which has spin-0 scalar and spin-1/2 fermion fields. We compare two different cases, which has or does not have a built-in cancelation mechanism. Our findings are as following. Chiral interactions are unable to generate large enough $\Delta a_e$ and $\Delta a_\mu$ to accommodate the experimental results. Although sizable $\Delta a_e$ and $\Delta a_\mu$ can be generated from non-chiral interactions, they are not contributed from the same source. Presently, the upper limit of $\mu\to e\gamma$ decay gives the most severe constraints on photonic penguin contributions in $\mu\to e$ transitions, but the situation may change in considering future experimental sensitivities. The $Z$-penguin diagrams can constrain chiral interaction better than photonic penguin diagrams in $\mu\to e$ transitions. In most of the parameter space, box contributions to $\mu\to 3e$ decay are subleading. The present bounds on $\Delta a_\tau$ and $d_\tau$ are unable to give useful constraints on parameters. In $\tau\to e$ $(\mu)$ transitions, the present $\tau\to e\gamma$ $(\mu \gamma)$ upper limit constrains the photonic penguin contribution better than the $\tau\to 3 e$ $(3\mu)$ upper limit, and $Z$-penguin amplitudes constrain chiral interaction better than photonic penguin amplitudes. Box contributions to $\tau\to 3e$ and $\tau\to 3\mu$ decays can sometime be comparable to $Z$-penguin contributions. The $\tau^-\to e^- \mu^+ e^-$ and $\tau^-\to \mu^- e^+ \mu^-$ rates are highly constrained by $\tau\to e\gamma$, $\mu\to e\gamma$ and $\tau\to \mu\gamma$, $\mu\to e\gamma$ upper limits, respectively. We compare the current experimental upper limits, future sensitivities and bounds from consistency on various muon and tau LFV processes.

\bigskip
\small

\pacs{Valid PACS appear here}

\maketitle

%
%

\section{Introduction}

The Large Hadron Collider completed run-2 in 2018 and is currently preparing for run-3. 
From the results of the searches, we see that New Physics (NP) signal is yet to be found (see, for example~\cite{LHC}, for a summery of the recent search results).  
It is therefore useful and timely to explore the high-precision frontier, where the NP at the scale beyond our reach may manifest in low energy processes via virtual effects.
Indeed, there are some interesting experimental activities in the lepton sector in recent years.

The muon's anomalous magnetic moment remains as a
hint of contributions from NP since 2001~\cite{Brown:2001mga}.
Presently the deviation of the experimental result $a^{\rm exp}_\mu$ from the Standard Model (SM) expectation $a^{\rm SM}_\mu$ is 3.7$\sigma$~\cite{Bennett:2006fi,PDG,Keshavarzi:2018mgv}:
\be
\Delta a_{\mu}=a^{\rm exp}_\mu-a^{\rm SM}_\mu=(27.06\pm7.26)\pm 10^{-10}.
\label{eq: muon g-2 expt} 
\en
For more details, see \cite{g-2 report, DHMZ, Campanario:2019mjh,Aoyama:2020ynm}.
New experiments in Fermilab and J-PARC are on their way to improve the sensitivities~\cite{g-2new}.

In addition, in 2018, a measurements of the fine-structure constant $\alpha$ using the recoil frequency of cesium-133 atoms in a matter-wave interferometer, infered a deviation on electron $g-2$ from the SM prediction,~\cite{Parker:2018vye}
\be
\Delta a_e=a^{\rm exp}_e-a^{\rm SM}_e=(-0.88\pm0.36)\pm 10^{-12}.
\label{eq: e g-2 expt}
\en

In the tau sector, the experimental and the theoretical results of the anomalous magnetic moment are given by
\be
-0.052<a^{\rm exp}_\tau<0.013,
\quad
a^{SM}_\tau=(1.17721\pm 0.00005)\times 10^{-3},
\en
respectively~\cite{PDG, Eidelman:2007sb}.
The experimental sensitivity is roughly one order of magnitude from the SM prediction.

Furthermore, it is known that the SM contributions to lepton electric dipole moments are at four-loop level and, consequently, are highly suppressed. 
For example, the electron electric dipole moment was estimated to be $d_e\simeq 8\times 10^{-41}$ e cm~\cite{Fukuyama:2012np}.
The present experimental bounds on electric dipole moment of $e,\mu$ and $\tau$ are given by~\cite{Andreev:2018ayy,Bennett:2008dy} 
\be
|d_e|<1.1\times 10^{-29}\, {\rm e\, cm},
\en
\be
|d_\mu|<1.9\times 10^{-19}\, {\rm e\,cm},
\en
and
\be
|d_\tau|<1.6\times 10^{-18}\, {\rm e \, cm},
\en
where the above limit on $d_e$ is used to constrain $d_\tau$ via $\Delta d_e=6.9\times 10^{-12} d_\tau$~\cite{Grozin:2008nw}.

It is known that SM prohibits charge lepton flavor violating (LFV) processes.
Hence, they are excellent probes of NP. 
Indeed, they are under intensive searches.
In 2016 the MEG collaboration reported the search result of $\mu\to e\gamma$ decay,~\cite{MEG}
\be
{\cal B}(\mu^+\to e^+\gamma)\leq 4.2\times 10^{-13},
\en
and the upgrade is on the way to improve the sensitivity by roughly one order of magnitude~\cite{Baldini:2018nnn}.
Interestingly, $\mu\to e\gamma$ decay may be closely related to lepton anomalous magnetic moments and other LFV processes, such as $\mu^+\to 3 e$ decays and muon to electron conversions, $\mu^- N\to e^- N$~\cite{review}.
See \cite{Lindner:2016bgg} for a review on $(g-2)_\mu$ and LFV processes.
Note that LFV processes can sometime be related to cosmological effects, see for example~\cite{Berezhiani:1989fp}.

Lepton flavor violating $\tau$ decays are also under intensive search.
Current bounds on $\tau\to e\gamma$, $\mu\gamma$, $3e$, $3 \mu$, $e\bar \mu e$, $\mu \bar e \mu$ decays was provided by $B$ factories. They are at the level of $10^{-8}$ and the sensitivities will be improved by two orders of magnitude in the updated $B$ factory~\cite{Amhis:2016xyh, Kou:2018nap}.

The current limits and future experimental sensitivities of various $\l'\to l\gamma$, $l\to l' \bar l'' l'$ and $l N\to l' N$ processes are summarized in Table~\ref{tab:expt bounds}.
%

\begin{table}[t!]
\caption{Present upper limits and future sensitivities of some muon and tau lepton flavor violating 
processes are listed~\cite{MEG,PDG,Baldini:2018nnn,Mihara2019,Amhis:2016xyh,Kou:2018nap}.}
 \label{tab:expt bounds}
\begin{ruledtabular}
\begin{tabular}{ l c  c }
~~~~~~
    & current limit 
    & future sensitivity
    \\
    \hline
${\cal B}(\mu^+\to e^+\gamma)$
    & $<4.2\times 10^{-13}$
    &  $6\times 10^{-14}$
    \\    
${\cal B}(\mu^+\to e^+e^+e^-)$
    & $<1.0\times 10^{-12}$
    &  $10^{-16}$ 
    \\ 
${\cal B}(\mu^- {\rm Ti}\to e^-{\rm Ti})$
    & $<4.3\times 10^{-12}$
    & $10^{-17}$ 
    \\   
${\cal B}(\mu^- {\rm Au}\to e^-{\rm Au})$
    & $<7\times 10^{-13}$
    & $10^{-16}$ 
    \\ 
${\cal B}(\mu^- {\rm Al}\to e^-{\rm Al})$
    & $\cdots$
    & $10^{-17}$ 
    \\ 
${\cal B}(\tau^-\to e^-\gamma)$
    & $<3.3\times 10^{-8}$ 
    & $3\times10^{-9}$ 
    \\ 
${\cal B}(\tau^-\to \mu^-\gamma)$
    & $<4.4\times 10^{-8}$ 
    & $1\times 10^{-9}$ 
    \\
${\cal B}(\tau^-\to e^-e^+ e^-)$
    & $<2.7\times 10^{-8}$ 
    & $4.3\times 10^{-10}$ 
    \\   
${\cal B}(\tau^-\to \mu^- e^+ \mu^-)$
    & $<1.7\times 10^{-8}$ 
    & $2.7\times 10^{-10}$ 
    \\  
${\cal B}(\tau^-\to e^-\mu^+ e^-)$
    & $<1.5\times 10^{-8}$ 
    & $2.4\times 10^{-10}$ 
    \\                  
${\cal B}(\tau^-\to \mu^-\mu^+\mu^-)$
    & $<2.1\times 10^{-8}$ 
    &  $3.3\times 10^{-10}$ 
    \\                            
\end{tabular}
\end{ruledtabular}
\end{table}

Many popular NP scenarios or models are disfavored or even closed to being ruled out by data (see, for example, ~\cite{LHC}).
Given the present situation, it is worthy to use a data driven approach.
It will be interesting to see where the present data lead us to.
As a working assumption, we consider a general class of models that lepton anomalous magnetic moment and various lepton flavor violating processes, such as $\mu\to e \gamma$, $\mu\to3e$, $\mu\to e$ conversions, $\tau\to e\gamma$, $\mu\gamma$, $3e$, $3 \mu$, $e\bar \mu e$ and $\mu \bar e \mu$ decays are induced by loop diagrams via exchanging spin-0 and spin-1/2 particles in this work.

Note that the above mentioned experimental results of $\Delta a_\mu$ and $\Delta a_e$ received a lot of attention.
There are studies involving leptoquark, two Higgs doublets, Supersymmetry particles, dark matters and so on  ~\cite{Davoudiasl:2018fbb,
Berlin:2018bsc,
Crivellin:2018qmi,
Zhang:2018fbm,
Dekens:2018pbu,
Liu:2018xkx,
Dutta:2018fge,
Han:2018znu,
Coy:2018bxr,
Dong:2019iaf,
Chen:2019wbk,
Mohlabeng:2019vrz,
Ibe:2019jbx,
CarcamoHernandez:2019pmy,
Crivellin:2019mvj,
Harigaya:2019shz,
Bigaran:2019bqv,
Endo:2019bcj,
Kawamura:2019rth,
Abdullah:2019ofw,
Bauer:2019gfk,
Badziak:2019gaf,
Mandal:2019gff,
CarcamoHernandez:2019ydc,
CarcamoHernandez:2019lhv,
Hiller:2019mou,
Keshavarzi:2019abf,
Bramante:2019exc,
Cornella:2019uxs,
Kawamura:2019hxp,
Calibbi:2019bay,
Krasnikov:2019dgh,
Altmannshofer:2020ywf,
Endo:2020mev,
CarcamoHernandez:2020pxw,
Haba:2020gkr,
Altmannshofer:2020axr,
Bigaran:2020jil,
Jana:2020pxx,
Calibbi:2020emz,
Chen:2020jvl,
Yang:2020bmh,
Hati:2020fzp,
Frank:2020kvp,
Dutta:2020scq,
Botella:2020xzf,
Abdallah:2020biq,
Chen:2020tfr,
Dorsner:2020aaz,
Keshavarzi:2020bfy,
Arbelaez:2020rbq,
Nomura:2020dzw,
Jana:2020joi,
Gherardi:2020qhc}.
It is interesting that many new physics models in these studies are similar to the framework adopted here. 
Furthermore, by considering simultaneously various processes or quantities involving different leptons, one can obtain useful information on new physics. For example, in \cite{Crivellin:2018qmi} by using Effective Field Theory (EFT) and some simplified models similar to the present framework, the authors found that the $\mu\to e\gamma$ bound requires the muon and electron sectors to be decoupled and, consequently, $\Delta a_\mu$ and $\Delta a_e$ cannot be explained from the same source, but as a bonus a large muon electric dipole moment is possible.
In addition, it is known in the literature that there are relations on $l'\to l \bar l'' l'''$ and $\l'\to l\gamma$ rates.
For example,  using an EFT approach \cite{Kuno:1999jp, Crivellin:2013hpa},
$l'\to l \bar l l$ and $\l'\to l\gamma$ rates are shown to be related as following,  
\be
\B(\mu\to 3e)\simeq \frac{1}{160} \B(\mu\to e\gamma),\,\, 
\B(\tau\to 3e)\simeq \frac{1}{95} \B(\tau\to e\gamma),\,\,
\B(\tau\to3\mu)\simeq \frac{1}{440} \B(\tau\to \mu\gamma),
\label{eq: 3l per lgamma ratio}
\en
if the photonic dipole penguins dominate in these $l'\to l \bar l l$ decays.
It is also known that the constraints on 4-lepton and $Z$-lepton-lepton contributions using $l'\to l \bar l l$ bounds are found to be less severe than the constraints of $\gamma$-lepton-lepton contributions using $\l'\to l\gamma$ bounds~\cite{Crivellin:2013hpa}.  
Studies involving different processes are useful to search for NP and to probe its properties as well.

It will be useful to compare the present approach to an EFT approach (see, for example, ~\cite{Buchmuller:1985jz,Crivellin:2013hpa,Crivellin:2018qmi}). For illustration, we use the above mentioned analysis on $\Delta a_\mu$, $\Delta a_e$ and the $\mu\to e\gamma$ decay as an example. As stated in \cite{Crivellin:2018qmi}, the relevant effective Hamiltonian is 
\be
H_{\rm eff}=c_R^{l_f l_i} \bar l_f\sigma_{\mu\nu} P_R l_i F^{\mu\nu}+{\rm H.c.},
\label{eq: dipole H}
\en
giving
\be
a_{l_i}=-\frac{4 m_{l_i}}{e} {\rm Re} \,c_R^{l_il_i},
\quad
{\cal B}(\mu\to e\gamma)=\frac{m^3_\mu}{4\pi\Gamma_\mu}(|c_R^{e\mu}|^2+|c_R^{\mu e}|^2).
\en
Note that there are in general no correlation between magnetic moments and lepton flavor violation~\cite{Crivellin:2018qmi}.
When NP particles couple to muon and electron simultaneously, one expects $c_R^{e\mu}=\sqrt{c_R^{ee} c_R^{\mu\mu}}$ and
the resulting $\mu\to e\gamma$ rate is 
\be
{\cal B}(\mu\to e\gamma)=\frac{\alpha m_\mu^2}{16 m_e\Gamma_\mu}|\Delta a_\mu \Delta a_e|\sim 8\times 10^{-5},
\label{eq: mu2egamma excess}
\en
which excesses the MEG bound by 8 order of magnitude~\cite{Crivellin:2018qmi}.
As an EFT approach only makes use of SM particles with all NP particles being integrated out, it is generic. 
For example, information on ${\rm Re}(c^{\mu\mu}_R)$, ${\rm Re}(c^{ee}_R)$ and $|c_R^{e\mu}|^2+|c_R^{\mu e}|^2$ can be extracted from data without referring to any specific NP model.
However, to correlate different quantities, such as $\Delta a_{\mu,e}$ and the $\mu\to e\gamma$ decay rate, one needs additional assumption on the underlying NP model. For example, the above $c_R^{e\mu}=\sqrt{c_R^{ee} c_R^{\mu\mu}}$ relation requires the NP particles to couple to muon and electron simultaneously~\cite{Crivellin:2018qmi}. 
The class of models adopted here provides a realization of this situation via one-loop diagrams in Fig.~\ref{fig:penguin&box}. 
In addition to the above discussion, note that the so-called $F_1$ photonic penguin and box contributions are usually lumped into the 4-lepton operators in an EFT approach. As a result, it will be difficult so separate them.
The present approach is less generic than an EFT approach, but it is more generic than a specific model, as we try to capture some common behaviors or ingredients of a class of models concerning the lepton sector. 
It is in between of a specify model and an EFT approach and it can be a bridge to link them.   
When comparing to an EFT approach, the limitation of the present approach is its less of generality, while the advantage of it is the ability to provide some correlations and detail informations, which are in general difficult to obtain in an EFT approach without introducing additional assumption.  

In this work two cases are considered
The first case does not have any built-in cancellation mechanism
and the second case has some built-in mechanism, such as Glashow-Iliopoulos-Maiani or super-Glashow-Iliopoulos-Maiani mechanism. 
These two cases are complementary to each other and it will be interesting to compare them.
This work is an updated and extended study of \cite{Chua:2012rn}, where only $\mu$ decays were considered.
Note that a similar setup, but in the quark sector, has been used in a study of the $b\to s\mu^+\mu^-$ decay \cite{Arnan:2016cpy}.

We briefly give the framework in the next section. 
In Sec. III, numerical results will be presented, 
where data on $g-2$, $d_l$ and upper limits of LFV rates will be used to constrain parameters
and the correlations between different processes will be investigated.
We give our conclusion in Sec.~IV, which is followed by two appendices.

\section{Framework}

\begin{figure}[t]
\centering
\subfigure[]{
  \includegraphics[width=6cm]{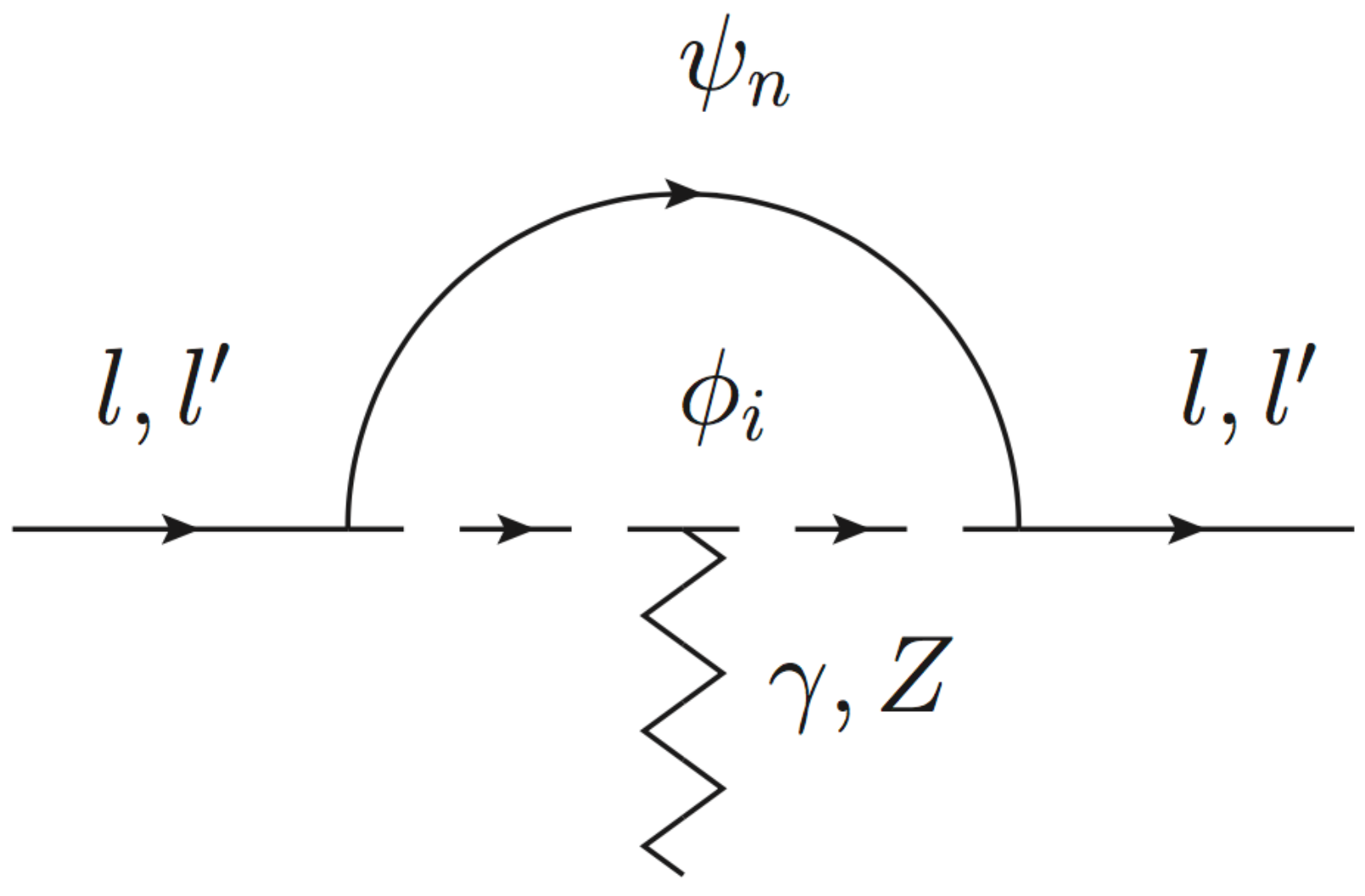}
}
\subfigure[]{
  \includegraphics[width=6cm]{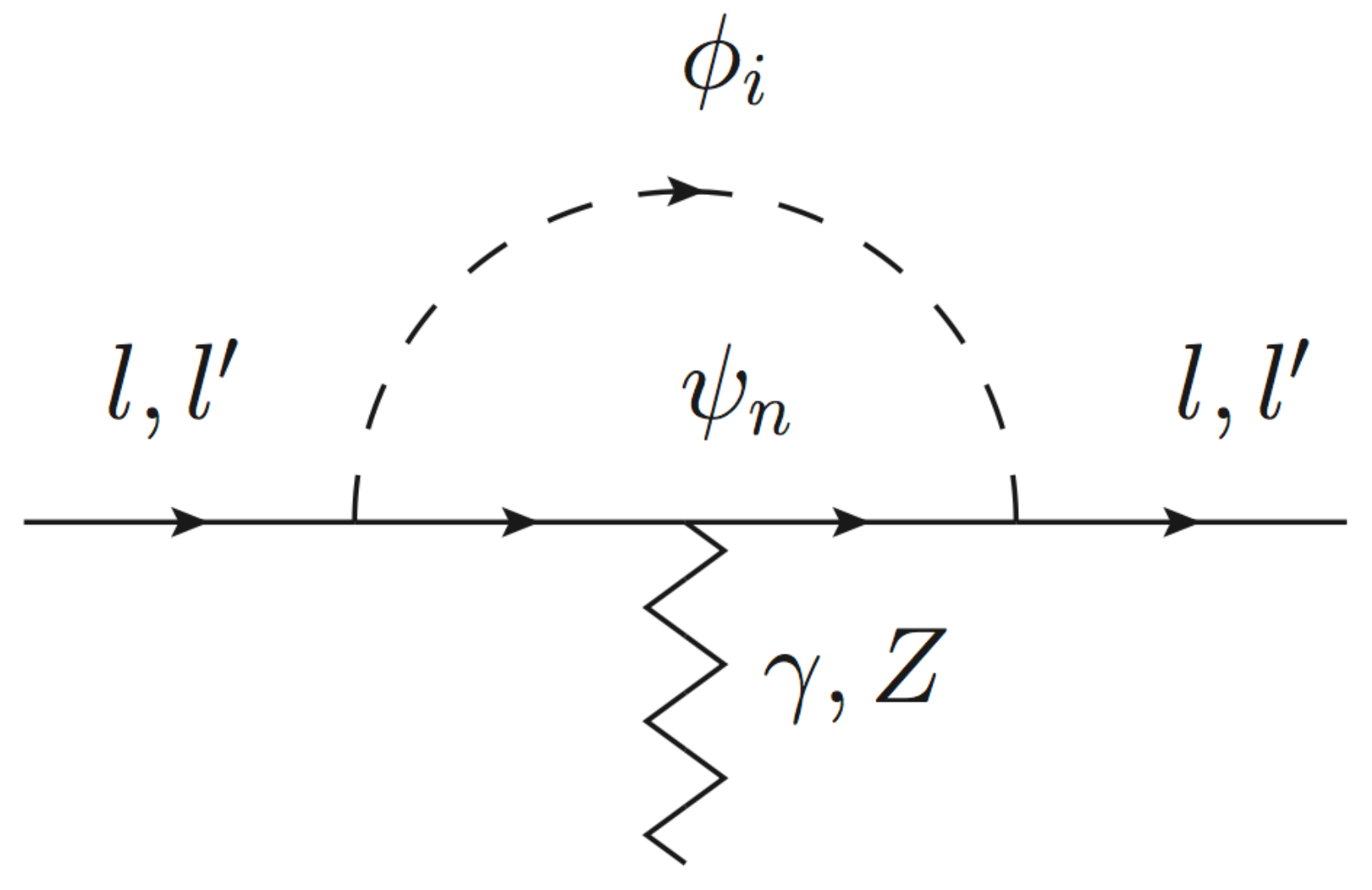}
}
\subfigure[]{
  \includegraphics[width=6cm]{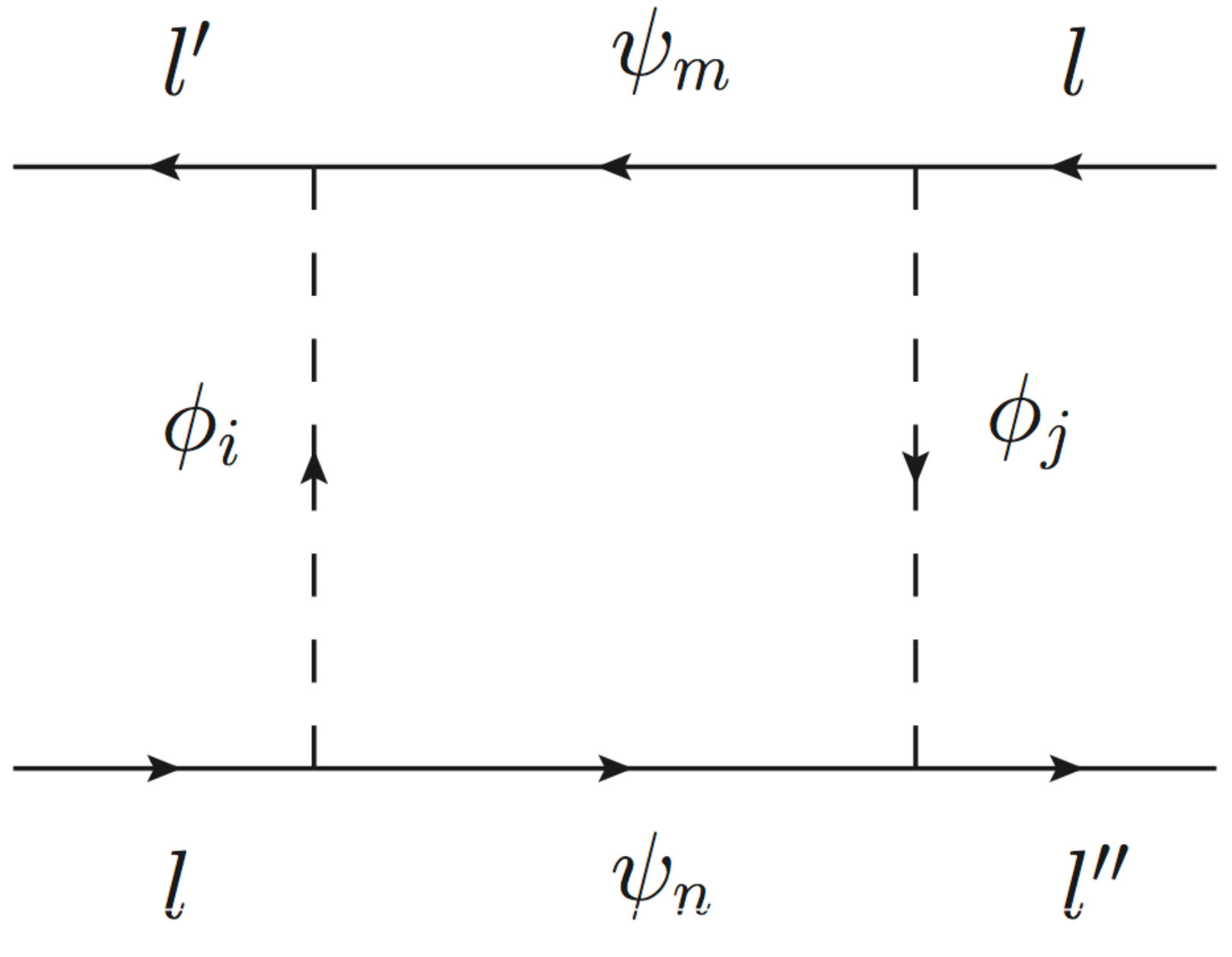}
}
\subfigure[]{
  \includegraphics[width=6cm]{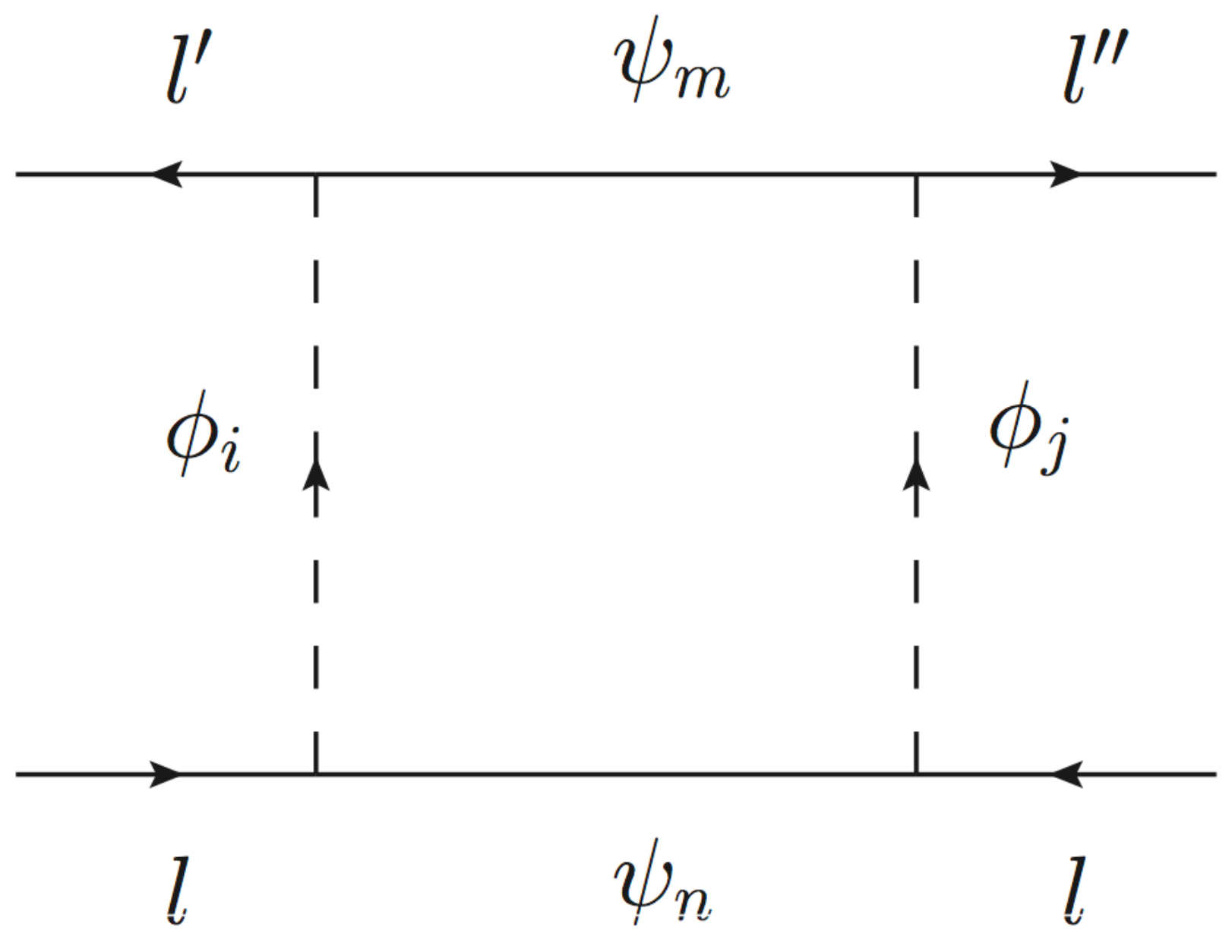}
}
\caption{ Diagrams contributing to various processes. Penguin diagrams contributing to $e$, $\mu$ and $\tau$, $g-2$, $d_l$, $l'\to l\gamma$, $\bar l'\to l\bar l l$ and $l' N\to l N$ processes are shown in Fig. 1 (a) and (b),
while box diagrams contributing to the $\bar l'\to l\bar l'' l$ process are shown in Fig. 1 (c) and (d). 
Note that we do not show diagrams involving self energy parts. Fig. 1 (d) is for the Majorana case.}
\label{fig:penguin&box}
\end{figure}

The generic interacting Lagrangian involving leptons ($l$), exotic spin-0 bosons ($\phi_i$) and spin-1/2 fermions ($\psi_n$) is given by
\be
{\cal L}_{\rm int}=\bar\psi_n(g^{ni}_{lL} P_L+g^{ni}_{lR} P_R) l\phi_i^*
+\bar l(g_{lL}^{ni*}P_R+g_{lR}^{ni*}P_L)\psi_n\phi_i,
\label{eq:Lint}
\en
where indices, $l$, $i$ and $n$, are summed and these fields are in their mass bases.
It can contribute to lepton $g-2$, $d_l$ and various LFV processes, such as $l'\to l\gamma$, $\bar l'\to l\bar l l$ decays and $l' N\to l N$ transitions, via diagrams shown in Fig.~\ref{fig:penguin&box}.  
Some useful formulas can be found in ref.~\cite{Chua:2012rn} and are collected in Appendix~\ref{app:Formula}.

As noted in the introduction, we consider two complementary cases.
In case I there is no any built-in cancellation mechanism.
A typical amplitude, $A$, may contain several sub-amplitudes, $A_j$, each comes from one of the loop diagrams (see Fig.~\ref{fig:penguin&box}) giving
\be
A=\sum_{j=1}^N A_j.
\label{eq: N A}
\en
To constrain these sub-amplitudes from data, we will switch them on one at a time.
Different sub-amplitudes are in principle independent from each other
as there is no any built-in cancellation mechanism. 
However, in a realistic model calculation, it is likely to have several amplitudes to appear at the same time and interfere. 
Nevertheless, it is well known that interference effects can be important only when the amplitudes are of similar size. 
For amplitudes of different sizes,
this analysis can constrain the most dominant amplitude.
On the other hand, through investigating the sizes of different sub-amplitudes 
the analysis can also identify the region, where several sub-amplitudes are of similar sizes, and, hence, identify where interference can be potentially important. 

The Wilson coefficients of a typical sub-amplitude can be obtained by using formulas in Appendix~\ref{app:Formula}, 
but with the following replacement, 
\be
g^{ni}_{lM}\to g_{lM}.
\en
Terms contributing to various processes in case I are shown in Table~\ref{tab: parameters case I}.  
Note that $\Delta T_{3\psi}$ is basically the difference of weak isospin quantum numbers of $\psi_R$ and $\psi_L$,
while $\kappa_{R,L}$ are defined in Eq.~(\ref{eq: kappa and Delta T3}).
Note that $\Delta T_{3\psi}$ is expected to be an order one quantity, while $\kappa_R$ is expected to be a small quantity.
See Appendix A for more informations.

\begin{figure}[t]
\centering
\subfigure[]{
  \includegraphics[width=7.5cm]{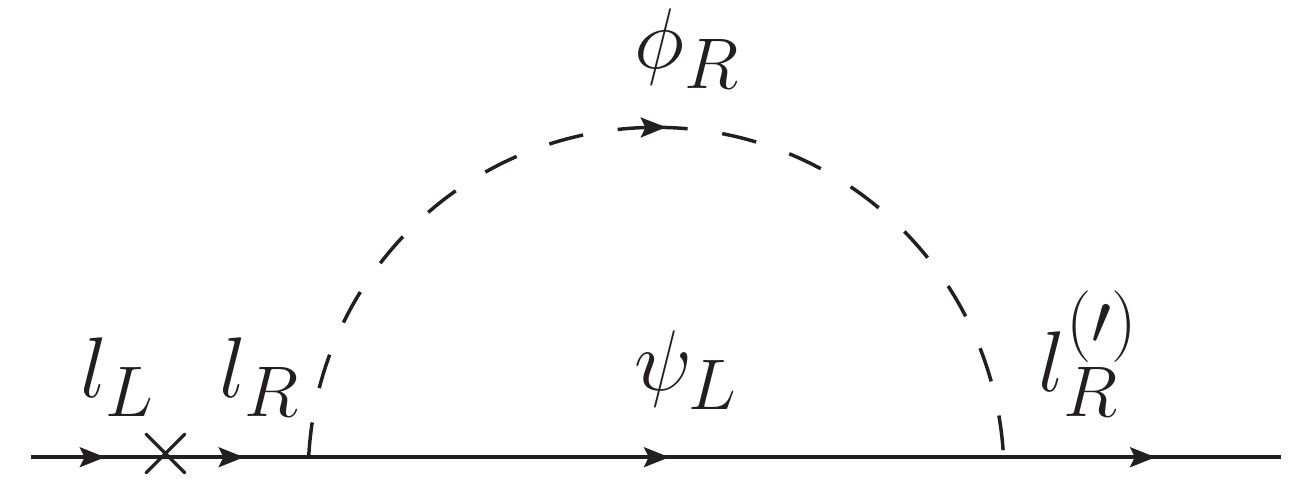}
}
\subfigure[]{
  \includegraphics[width=7.5cm]{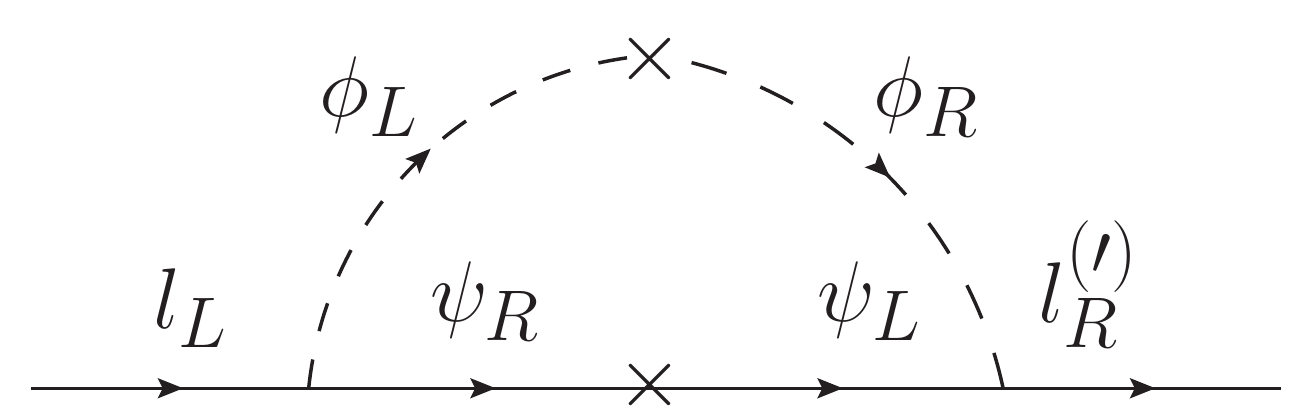}
}
\caption{Two types of photonic dipole penguin diagrams (photon line not shown). The diagram in the left panel does not have the so-called chiral enhancement, while the one in the right panel has. The crosses denote mass insertions or mixings that connect $l_L$ and $l_R$, $\psi_L$ and $\psi_R$ fields, $\phi_L$ and $\phi_R$ fields. 
Since $l_L$ and $l_R$ have different weak quantum numbers, the cross in the left diagram and 
one of the crosses in the right diagram need to couple to the Higgs VEV, while the other cross in the right diagram connects fields with identical weak quantum numbers.
See text for more details.}
\label{fig:chiralenhancement}
\end{figure}

In Fig.~\ref{fig:chiralenhancement} we gives two typical diagrams contributing to the photonic dipole penguins. 
The left diagram can occur in a chiral interaction, while the right diagram is possible only for the so-called non-chiral interaction, where $\phi$ and $\psi$ couple to both $l_L$ and $l^{(\prime)}_R$ at the same time. It is well known in the literature, see, for example, ref.~\cite{Gabbiani:1996hi},  that a non-chiral interaction can provide chiral enhancement in photonic dipole penguin amplitudes resulting sizable effects in quantities and processes such as $\Delta a_l$, $l\to l'\gamma$ decays and so on.
A way to see this is by using the EFT approach. Before spontaneous symmetry breaking, we have the following dipole operators,~\cite{Crivellin:2013hpa}
\be
Q_{eW}=(\bar L_L\sigma^{\mu\nu} l_R)\tau^I H W^I_{\mu\nu},
\quad
Q_{eB}=(\bar L_L\sigma^{\mu\nu} l_R) H B_{\mu\nu},
\en
where $L_L$, $l_R$ are the iso-doublet and singlet lepton fields, $W_{\mu\nu}$ and $B_{\mu\nu}$ are the $SU(2)$ and $U(1)$ field strengths and $H$ is the Higgs field. 
To obtain a photonic dipole interaction term as shown in Eq. (\ref{eq: dipole H}), one needs the above operators, but with the Higgs field replaced by its vacuum expectation value (VEV), i.e. $H\to \la H\ra$. 
In Fig.~\ref{fig:chiralenhancement}~(a), the Higgs-lepton-lepton Yukawa vertex applies to the external lepton line with the Higgs field replaced by its VEV, while in Fig.~\ref{fig:chiralenhancement}~(b), it is required to have either $\psi_L$ and $\psi_R$ or $\phi_L$ and $\phi_R$ being mixed due to a Higgs VEV, and the other pair of fields with identical quantum numbers. 
It is clear that the left and right diagrams in Fig.~\ref{fig:chiralenhancement} are associated with $m_l$ and $m_\psi$, respectively, and the mass ratio gives rise to the chiral enhancement.
It is important to note that only fields with suitable weak quantum numbers that mix due to a Higgs VEV can have non-chiral interaction generating chiral enhancement. 
Hence, it is non-trivial to have chiral enhancement in a new physics model. 
For the possible $SU(2)\times U(1)$ quantum numbers of $\psi$ and $\phi$, and the combinations that can generate the non-chiral interaction resulting chiral enhancement, see Appendix~\ref{app: QN}.

\begin{table}[t]
\caption{Terms contributing to various processes in case I.}
 \label{tab: parameters case I}
\footnotesize{
\begin{center}
\begin{tabular}{ l c  c c c c}
\hline
Processes
    & $\gamma$-penguin
    & $\gamma$-penguin
    & $Z$-penguin
    & Box
    \\
    \hline
$\Delta a_l$
    & $ Q_{\phi,\psi}|g_{l L(R)}|^2$
    &  $Q_{\phi,\psi}{\rm  Re}(g^*_{l R} g_{l L})$
    &  
    & 
    \\ 
$d_l$
    & 
    &  $Q_{\phi,\psi}{\rm  Im}(g^*_{l R} g_{l L})$
    &  
    & 
    \\             
$\mu^+\to e^+ \gamma$ 
    & $Q_{\phi,\psi} g^*_{\mu L(R)} g_{e L(R)}$
    &  $Q_{\phi,\psi} g^*_{\mu R(L)} g_{e L(R)}$
    &  
    &
    \\    
$\mu^+\to e^- e^+ e^-$
    & $ Q_{\phi,\psi} g^*_{\mu L(R)} g_{e L(R)} $
    &  $ Q_{\phi,\psi} g^*_{\mu R(L)} g_{e L(R)}$
    & $ g^*_{\mu R(L)} g_{e R(L)} \Delta T_{3\psi}(\kappa_{R(L)})$
    &  $g^*_{\mu M} g_{e N} g^*_{e O} g_{e P}$
    \\   
$\mu^- N\to e^- N$
    & $Q_{\phi,\psi} g_{\mu L(R)} g^*_{e L(R)}$
    &  $Q_{\phi,\psi} g_{\mu R(L)} g^*_{e L(R)}$
    &  $g_{\mu R(L)} g^*_{e R(L)} \Delta T_{3\psi}(\kappa_{R(L)})$
    &  $g_{\mu M} g^*_{e N} g_{e O} g^*_{eP}$
    \\
$\tau^-\to e^- \gamma$ 
    & $Q_{\phi,\psi} g_{\tau L(R)} g^*_{e L(R)}$
    &  $Q_{\phi,\psi} g_{\tau R(L)} g^*_{e L(R)}$
    &  
    &
    \\           
$\tau^-\to e^- e^+ e^-$
    &  $Q_{\phi,\psi} g_{\tau L(R)} g^*_{e L(R)}$
    &  $Q_{\phi,\psi} g_{\tau R(L)} g^*_{e L(R)}$
    &  $g_{\tau R(L)} g^*_{e R(L)} \Delta T_{3\psi}(\kappa_{R(L)})$
    &  $g_{\tau M} g^*_{e N} g_{e M} g^*_{e N}$
    \\
$\tau^-\to \mu^- \gamma$ 
    &  $Q_{\phi,\psi} g_{\tau L(R)} g^*_{\mu L(R)}$
    &  $Q_{\phi,\psi} g_{\tau R(L)} g^*_{\mu L(R)}$
    &  
    &
    \\          
$\tau^-\to \mu^-\mu^+\mu^-$
    &  $Q_{\phi,\psi} g_{\tau L(R)} g^*_{\mu L(R)}$
    &  $Q_{\phi,\psi} g_{\tau R(L)} g^*_{\mu L(R)}$
    &  $g_{\tau R(L)} g^*_{\mu R(L)} \Delta T_{3\psi}(\kappa_{R(L)})$
    &  $g_{\tau M} g^*_{\mu N} g_{\mu M} g^*_{\mu N}$
     \\ 
$\tau^-\to e^-\mu^+ e^-$
    &
    &
    &
    &  $g_{\tau M} g^*_{e N} g_{\mu M} g^*_{e N}$
    \\   
$\tau^-\to \mu^- e^+\mu^+$
    &  
    &
    &
    &  $g_{\tau M} g^*_{\mu N} g_{e M} g^*_{\mu N}$
   \\  
    \hline                  
\end{tabular}
\end{center}
}
\end{table}

In the second case (case II), there is a built-in cancellation mechanism.
Now some sub-amplitudes in Eq.~(\ref{eq: N A}) are related intimately. 
They need to be grouped together to allow the cancellation mechanism to take place, 
and the resulting group of amplitudes should be viewed as a new sub-amplitude. 
To constrain these new sub-amplitudes from data, we will turn them on one at a time.
To be specify, we consider the following replacement,
\be
g^{ni}_{lM}\to g^{i}_{lM}=g_{lM}\Gamma^{il}_M,
\en
where $g_{lM}$ is real (as the phase is absorbed into $\Gamma_M$) and we have $M=L,R$.
These $\Gamma$ satisfy the following relations:
\be
\Gamma^{\dagger li}_M m^2_i\Gamma^{il'}_N=(m^2_\phi) ^{ll'}_{MN},
\quad
\Gamma^{\dagger li}_M\Gamma^{il'}_N=\delta^{ll'}\delta_{MN},
\en
where the $\delta$s are Kronecker deltas.
Typical terms in a Wilson coefficient given in Appendix~\ref{app:Formula} should now be replaced accordingly:
\be
\sum_i g^{i*}_{l' M} f(m^2_{\psi}, m^2_{\phi_i}) g^{i}_{l N}
\to m^2_\phi\frac{\partial }{\partial m^2_\phi}f(m^2_{\psi}, m^2_{\phi}) g_{\mu M} g_{e N} \delta^{MN}_{l' l},
\en
where $m^2_\phi$ is the average of the mass squared of $\phi_i$ and $\delta^{MN}_{l' l}$ is the mixing angle defined in the usual way (do not confuse it with the Kronecker delta):~\cite{Gabbiani:1996hi}
\be
\delta^{MN}_{l' l}
\equiv \frac{1}{m^2_\phi}
\Gamma^{M\dagger}_{l' i} (m^2_{\phi_i}-m^2_{\phi}) \Gamma^N_{i l}
=\frac{(m^{2 }_\phi)^{MN}_{l' l}}{m^2_\phi}.
\en
Note that as a common practice only the leading terms of $\delta$ are kept in the amplitudes. 
Therefore, to employ the ``$\delta$" parameterization, 
one needs to assume a large degree of flavor-alignment of the new fields to the SM leptons,
i.e. the mass matrices of these new fields are almost diagonal in the mass basis of the SM leptons, 
and the small misalignment can be encoded in these $\delta$s.
Usually this requires introducing additional symmetry to the model.

Terms contributing to various processes in case II are shown in Table~\ref{tab: parameters case II}.

\begin{table}[t]
\caption{Terms contributing to various processes in case II.}
 \label{tab: parameters case II}
\footnotesize{
\begin{center}
\begin{tabular}{ l c  c c c c}
\hline
Processes
    & $\gamma$-penguin
    & $\gamma$-penguin
    & $Z$-penguin
    & Box
    \\
    \hline
$\Delta a_l$
    & $ Q_{\phi,\psi}|g_{l L(R)}|^2$
    &  $Q_{\phi,\psi}{\rm  Re}(g^*_{l R} g_{l L}\delta^{ll}_{RL})$
    &  
    & 
    \\ 
$d_l$
    & 
    &  $Q_{\phi,\psi}{\rm  Im}(g^*_{l R} g_{l L}\delta^{ll}_{RL})$
    &  
    & 
    \\          
$\mu^+\to e^+ \gamma$ 
    & $Q_{\phi,\psi} g^*_{\mu M} g_{e M}\delta^{\mu e}_{MM}$
    &  $Q_{\phi,\psi} g^*_{\mu R(L)} g_{e L(R)}\delta^{\mu e}_{RL(LR)}$
    &  
    &
    \\    
$\mu^+\to e^- e^+ e^-$
    & $ Q_{\phi,\psi} g^*_{\mu M} g_{e M}\delta^{\mu e}_{MM} $
    &  $ Q_{\phi,\psi} g^*_{\mu R(L)} g_{e L(R)}\delta^{\mu e}_{RL(LR)}$
    & $ g^*_{\mu M} g_{e M} \Delta T_{3\psi}\delta^{\mu e}_{MM}$
    &  $g^*_{\mu M} g_{e M} g^*_{e N} g_{e N}\delta^{\mu e}_{M M}$
    \\   
$\mu^- N\to e^- N$
    & $Q_{\phi,\psi} g_{\mu M} g^*_{e M}\delta^{e\mu}_{MM}$
    &  $Q_{\phi,\psi} g_{\mu R(L)} g^*_{e L(R)}\delta^{e\mu}_{LR(RL)}$
    &  $g_{\mu M} g^*_{e M} \Delta T_{3\psi}\delta^{e\mu}_{MM}$
    &  $g_{\mu M} g^*_{e M} g_{e N} g^*_{e N}\delta^{e\mu}_{MM}$
    \\
$\tau^-\to e^- \gamma$ 
    & $Q_{\phi,\psi} g_{\tau M} g^*_{e M}\delta^{e\tau}_{MM}$
    &  $Q_{\phi,\psi} g_{\tau R(L)} g^*_{e L(R)}\delta^{e\tau}_{LR(RL)}$
    &  
    &
    \\           
$\tau^-\to e^- e^+ e^-$
    &  $Q_{\phi,\psi} g_{\tau M} g^*_{e M}\delta^{e\tau}_{MM}$
    &  $Q_{\phi,\psi} g_{\tau R(L)} g^*_{e L(R)}\delta^{e\tau}_{LR(RL)}$
    &  $g_{\tau M} g^*_{e M} \Delta T_{3\psi}\delta^{e\tau}_{MM}$
    &  $g_{\tau M} g^*_{e M} g_{e N} g^*_{e N}\delta^{e\tau}_{MM}$
    \\
$\tau^-\to \mu^- \gamma$ 
    &  $Q_{\phi,\psi} g_{\tau M} g^*_{\mu M}\delta^{e\tau}_{MM}$
    &  $Q_{\phi,\psi} g_{\tau R(L)} g^*_{\mu L(R)}\delta^{e\tau}_{LR(RL)}$
    &  
    &
    \\          
$\tau^-\to \mu^-\mu^+\mu^-$
    &  $Q_{\phi,\psi} g_{\tau M} g^*_{\mu M}\delta^{\mu\tau}_{MM}$
    &  $Q_{\phi,\psi} g_{\tau R(L)} g^*_{\mu L(R)}\delta^{\mu\tau}_{LR(RL)}$
    &  $g_{\tau M} g^*_{\mu M} \Delta T_{3\psi}\delta^{\mu\tau}_{MM}$
    &  $g_{\tau M} g^*_{\mu M} g_{\mu N} g^*_{\mu N}\delta^{\mu\tau}_{MM}$
     \\ 
$\tau^-\to e^-\mu^+ e^-$
    &
    &
    &
    &  $g_{\tau M} g^*_{e N} g_{\mu O} g^*_{e P}\delta^{e\tau}_{NM}\delta^{e\mu}_{PO}$
    \\   
$\tau^-\to \mu^- e^+\mu^+$
    &  
    &
    &
    &  $g_{\tau M} g^*_{\mu M} g_{e O} g^*_{\mu P}\delta^{\mu\tau}_{NM}\delta^{\mu e}_{PO}$
   \\  
    \hline                  
\end{tabular}
\end{center}
}
\end{table}


\section{Results}

\begin{table}[hb!]
\caption{\footnotesize{Constraints on parameters in case I using $x\equiv m_\phi/m_\psi=1$ and $m_\psi=500$~GeV from various processes are shown. Results are applicable with $L$ and $R$ interchanged. 
Results for other $m_\psi$ can be obtained by scaling with a $(\frac{m_\psi}{500 {\rm GeV}})^2$ or $\frac{m_\psi}{500 {\rm GeV}}$ factor, where the latter is for $Q_{\phi,\psi} g^*_{l^{(\prime)} R} g_{l L}$.
Results in $[...]$ are obtained by using the future experimental sensitivities, results in $\{...\}$ are for the Majorana case.}}
 \label{tab: results case I x=1}
\footnotesize{
\begin{center}
\begin{tabular}{ l |c  c c c c}
\hline
Processes
    & constraints
    & constraints
    & constraints
    & constraints
    \\
    \hline
    & $ Q_{\phi}|g_{e R}|^2$
    &  $ Q_{\psi}|g_{e R}|^2$
    &  $Q_{\phi}{\rm  Re}(g^*_{e R} g_{e L})$
    &  $Q_{\psi}{\rm  Re}(g^*_{e R} g_{e L})$
    \\ 
$\Delta a_e$
   & $-1597\pm653$
   & $1597\mp653$
   & $(-4.1\pm1.6)\times 10^{-4}$
   & $(2.0\mp 0.8)\times 10^{-4}$
   \\  
   \hline  
    & $Q_{\phi}|g_{\mu R}|^2$
    &  $Q_{\psi}|g_{\mu R}|^2$
    &  $Q_{\phi}{\rm  Re}(g^*_{\mu R} g_{\mu L})$
    & $Q_{\psi}{\rm  Re}(g^*_{\mu R} g_{\mu L})$
    \\ 
$\Delta a_\mu$
   & $115\pm31$
   & $-115\mp31$
   & $(6.1\pm1.6)\times 10^{-3}$
   & $(-3.0\mp 0.8)\times 10^{-3}$
   \\
   \hline        
    & $Q_{\phi} |g_{\tau R}|^2$
    &  $Q_{\psi} |g_{\tau R}|^2$
    &  $Q_{\phi}{\rm  Re}(g^*_{\tau R} g_{\tau L})$
    & $Q_{\psi}{\rm  Re}(g^*_{\tau R} g_{\tau L})$
    \\ 
$\Delta a_\tau$
   & $(-7\sim2)\times 10^6$
   & $(-2\sim7)\times 10^6$
   & $(-7\sim 2)\times 10^3$
   & $(-0.8\sim 3)\times 10^3$
   \\   
   \hline  
    &  $|Q_{\phi}{\rm  Im}(g^*_{e R} g_{e L})|$
    &  $|Q_{\phi}{\rm  Im}(g^*_{e R} g_{e L})|$
    &  $|Q_{\phi}{\rm  Im}(g^*_{\mu(\tau) R} g_{\mu(\tau) L})|$
    &  $|Q_{\psi}{\rm  Im}(g^*_{\mu(\tau) R} g_{\mu(\tau) L})|$
    \\ 
$d_e$, $d_\mu$, $d_\tau$ 
   & $2.6\times 10^{-10}$
   & $1.3\times 10^{-10}$
   & $4.6\,(38.3)$
   & $2.3\,(19.1)$
   \\  
   \hline                
    &  $|Q_{\phi} g^*_{\mu R} g_{e R}|$
    &  $|Q_{\psi} g^*_{\mu R} g_{e R}|$
    &  $|Q_{\phi} g^*_{\mu R} g_{e L}|$
    &  $|Q_{\psi} g^*_{\mu R} g_{e L}|$
    \\  
$\mu^+\to e^+ \gamma$ 
    & $0.002\,[0.0008]$
    &  $0.002\,[0.0008]$
    &  $11\,[4]\times 10^{-8}$
    &  $6\, [2]\times 10^{-8}$ 
    \\        
$\mu^+\to e^- e^+ e^-$
    & $0.046\,[0.0005]$
    &  $0.030\,[0.0003]$
    &  $224\, [2]\times 10^{-8}$
    &  $112\, [1]\times 10^{-8}$ 
    \\  
$\mu^- {\rm Au}\to e^- {\rm Au}$
    &  $0.020\,[0.0002]$
    &  $0.016\,[0.0002]$
    &  $236\,[3]\times 10^{-8}$
    &  $118\,[1]\times 10^{-8}$ 
    \\    
$\mu^- {\rm Ti}\to e^- {\rm Ti}$
    &  $0.051\,[0.00008]$
    &  $0.046\,[0.00007]$
    &  $569\,[0.9]\times 10^{-8}$
    &  $284\,[0.4]\times 10^{-8}$ 
    \\ 
$\mu^- {\rm Al}\to e^- {\rm Al}$
    &  $[0.00010]$
    &  $[0.00009]$
    &  $[1.1\times 10^{-8}]$
    &  $[0.5\times 10^{-8}]$ 
    \\  
    \hline             
    &  $|g^*_{\mu R} g_{e R} \Delta T_{3\psi}|$
    &  $|g^*_{\mu R} g_{e R} \kappa_{R}|$
    &  $|g^*_{\mu R} g_{e R} g^*_{e R} g_{e R}|$
    &  $|g^*_{\mu R} g_{e R} g^*_{e L} g_{e L}|$
    \\        
$\mu^+\to e^- e^+ e^-$
    & $393\,[4]\times 10^{-6}$
    & $115\,[1]\times 10^{-6}$
    & $0.01\,\{-\}[1\times 10^{-4}\,\{-\}]$
    & $7\,\{7\}\times 10^{-3}[7\{7\}\times 10^{-5}]$ 
    \\ 
$\mu^- {\rm Au}\to e^- {\rm Au}$
    &  $492\,[6] \times 10^{-7}$
    &  $145\,[2]\times 10^{-7}$
    &  
    &  
    \\    
$\mu^- {\rm Ti}\to e^- {\rm Ti}$
    &  $1718\,[3]\times 10^{-7}$
    &  $5049\,[8]\times 10^{-8}$
    &  
    &   
    \\ 
$\mu^- {\rm Al}\to e^- {\rm Al}$
    &  $[4\times 10^{-7}]$
    &  $[1\times 10^{-7}]$
    &  
    &   
    \\   
    \hline                             
    & $|Q_{\phi} g_{\tau R} g^*_{e R}|$
    & $|Q_{\psi} g_{\tau R} g^*_{e R}|$ 
    & $|Q_{\phi} g_{\tau R} g^*_{e L}|$ 
    & $|Q_{\psi} g_{\tau R} g^*_{e L}|$
    \\   
$\tau^-\to e^- \gamma$ 
    & $1.4\,[0.4]$
    &  $1.4\,[0.4]$
    &  $13\,[4]\times 10^{-4}$
    & $6\,[2]\times 10^{-4}$
    \\               
$\tau^-\to e^- e^+ e^-$
    &  $13.2\,[1.7]$
    &  $10.0\,[1.3]$
    &  $11\,[1]\times 10^{-3}$
    &  $56\,[7]\times 10^{-4}$
    \\
    \hline
    &  $|g_{\tau R} g^*_{e R} \Delta T_{3\psi}|$
    &  $|g_{\tau R} g^*_{e R} \kappa_{R}|$
    &  $|g_{\tau R} g^*_{e R} g_{e R} g^*_{e R}|$
    &  $|g_{\tau R} g^*_{e R} g_{e L} g^*_{e L}|$ 
    \\ 
$\tau^-\to e^- e^+ e^-$
    & $0.15\,[0.02]$
    & $0.05\,[0.006]$
    & $4.3\,\{-\}[0.5\,\{-\}]$
    & $2.9\,\{2.9\}[0.4\,\{0.4\}]$
    \\  
    \hline         
    &  $|Q_{\phi} g_{\tau R} g^*_{\mu R}|$
    &  $|Q_{\psi} g_{\tau R} g^*_{\mu R}|$
    &  $|Q_{\phi} g_{\tau R} g^*_{\mu L}|$
    &  $|Q_{\psi} g_{\tau R} g^*_{\mu L}|$
    \\
$\tau^-\to \mu^- \gamma$ 
    &  $1.7\,[0.3]$
    &  $1.7\,[0.3]$
    &  $15\,[2]\times 10^{-4}$
    &  $7\,[1]\times 10^{-4}$
    \\   
$\tau^-\to \mu^-\mu^+\mu^-$
    & $30.7\,[3.9]$
    & $12.5\,[1.6]$
    & $21\,[3]\times 10^{-3}$
    & $11\,[1]\times 10^{-3}$
    \\  
    \hline             
    &  $|g_{\tau R} g^*_{\mu R} \Delta T_{3\psi}|$
    &  $|g_{\tau R} g^*_{\mu R} \kappa_{R}|$
    &  $|g_{\tau R} g^*_{\mu R} g_{\mu R} g^*_{\mu R}|$
    &  $|g_{\tau R} g^*_{\mu R} g_{\mu L} g^*_{\mu L}|$
     \\ 
$\tau^-\to \mu^-\mu^+\mu^-$
    & $0.14\,[0.02]$
    & $0.04\,[0.005]$
    & $3.8\,\{-\}\,[0.5\,\{-\}]$ 
    & $2.5\,\{2.5\}\,[0.3\,\{0.3\}]$ 
     \\     
     \hline      
    &  $|g_{\tau R} g^*_{e R} g_{\mu R} g^*_{e R}|$
    &  $|g_{\tau R} g^*_{e R} g_{\mu L} g^*_{e L}|$
    &  $|g_{\tau R} g^*_{e L} g_{\mu R} g^*_{e L}|$
    &  
    \\  
$\tau^-\to e^-\mu^+ e^-$
    & $3.2\,\{-\}\,[0.4\,\{-\}]$
    & $2.3\,\{2.3\}\,[0.3\,\{0.3\}]$
    & $6.4\,\{6.4\}\,[0.8\,\{0.8\}]$
    &  
    \\   
    \hline    
    &  $|g_{\tau R} g^*_{\mu R} g_{e R} g^*_{\mu R}|$
    &  $|g_{\tau R} g^*_{\mu R} g_{e L} g^*_{\mu L}|$
    &  $|g_{\tau R} g^*_{\mu L} g_{e R} g^*_{\mu L}|$
    &  
   \\  
$\tau^-\to \mu^- e^+\mu^+$
    & $3.4\,\{-\}\,[0.4\,\{-\}]$
    & $2.4\,\{2.4\}\,[0.3\,\{0.3\}]$
    & $6.8\,\{6.8\}\,[0.9\,\{0.9\}]$
    &  
    \\      
    \hline                  
\end{tabular}
\end{center}
}
\end{table}

In this section we present the numerical results for cases I and II. 
Experimental inputs are from refs.~\cite{MEG,PDG,Baldini:2018nnn,Mihara2019,Amhis:2016xyh,Kou:2018nap} and are shown in Table I.
Further inputs not listed in the table are from ref.~\cite{PDG}.

\subsection{Case I}\label{sec: case I}

\begin{table}[ht!]
\caption{Same as Table~\ref{tab: results case I x=1}, but with $x\equiv m_\phi/m_\psi=0.5$.}
 \label{tab: results case I x=0.5}
\footnotesize{
\begin{center}
\begin{tabular}{ l |c  c c c c}
\hline
Processes
    & constraints
    & constraints
    & constraints
    & constraints
    \\
    \hline
    & $ Q_{\phi}|g_{e R}|^2$
    &  $ Q_{\psi}|g_{e R}|^2$
    &  $Q_{\phi}{\rm  Re}(g^*_{e R} g_{e L})$
    &  $Q_{\psi}{\rm  Re}(g^*_{e R} g_{e L})$
    \\ 
$\Delta a_e$
   & $-811\pm332$ 
   & $1059\mp433$ 
   & $(-2.3\pm1.0)\times 10^{-4}$ 
   & $(1.5\mp 0.7)\times 10^{-4}$ 
   \\  
   \hline  
    & $Q_{\phi}|g_{\mu R}|^2$
    &  $Q_{\psi}|g_{\mu R}|^2$
    &  $Q_{\phi}{\rm  Re}(g^*_{\mu R} g_{\mu L})$
    & $Q_{\psi}{\rm  Re}(g^*_{\mu R} g_{\mu L})$
    \\ 
$\Delta a_\mu$
   & $58\pm16$ 
   & $-76\mp20$ 
   & $(3.5\pm0.9)\times 10^{-3}$ 
   & $(-2.3\mp 0.6)\times 10^{-3}$ 
   \\
   \hline        
    & $Q_{\phi} |g_{\tau R}|^2$
    &  $Q_{\psi} |g_{\tau R}|^2$
    &  $Q_{\phi}{\rm  Re}(g^*_{\tau R} g_{\tau L})$
    & $Q_{\psi}{\rm  Re}(g^*_{\tau R} g_{\tau L})$
    \\ 
$\Delta a_\tau$
   & $(-4\sim1)\times 10^6$ 
   & $(-1\sim5)\times 10^6$ 
   & $(-4\sim 1)\times 10^3$ 
   & $(-0.7\sim 3)\times 10^3$ 
   \\   
   \hline  
    &  $|Q_{\phi}{\rm  Im}(g^*_{e R} g_{e L})|$
    &  $|Q_{\phi}{\rm  Im}(g^*_{e R} g_{e L})|$
    &  $|Q_{\phi}{\rm  Im}(g^*_{\mu(\tau) R} g_{\mu(\tau) L})|$
    &  $|Q_{\psi}{\rm  Im}(g^*_{\mu(\tau) R} g_{\mu(\tau) L})|$
    \\ 
$d_e$, $d_\mu$, $d_\tau$ 
   & $1.5\times 10^{-10}$
   & $1.0\times 10^{-10}$
   & $2.6\,(22.0)$
   & $1.8\,(14.9)$
   \\  
   \hline                           
    &  $|Q_{\phi} g^*_{\mu R} g_{e R}|$
    &  $|Q_{\psi} g^*_{\mu R} g_{e R}|$
    &  $|Q_{\phi} g^*_{\mu R} g_{e L}|$
    &  $|Q_{\psi} g^*_{\mu R} g_{e L}|$
    \\  
$\mu^+\to e^+ \gamma$ 
    & $0.001\,[0.0004]$ 
    &  $0.001\,[0.0005]$ 
    &  $7\,[2]\times 10^{-8}$ 
    &  $4\, [2]\times 10^{-8}$  
    \\        
$\mu^+\to e^- e^+ e^-$
    & $0.024\,[0.0002]$ 
    &  $0.021\,[0.0002]$ 
    &  $129\, [1]\times 10^{-8}$ 
    &  $87\, [0.9]\times 10^{-8}$  
    \\  
$\mu^- {\rm Au}\to e^- {\rm Au}$
    &  $0.008\,[0.0001]$ 
    &  $0.013\,[0.0002]$ 
    &  $136\,[2]\times 10^{-8}$ 
    &  $92\,[1]\times 10^{-8}$  
    \\    
$\mu^- {\rm Ti}\to e^- {\rm Ti}$
    &  $0.022\,[0.00003]$ 
    &  $0.038\,[0.00006]$ 
    &  $327\,[0.5]\times 10^{-8}$ 
    &  $222\,[0.3]\times 10^{-8}$ 
    \\ 
$\mu^- {\rm Al}\to e^- {\rm Al}$
    &  $[4\times 10^{-5}]$ 
    &  $[7\times 10^{-5}]$ 
    &  $[6.2\times 10^{-9}]$ 
    &  $[4.2\times 10^{-9}]$  
    \\  
    \hline             
    &  $|g^*_{\mu R} g_{e R} \Delta T_{3\psi}|$
    &  $|g^*_{\mu R} g_{e R} \kappa_{R}|$
    &  $|g^*_{\mu R} g_{e R} g^*_{e R} g_{e R}|$
    &  $|g^*_{\mu R} g_{e R} g^*_{e L} g_{e L}|$
    \\        
$\mu^+\to e^- e^+ e^-$
    & $274\,[3]\times 10^{-6}$ 
    & $148\,[1]\times 10^{-6}$ 
    & $6\,\{7\}\times 10^{-3}[6\{7\}\times 10^{-5}]$ 
    & $3\,\{3\}\times 10^{-3}[3\{3\}\times 10^{-5}]$ 
    \\ 
$\mu^- {\rm Au}\to e^- {\rm Au}$
    &  $343\,[4] \times 10^{-7}$ 
    &  $186\,[2]\times 10^{-7}$ 
    &  
    &   
    \\    
$\mu^- {\rm Ti}\to e^- {\rm Ti}$
    &  $1120\,[2]\times 10^{-7}$ 
    &  $649\,[1]\times 10^{-7}$ 
    &  
    &   
    \\ 
$\mu^- {\rm Al}\to e^- {\rm Al}$
    &  $[3\times 10^{-7}]$ 
    &  $[1\times 10^{-7}]$ 
    &  
    &  
    \\   
    \hline                             
    & $|Q_{\phi} g_{\tau R} g^*_{e R}|$
    & $|Q_{\psi} g_{\tau R} g^*_{e R}|$ 
    & $|Q_{\phi} g_{\tau R} g^*_{e L}|$ 
    & $|Q_{\psi} g_{\tau R} g^*_{e L}|$
    \\   
$\tau^-\to e^- \gamma$ 
    & $0.7\,[0.2]$ 
    &  $1.0\,[0.3]$ 
    &  $7\,[2]\times 10^{-4}$ 
    & $5\,[1]\times 10^{-4}$ 
    \\               
$\tau^-\to e^- e^+ e^-$
    &  $6.8\,[0.9]$ 
    &  $6.9\,[0.9]$ 
    &  $65\,[8]\times 10^{-4}$ 
    &  $44\,[6]\times 10^{-4}$ 
    \\
    \hline
    &  $|g_{\tau R} g^*_{e R} \Delta T_{3\psi}|$
    &  $|g_{\tau R} g^*_{e R} \kappa_{R}|$
    &  $|g_{\tau R} g^*_{e R} g_{e R} g^*_{e R}|$
    &  $|g_{\tau R} g^*_{e R} g_{e L} g^*_{e L}|$ 
    \\ 
$\tau^-\to e^- e^+ e^-$
    & $0.11\,[0.01]$ 
    & $0.06\,[0.007]$ 
    & $2.5\,\{2.7\}[0.3\,\{0.3\}]$ 
    & $1.1\,\{1.1\}[0.1\,\{0.1\}]$ 
    \\  
    \hline         
    &  $|Q_{\phi} g_{\tau R} g^*_{\mu R}|$
    &  $|Q_{\psi} g_{\tau R} g^*_{\mu R}|$
    &  $|Q_{\phi} g_{\tau R} g^*_{\mu L}|$
    &  $|Q_{\psi} g_{\tau R} g^*_{\mu L}|$
    \\
$\tau^-\to \mu^- \gamma$ 
    &  $0.8\,[0.1]$ 
    &  $1.1\,[0.2]$ 
    &  $9\,[1]\times 10^{-4}$ 
    &  $58\,[9]\times 10^{-5}$ 
    \\   
$\tau^-\to \mu^-\mu^+\mu^-$
    & $16.7\,[2.1]$ 
    & $8.9\,[1.1]$ 
    & $12\,[2]\times 10^{-3}$ 
    & $8\,[1]\times 10^{-3}$ 
    \\  
    \hline             
    &  $|g_{\tau R} g^*_{\mu R} \Delta T_{3\psi}|$
    &  $|g_{\tau R} g^*_{\mu R} \kappa_{R}|$
    &  $|g_{\tau R} g^*_{\mu R} g_{\mu R} g^*_{\mu R}|$
    &  $|g_{\tau R} g^*_{\mu R} g_{\mu L} g^*_{\mu L}|$
     \\ 
$\tau^-\to \mu^-\mu^+\mu^-$
    & $0.09\,[0.01]$ 
    & $0.05\,[0.006]$ 
    & $2.2\,\{2.4\}\,[0.3\,\{0.3\}]$ 
    & $1.0\,\{1.0\}\,[0.1\,\{0.1\}]$ 
     \\     
     \hline      
    &  $|g_{\tau R} g^*_{e R} g_{\mu R} g^*_{e R}|$
    &  $|g_{\tau R} g^*_{e R} g_{\mu L} g^*_{e L}|$
    &  $|g_{\tau R} g^*_{e L} g_{\mu R} g^*_{e L}|$
    &  
    \\  
$\tau^-\to e^-\mu^+ e^-$
    & $1.8\,\{2.0\}\,[0.2\,\{0.3\}]$ 
    & $0.9\,\{0.9\}\,[0.1\,\{0.1\}]$ 
    & $1.9\,\{1.9\}\,[0.2\,\{0.2\}]$ 
    &  
    \\   
    \hline    
    &  $|g_{\tau R} g^*_{\mu R} g_{e R} g^*_{\mu R}|$
    &  $|g_{\tau R} g^*_{\mu R} g_{e L} g^*_{\mu L}|$
    &  $|g_{\tau R} g^*_{\mu L} g_{e R} g^*_{\mu L}|$
    &  
   \\  
$\tau^-\to \mu^- e^+\mu^+$
    & $2.0\,\{2.2\}\,[0.2\,\{0.3\}]$ 
    & $1.0\,\{1.0\}\,[0.1\,\{0.1\}]$ 
    & $2.1\,\{2.1\}\,[0.3\,\{0.3\}]$ 
    &  
    \\      
    \hline                  
\end{tabular}
\end{center}
}
\end{table}

\begin{table}[ht!]
\caption{Same as Table~\ref{tab: results case I x=1}, but with $x\equiv m_\phi/m_\psi=2$.}
 \label{tab: results case I x=2}
\footnotesize{
\begin{center}
\begin{tabular}{ l |c  c c c c}
\hline
Processes
    & constraints
    & constraints
    & constraints
    & constraints
    \\
    \hline
    & $ Q_{\phi}|g_{e R}|^2$
    &  $ Q_{\psi}|g_{e R}|^2$
    &  $Q_{\phi}{\rm  Re}(g^*_{e R} g_{e L})$
    &  $Q_{\psi}{\rm  Re}(g^*_{e R} g_{e L})$
    \\ 
$\Delta a_e$
   & $-4234\pm 1732$ 
   & $3247\mp 1328$ 
   & $(-9.4\pm3.8)\times 10^{-4}$ 
   & $(3.2\mp 1.3)\times 10^{-4}$ 
   \\  
   \hline  
    & $Q_{\phi}|g_{\mu R}|^2$
    &  $Q_{\psi}|g_{\mu R}|^2$
    &  $Q_{\phi}{\rm  Re}(g^*_{\mu R} g_{\mu L})$
    & $Q_{\psi}{\rm  Re}(g^*_{\mu R} g_{\mu L})$
    \\ 
$\Delta a_\mu$
   & $305\pm82$ 
   & $-234\mp63$ 
   & $(14.0\pm3.7)\times 10^{-3}$ 
   & $(-4.8\mp 1.3)\times 10^{-3}$ 
   \\
   \hline        
    & $Q_{\phi} |g_{\tau R}|^2$
    &  $Q_{\psi} |g_{\tau R}|^2$
    &  $Q_{\phi}{\rm  Re}(g^*_{\tau R} g_{\tau L})$
    & $Q_{\psi}{\rm  Re}(g^*_{\tau R} g_{\tau L})$
    \\ 
$\Delta a_\tau$
   & $(-20\sim5)\times 10^6$ 
   & $(-4\sim16)\times 10^6$ 
   & $(-16\sim 4)\times 10^3$ 
   & $(-1\sim 5)\times 10^3$ 
   \\   
   \hline   
    &  $|Q_{\phi}{\rm  Im}(g^*_{e R} g_{e L})|$
    &  $|Q_{\phi}{\rm  Im}(g^*_{e R} g_{e L})|$
    &  $|Q_{\phi}{\rm  Im}(g^*_{\mu(\tau) R} g_{\mu(\tau) L})|$
    &  $|Q_{\psi}{\rm  Im}(g^*_{\mu(\tau) R} g_{\mu(\tau) L})|$
    \\ 
$d_e$, $d_\mu$, $d_\tau$ 
   & $6.1\times 10^{-10}$
   & $2.1\times 10^{-10}$
   & $10.5\,(88.1)$
   & $3.6\,(30.3)$
   \\  
   \hline               
    &  $|Q_{\phi} g^*_{\mu R} g_{e R}|$
    &  $|Q_{\psi} g^*_{\mu R} g_{e R}|$
    &  $|Q_{\phi} g^*_{\mu R} g_{e L}|$
    &  $|Q_{\psi} g^*_{\mu R} g_{e L}|$
    \\  
$\mu^+\to e^+ \gamma$ 
    & $0.006\,[0.002]$ 
    &  $0.004\,[0.002]$ 
    &  $26\,[10]\times 10^{-8}$ 
    &  $9\, [3]\times 10^{-8}$ 
    \\        
$\mu^+\to e^- e^+ e^-$
    & $0.120\,[0.001]$ 
    &  $0.056\,[0.0006]$ 
    &  $516\, [5]\times 10^{-8}$ 
    &  $177\, [2]\times 10^{-8}$ 
    \\  
$\mu^- {\rm Au}\to e^- {\rm Au}$
    &  $0.059\,[0.0007]$ 
    &  $0.024\,[0.0003]$ 
    &  $542\,[6]\times 10^{-8}$ 
    &  $187\,[2]\times 10^{-8}$ 
    \\    
$\mu^- {\rm Ti}\to e^- {\rm Ti}$
    &  $0.151\,[0.0002]$ 
    &  $0.069\,[0.0001]$ 
    &  $1309\,[2]\times 10^{-8}$ 
    &  $450\,[0.7]\times 10^{-8}$  
    \\ 
$\mu^- {\rm Al}\to e^- {\rm Al}$
    &  $[0.0003]$ 
    &  $[0.0001]$ 
    &  $[2.5\times 10^{-8}]$ 
    &  $[0.9\times 10^{-8}]$ 
    \\  
    \hline             
    &  $|g^*_{\mu R} g_{e R} \Delta T_{3\psi}|$
    &  $|g^*_{\mu R} g_{e R} \kappa_{R}|$
    &  $|g^*_{\mu R} g_{e R} g^*_{e R} g_{e R}|$
    &  $|g^*_{\mu R} g_{e R} g^*_{e L} g_{e L}|$
    \\        
$\mu^+\to e^- e^+ e^-$
    & $695\,[7]\times 10^{-6}$ 
    & $879\,[9]\times 10^{-7}$ 
    & $0.03\,\{0.05\}[2\{5\}\times 10^{-4}]$ 
    & $0.02\,\{0.02\}[2\{2\}\times 10^{-4}]$ 
    \\ 
$\mu^- {\rm Au}\to e^- {\rm Au}$
    &  $87\,[1] \times 10^{-6}$ 
    &  $110\,[1]\times 10^{-7}$ 
    &  
    &   
    \\    
$\mu^- {\rm Ti}\to e^- {\rm Ti}$
    &  $3038\,[5]\times 10^{-7}$ 
    &  $3845\,[6]\times 10^{-8}$ 
    &  
    &   
    \\ 
$\mu^- {\rm Al}\to e^- {\rm Al}$
    &  $[7\times 10^{-7}]$ 
    &  $[8\times 10^{-8}]$ 
    &  
    &  
    \\   
    \hline                             
    & $|Q_{\phi} g_{\tau R} g^*_{e R}|$
    & $|Q_{\psi} g_{\tau R} g^*_{e R}|$ 
    & $|Q_{\phi} g_{\tau R} g^*_{e L}|$ 
    & $|Q_{\psi} g_{\tau R} g^*_{e L}|$
    \\   
$\tau^-\to e^- \gamma$ 
    &  $3.8\,[1.1]$ 
    &  $2.9\,[0.9]$ 
    &  $29\,[8]\times 10^{-4}$ 
    & $10\,[3]\times 10^{-4}$ 
    \\               
$\tau^-\to e^- e^+ e^-$
    &  $34.7\,[4.4]$ 
    &  $19.1\,[2.4]$ 
    &  $26\,[3]\times 10^{-3}$ 
    &  $9\,[1]\times 10^{-3}$ 
    \\
    \hline
    &  $|g_{\tau R} g^*_{e R} \Delta T_{3\psi}|$
    &  $|g_{\tau R} g^*_{e R} \kappa_{R}|$
    &  $|g_{\tau R} g^*_{e R} g_{e R} g^*_{e R}|$
    &  $|g_{\tau R} g^*_{e R} g_{e L} g^*_{e L}|$ 
    \\ 
$\tau^-\to e^- e^+ e^-$
    & $0.27\,[0.03]$ 
    & $0.03\,[0.004]$ 
    & $9.9\,\{18.9\}[1.2\,\{2.4\}]$ 
    & $9.2\,\{9.2\}[1.2\,\{1.2\}]$ 
    \\  
    \hline         
    &  $|Q_{\phi} g_{\tau R} g^*_{\mu R}|$
    &  $|Q_{\psi} g_{\tau R} g^*_{\mu R}|$
    &  $|Q_{\phi} g_{\tau R} g^*_{\mu L}|$
    &  $|Q_{\psi} g_{\tau R} g^*_{\mu L}|$
    \\
$\tau^-\to \mu^- \gamma$ 
    &  $4.4\,[0.7]$ 
    &  $3.4\,[0.5]$ 
    &  $34\,[5]\times 10^{-4}$ 
    &  $11\,[2]\times 10^{-4}$ 
    \\   
$\tau^-\to \mu^-\mu^+\mu^-$
    & $78.0\,[9.8]$ 
    & $22.4\,[2.8]$ 
    & $49\,[6]\times 10^{-3}$ 
    & $17\,[2]\times 10^{-3}$ 
    \\  
    \hline             
    &  $|g_{\tau R} g^*_{\mu R} \Delta T_{3\psi}|$
    &  $|g_{\tau R} g^*_{\mu R} \kappa_{R}|$
    &  $|g_{\tau R} g^*_{\mu R} g_{\mu R} g^*_{\mu R}|$
    &  $|g_{\tau R} g^*_{\mu R} g_{\mu L} g^*_{\mu L}|$
     \\ 
$\tau^-\to \mu^-\mu^+\mu^-$
    & $0.24\,[0.03]$ 
    & $0.03\,[0.004]$ 
    & $8.7\,\{16.7\}\,[1.1\,\{2.1\}]$ 
    & $8.1\,\{8.1\}\,[1.0\,\{1.0\}]$ 
     \\     
     \hline      
    &  $|g_{\tau R} g^*_{e R} g_{\mu R} g^*_{e R}|$
    &  $|g_{\tau R} g^*_{e R} g_{\mu L} g^*_{e L}|$
    &  $|g_{\tau R} g^*_{e L} g_{\mu R} g^*_{e L}|$
    &  
    \\  
$\tau^-\to e^-\mu^+ e^-$
    & $7.4\,\{14.1\}\,[0.9\,\{1.8\}]$ 
    & $7.1\,\{7.1\}\,[0.9\,\{0.9\}]$ 
    & $31.0\,\{31.0\}\,[3.9\,\{3.9\}]$ 
    &  
    \\   
    \hline    
    &  $|g_{\tau R} g^*_{\mu R} g_{e R} g^*_{\mu R}|$
    &  $|g_{\tau R} g^*_{\mu R} g_{e L} g^*_{\mu L}|$
    &  $|g_{\tau R} g^*_{\mu L} g_{e R} g^*_{\mu L}|$
    &  
   \\  
$\tau^-\to \mu^- e^+\mu^+$
    & $7.9\,\{15.0\}\,[1.0\,\{1.9\}]$ 
    & $7.5\,\{7.5\}\,[0.9\,\{0.9\}]$ 
    & $33.0\,\{33.0\}\,[4.2\,\{4.2\}]$ 
    &  
    \\      
    \hline                  
\end{tabular}
\end{center}
}
\end{table}

\begin{figure}[t!]
\centering
\subfigure[]{
  \includegraphics[width=6.7cm]{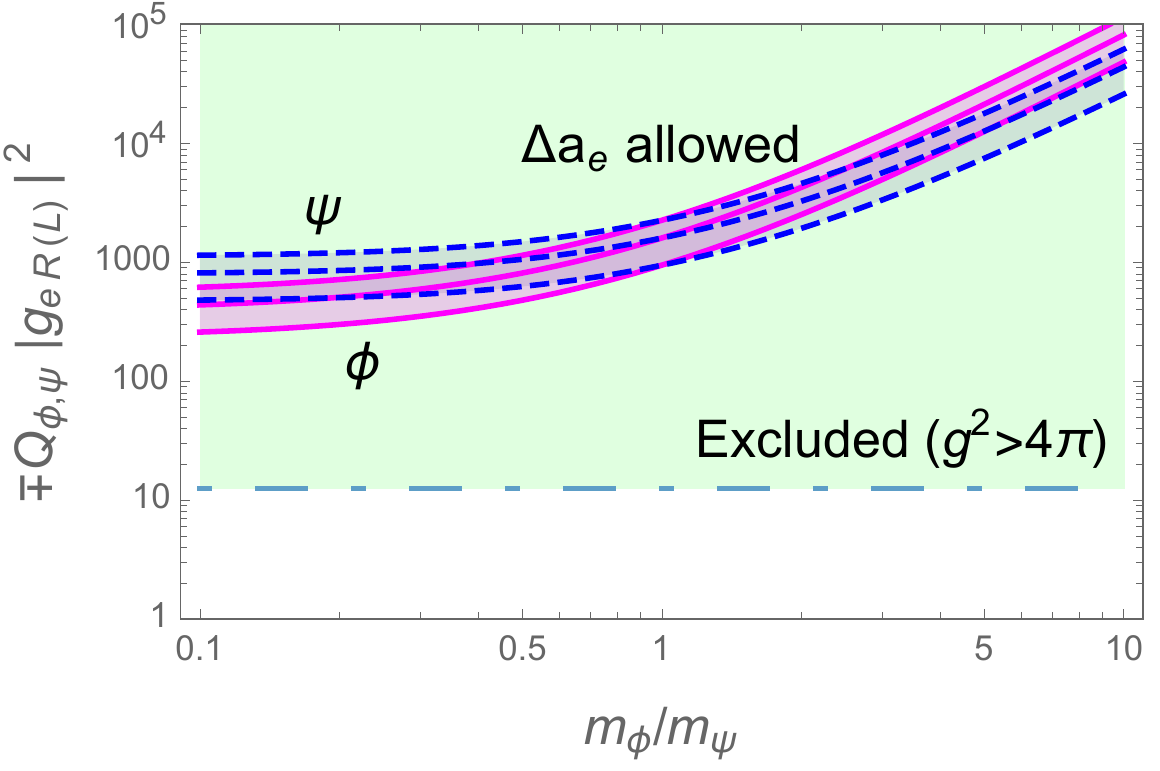}
}
\hspace{0.5cm}
\subfigure[]{
  \includegraphics[width=7cm]{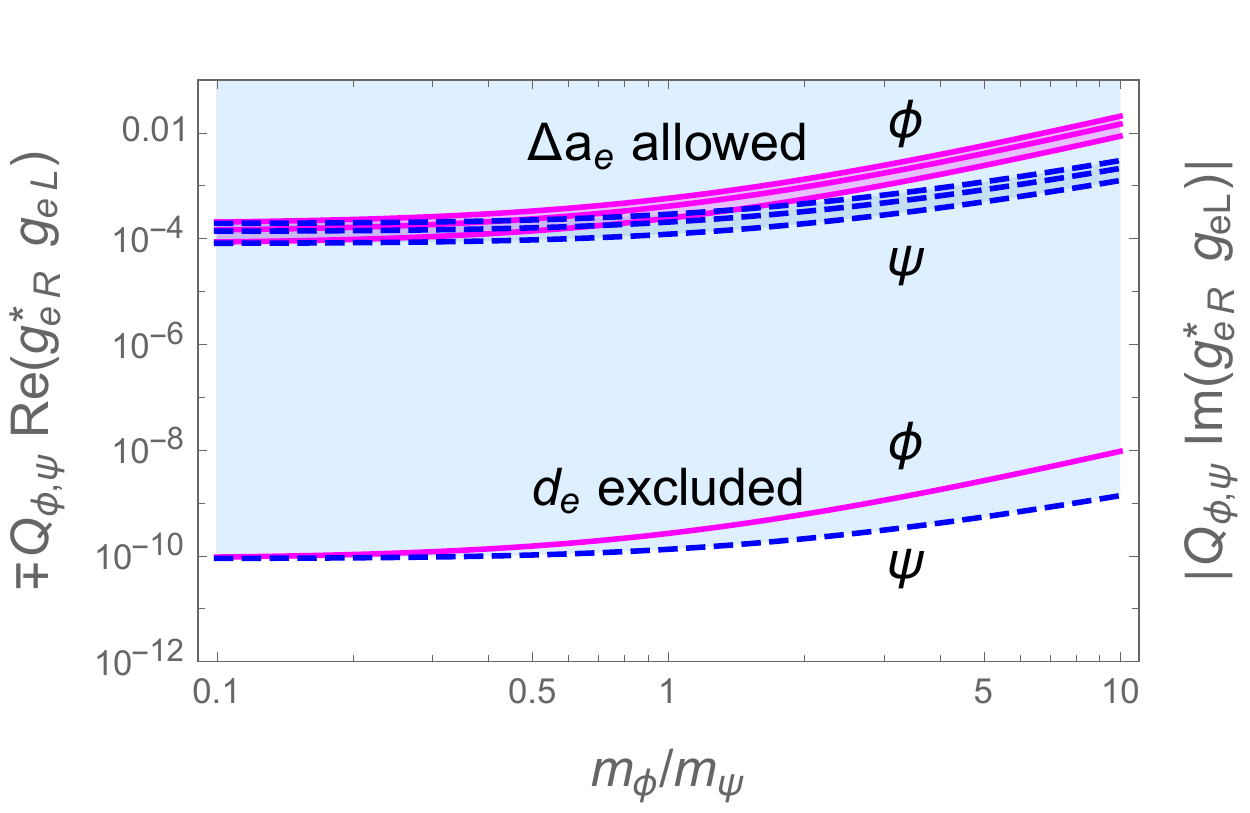}
}
\subfigure[]{
  \includegraphics[width=6.7cm]{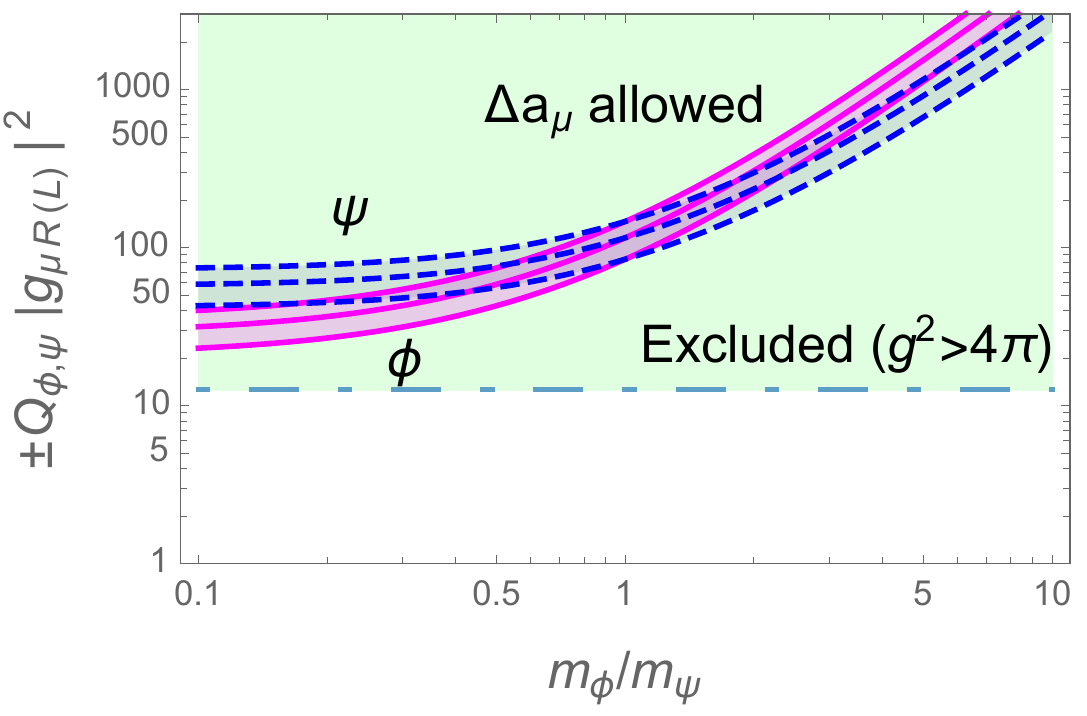}
}
\hspace{0.5cm}
\subfigure[]{
  \includegraphics[width=7cm]{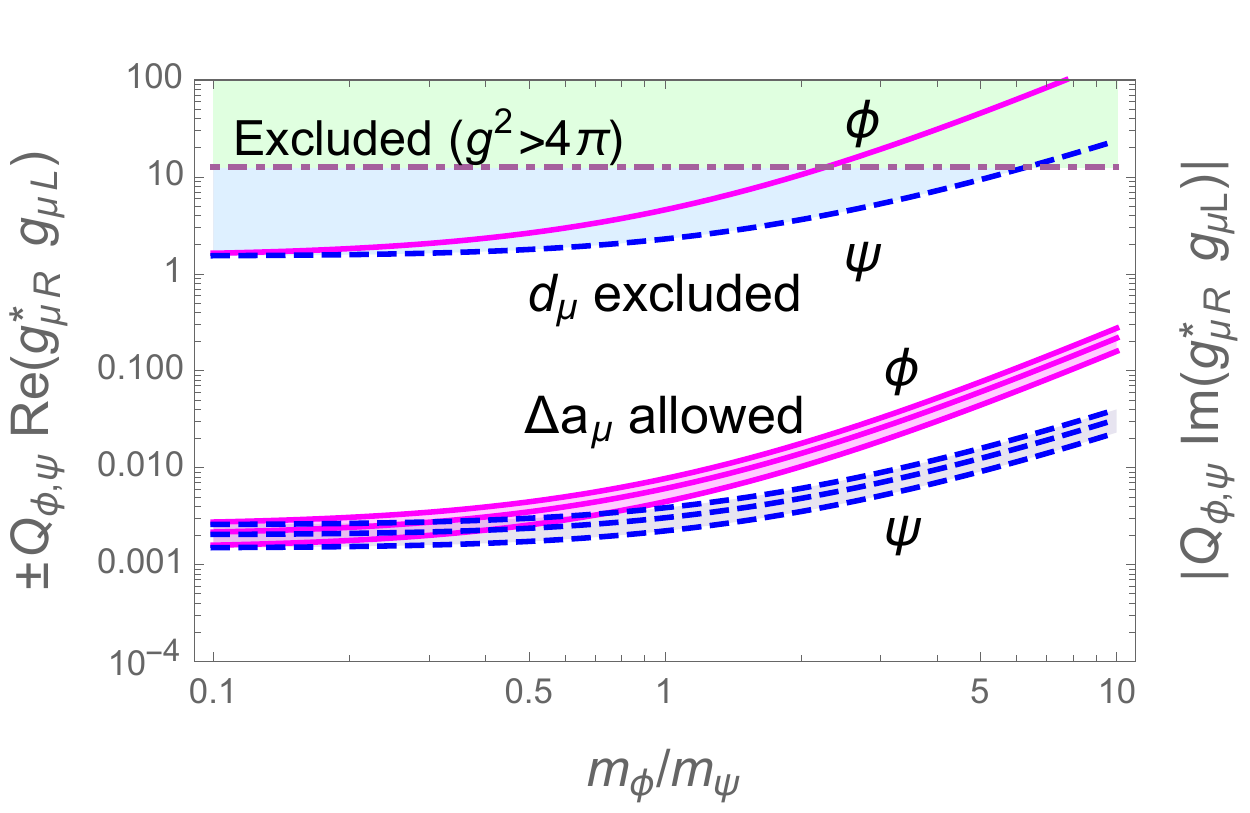}
}
\subfigure[]{
  \includegraphics[width=7cm]{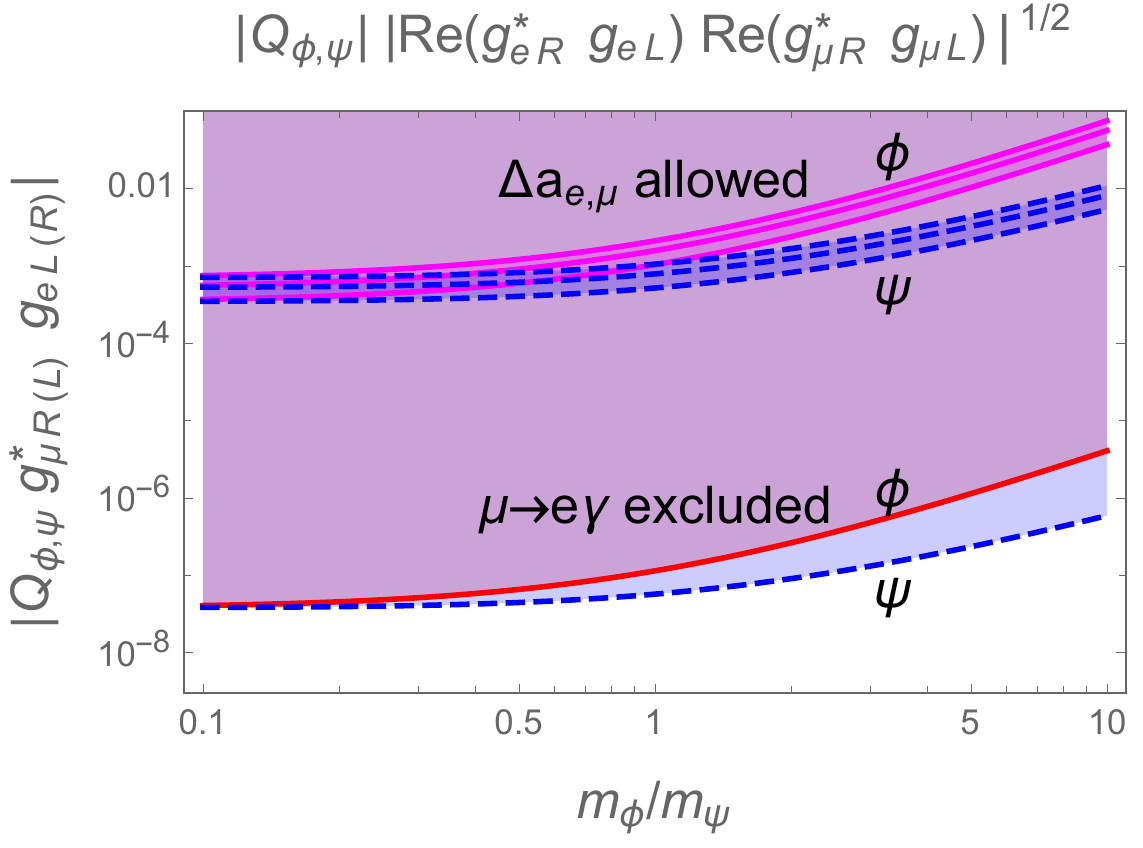}
}
\caption{We show in (a) and (b), allowed parameter space for $\mp Q_{\phi,\psi} |g_{e L(R)}|^2$, $\mp Q_{\phi,\psi}{\rm  Re}(g^*_{e R} g_{e L})$ and $|Q_{\phi,\psi} {\rm Im}(g^*_{e R} g_{eL})|$ constrained by $\Delta a_e$ and $d_e$,
in (c) and (d), allowed parameter space for $\pm Q_{\phi,\psi} |g_{\mu L(R)}|^2$, $\pm Q_{\phi,\psi}{\rm  Re}(g^*_{\mu R} g_{\mu L})$ and $|Q_{\phi,\psi} {\rm Im}(g^*_{\mu R} g_{\mu L})|$ constrained by $\Delta a_\mu$ and $d_\mu$,
in (e) allowed parameter space for $|Q_{\phi,\psi} g^*_{\mu R(L)} g_{e L(R)}|$ constrained by $\mu\to e\gamma$ and the parameter space on $|Q_{\phi,\psi}| |{\rm Re}(g^*_{eR} g_{eL}) {\rm Re}(g^*_{\mu R} g_{\mu L})|^{1/2}$ to produce $\Delta a_e$ and $\Delta a_\mu$.
These results are given for $m_\psi=500$~GeV.
For other $m_\psi$, plots in (a) and (c) scale with $(500\,{\rm GeV}/m_\psi)^2$, while plots in (b), (d) and (e) scale with $500\,{\rm GeV}/m_\psi$.
}
\label{fig:e muon g-2 1}
\end{figure}

\begin{figure}[ht!]
\centering
\subfigure[]{
  \includegraphics[width=6.5cm]{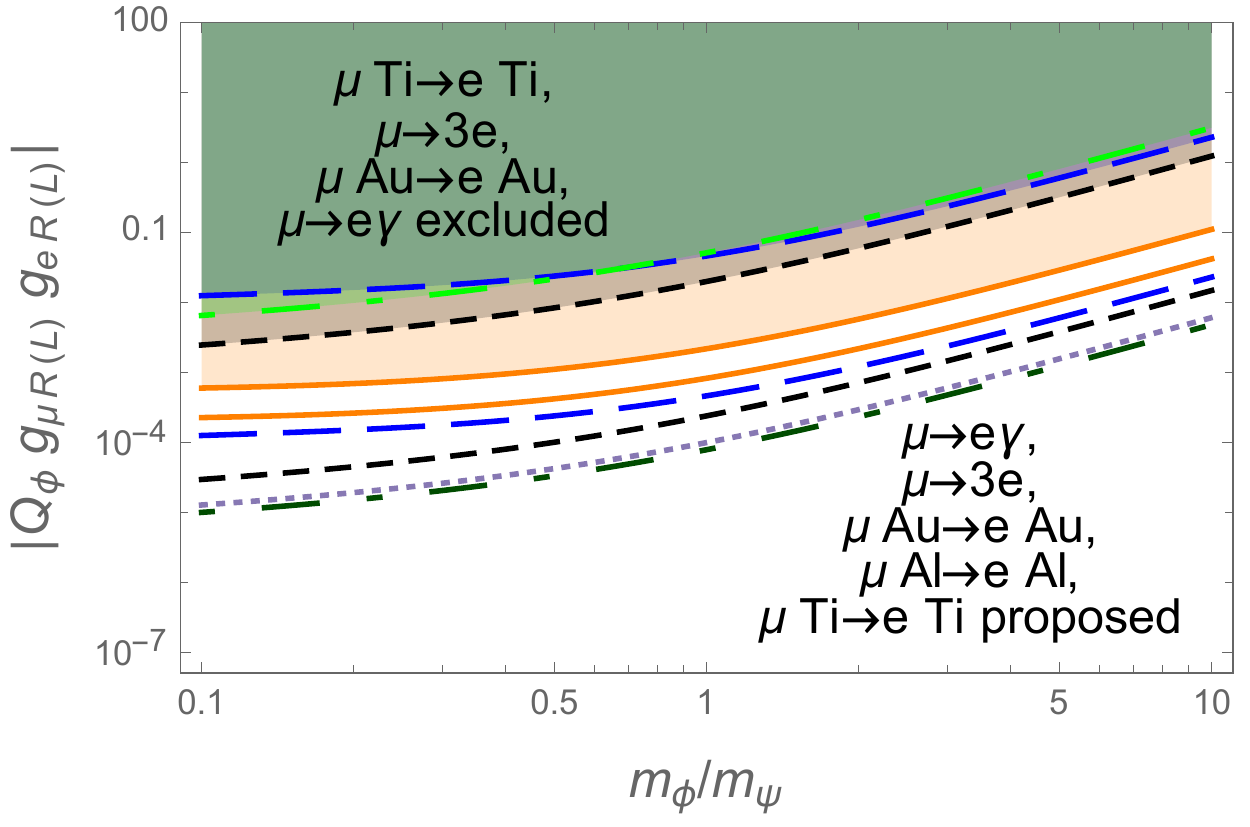}
}
\hspace{0.5cm}
\subfigure[]{
  \includegraphics[width=6.5cm]{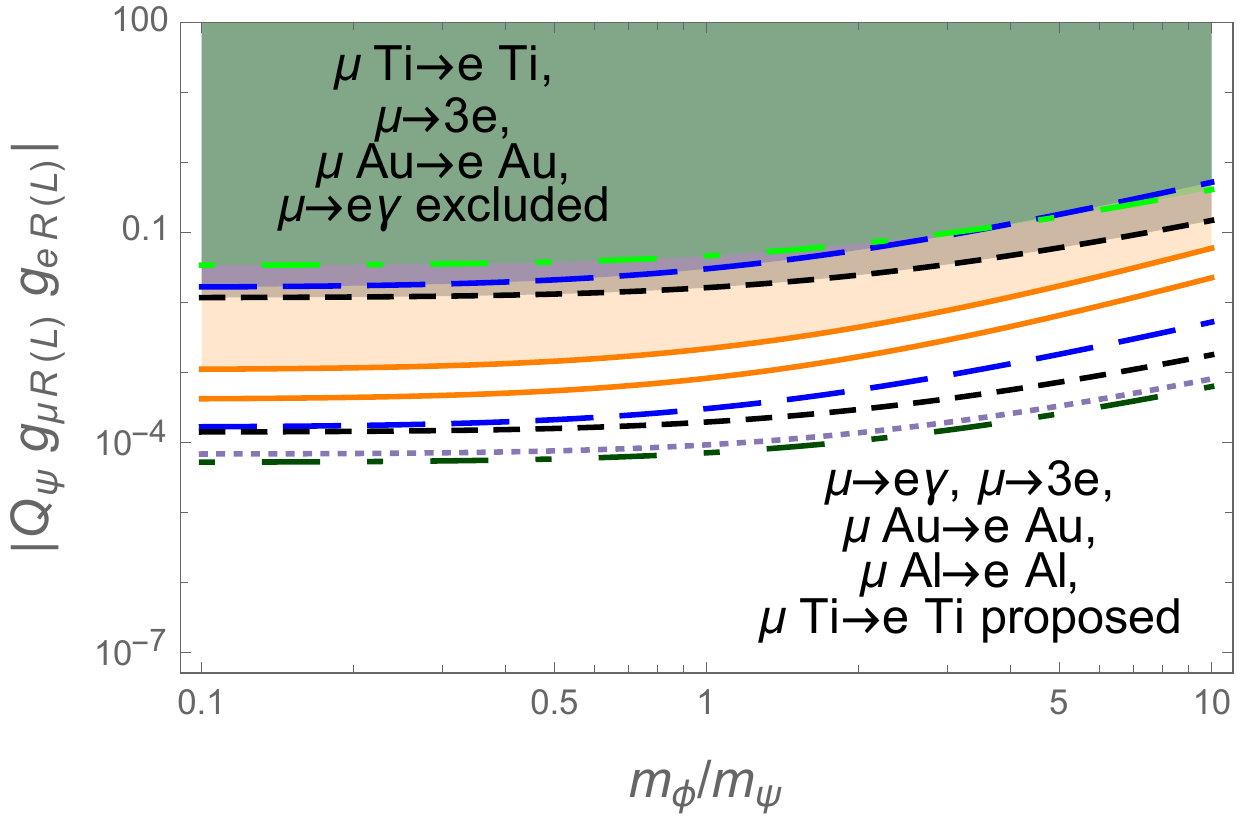}
}
\subfigure[]{
  \includegraphics[width=6.5cm]{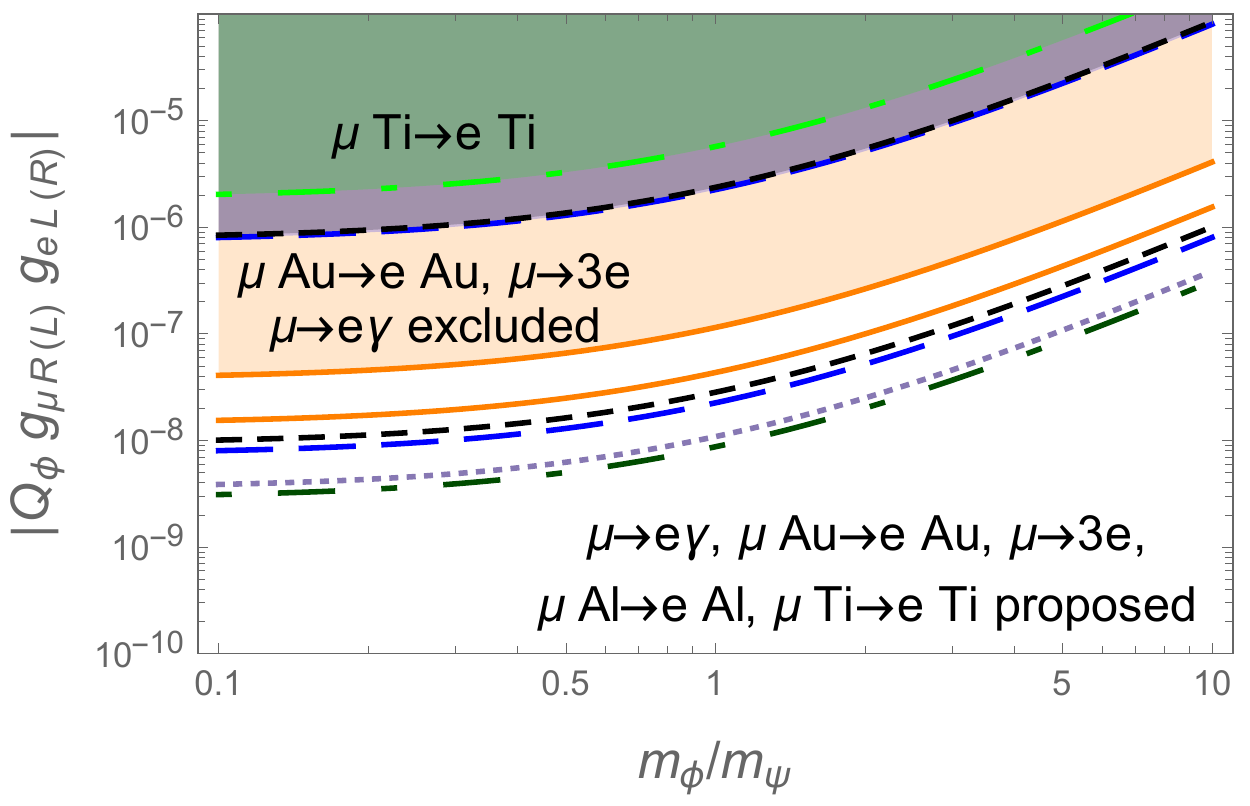}
}
\hspace{0.5cm}
\subfigure[]{
  \includegraphics[width=6.5cm]{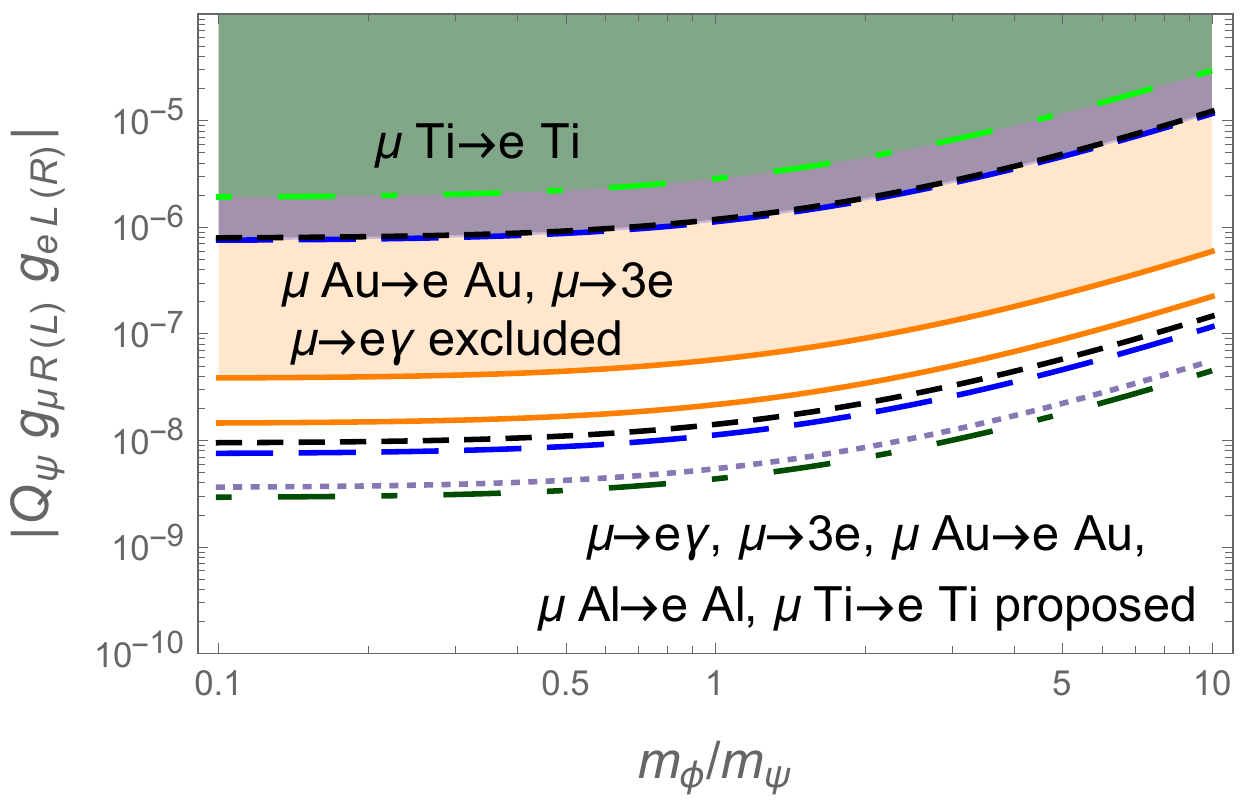}
}
\subfigure[]{
  \includegraphics[width=6.5cm]{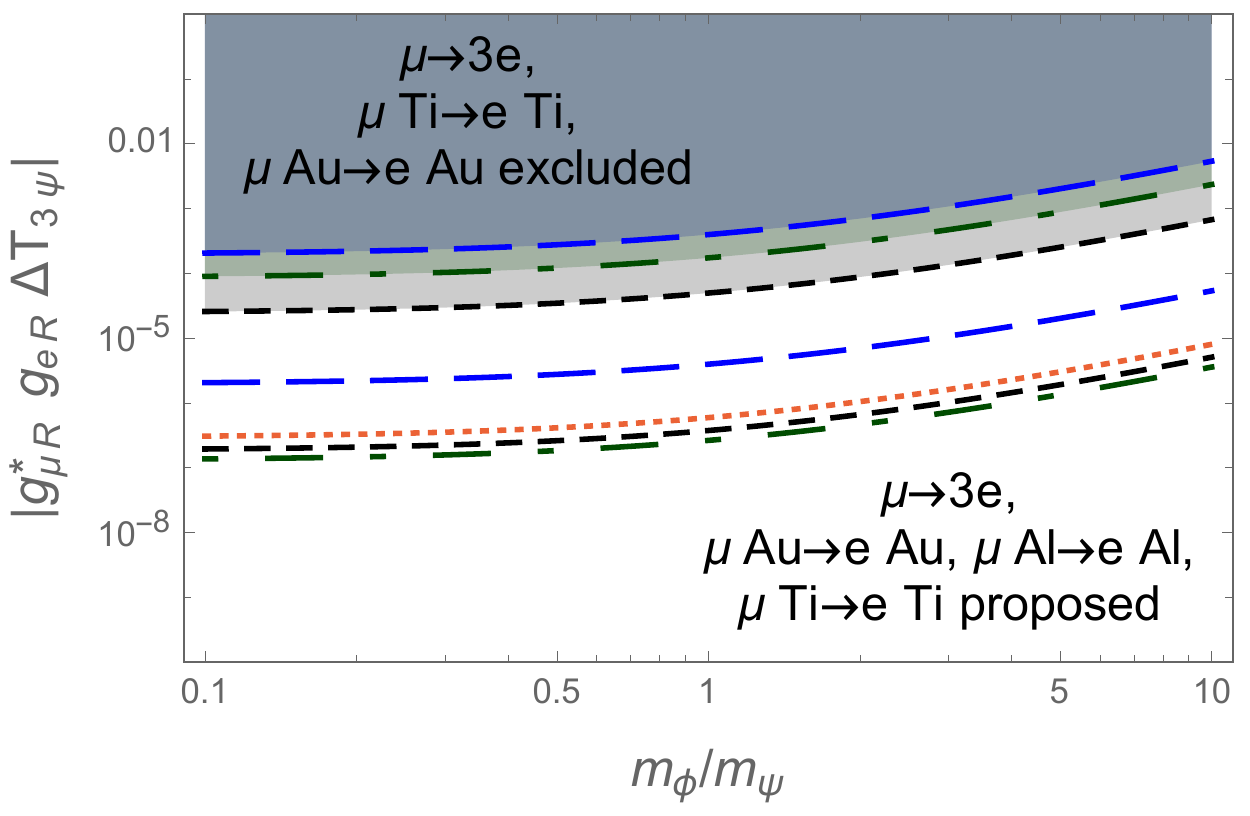}
}
\hspace{0.5cm}
\subfigure[]{
  \includegraphics[width=6.5cm]{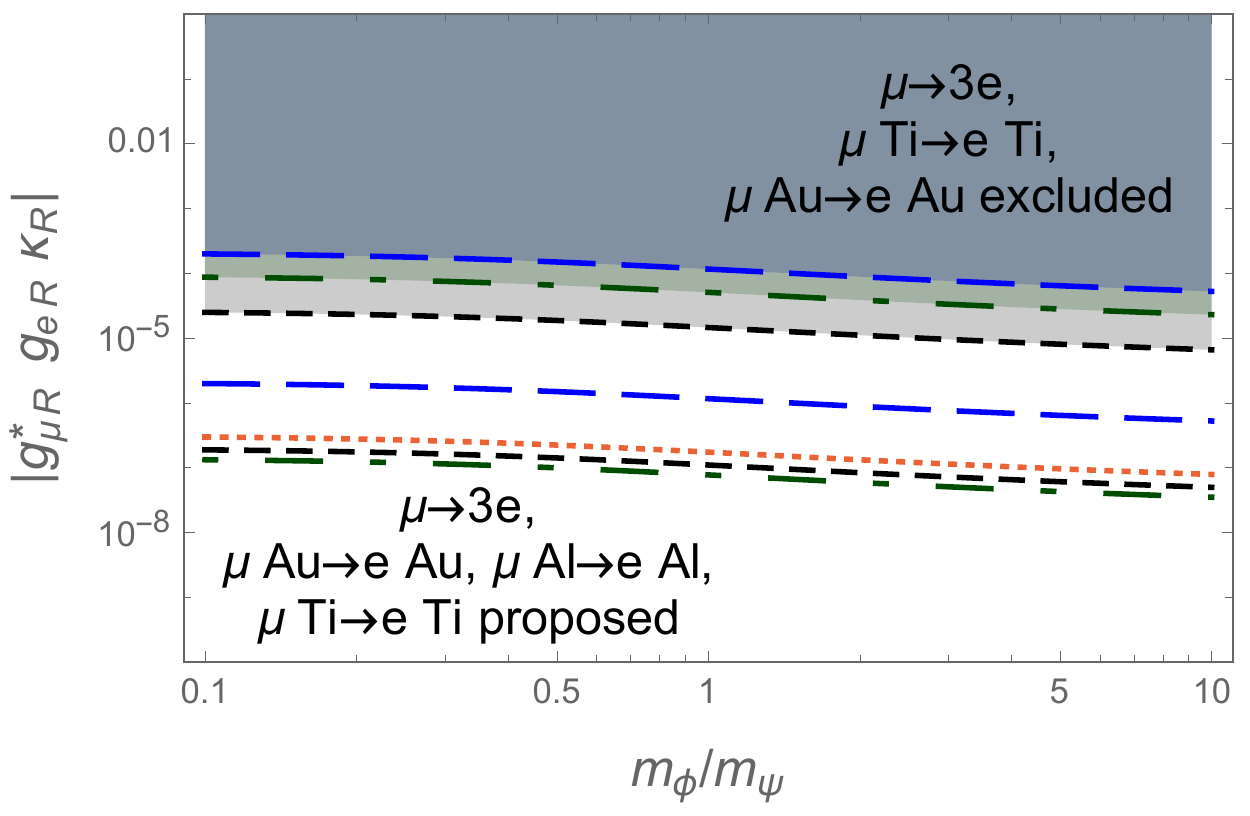}
}
\subfigure[]{
  \includegraphics[width=6.5cm]{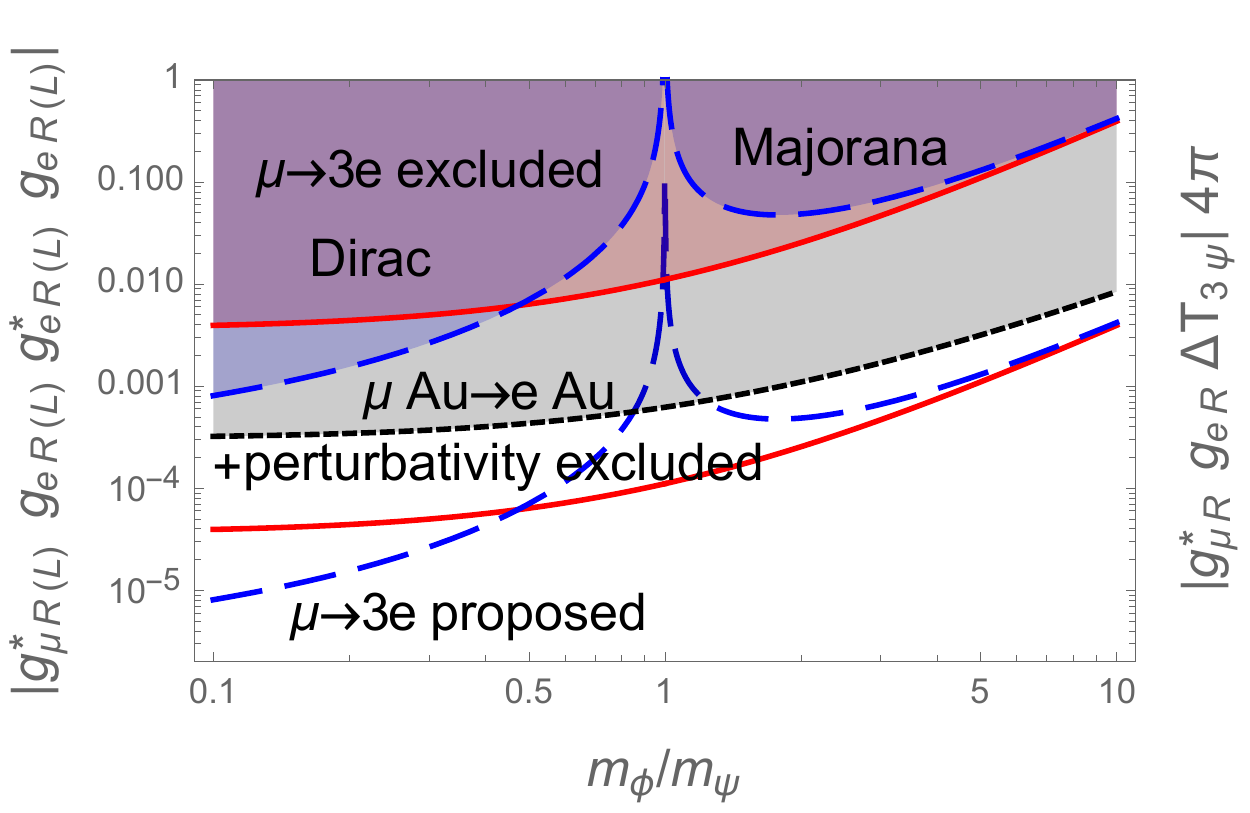}
}
\hspace{0.5cm}
\subfigure[]{
  \includegraphics[width=6.5cm]{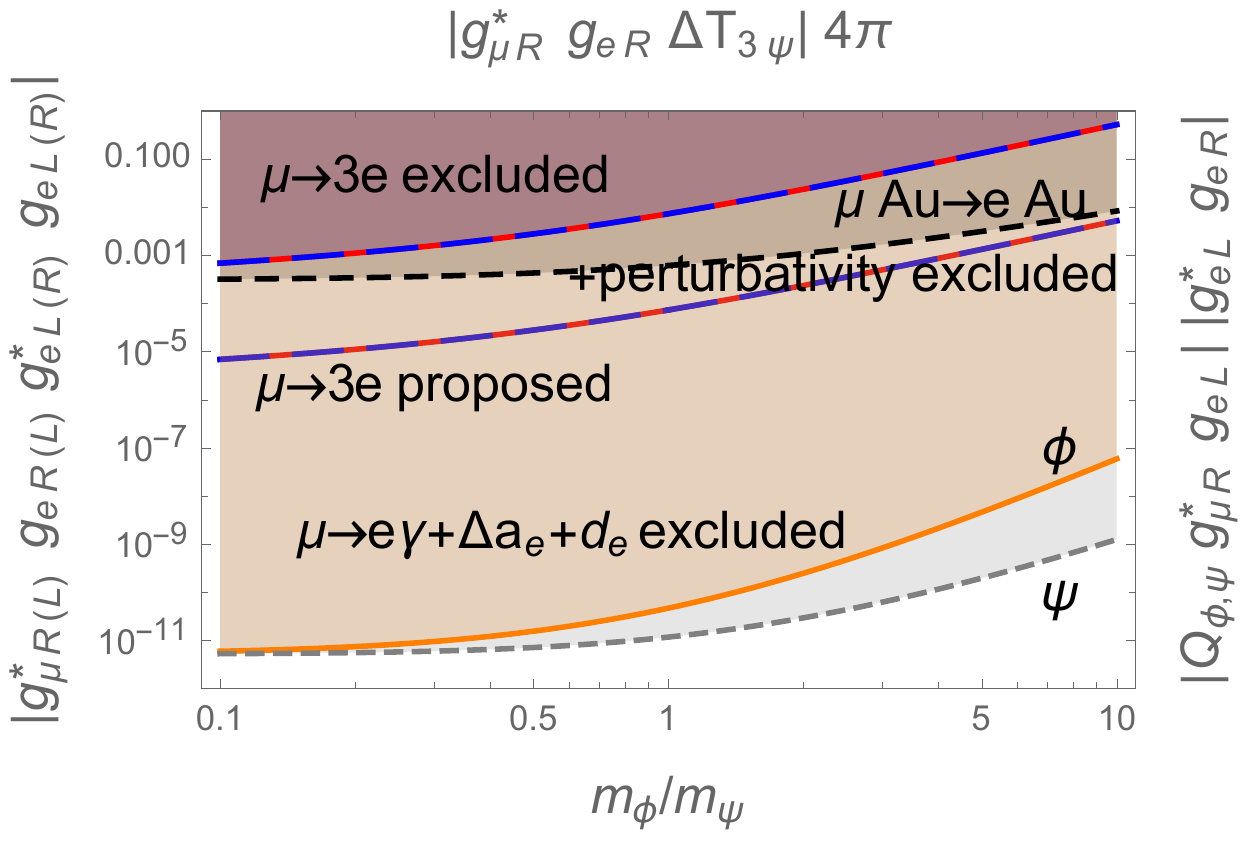}
}
\caption{Parameter space excluded or projected by various experimental bounds or expected sensitivities on $\mu\to e$ LFV processes from photonic penguin, $Z$-penguin and box contributions.} 
\label{fig:mu2eLFV}
\end{figure}

\begin{figure}[ht!]
\centering
\subfigure[]{
  \includegraphics[width=6.5cm]{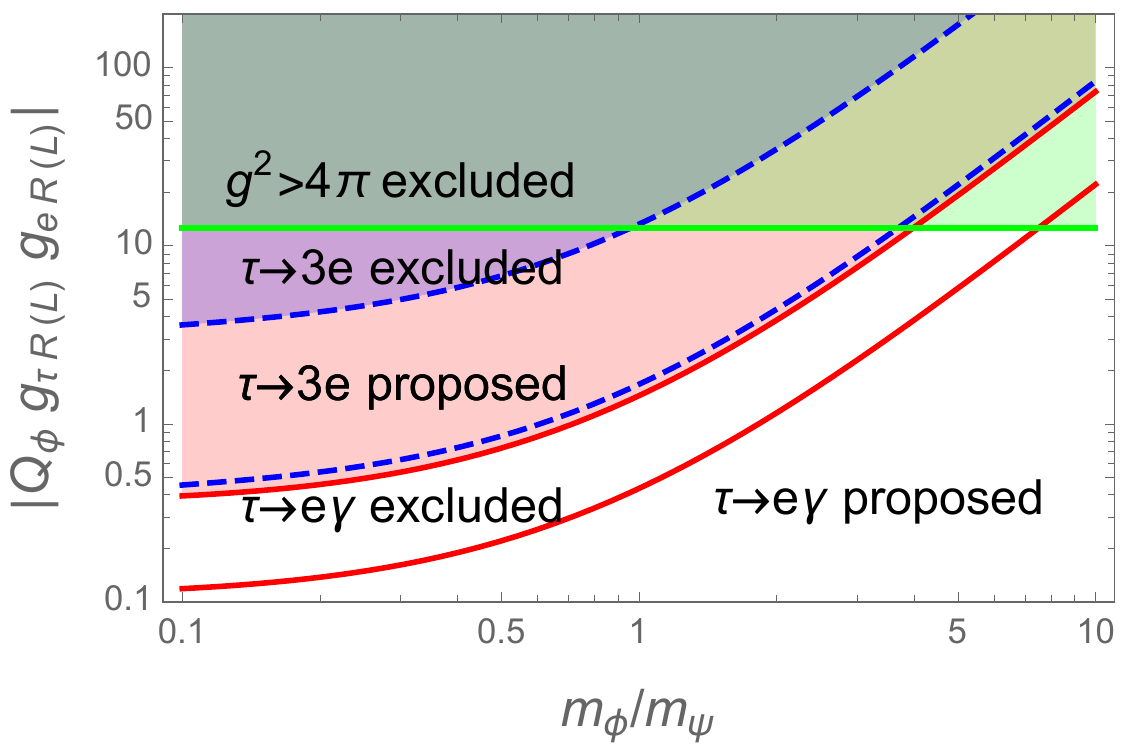}
}
\hspace{0.5cm}
\subfigure[]{
  \includegraphics[width=6.5cm]{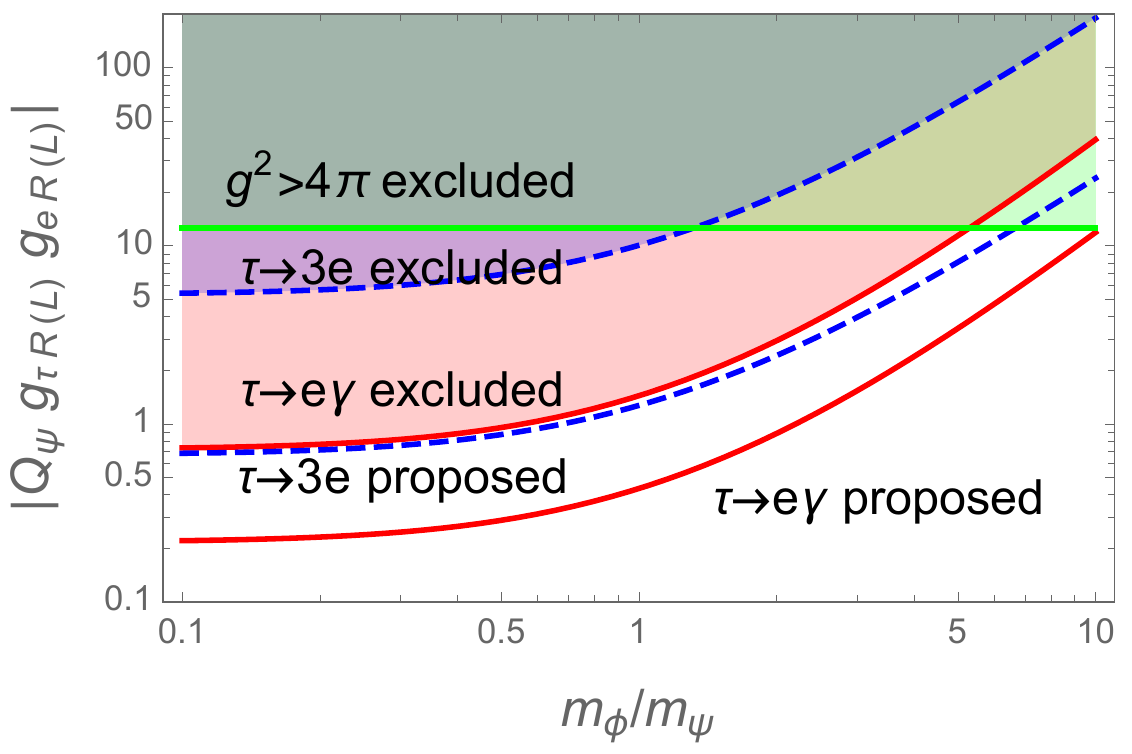}
}
\subfigure[]{
  \includegraphics[width=6.5cm]{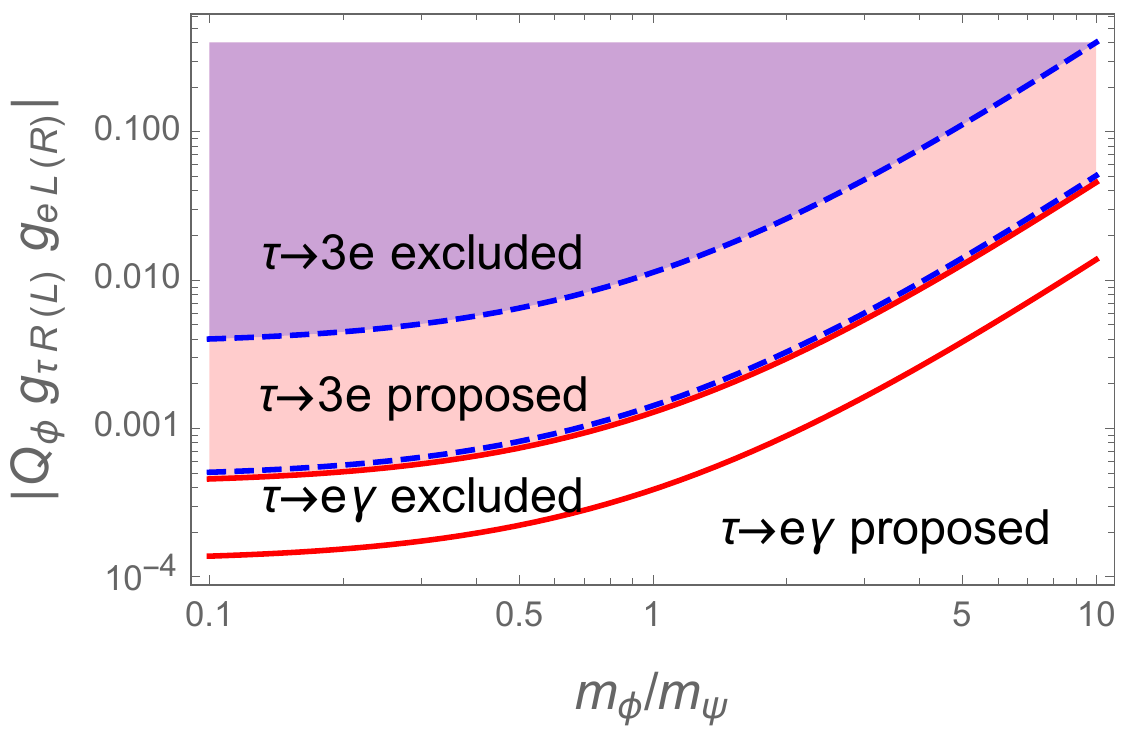}
}
\hspace{0.5cm}
\subfigure[]{
  \includegraphics[width=6.5cm]{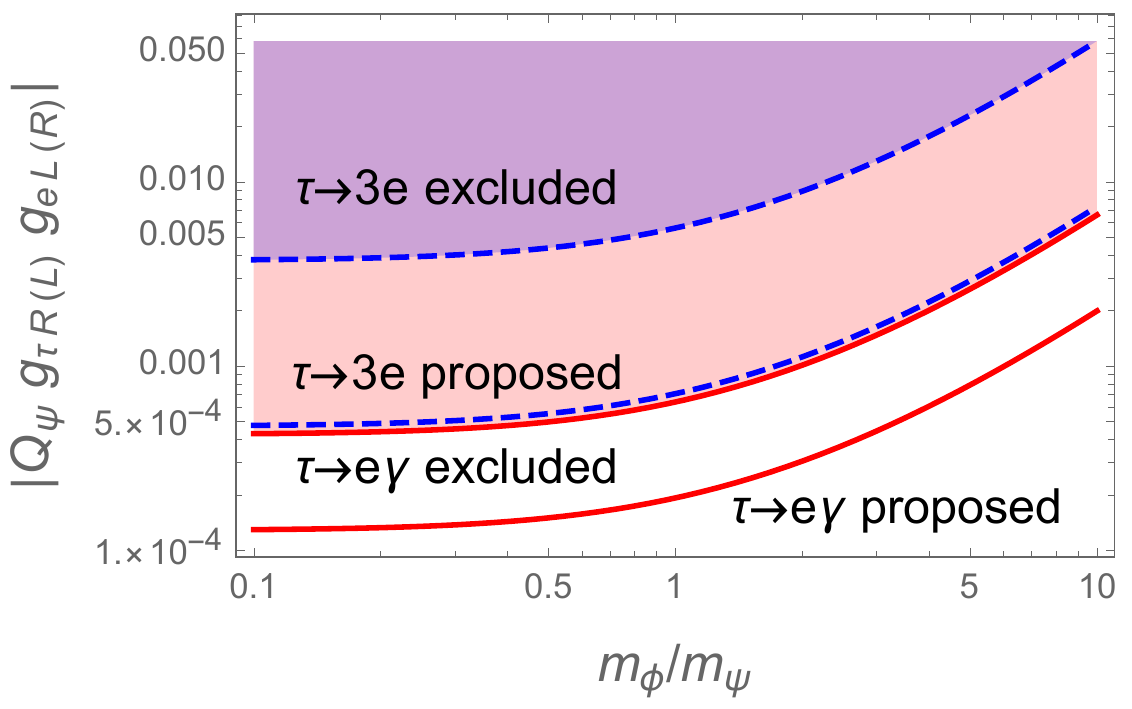}
}
\subfigure[]{
  \includegraphics[width=6.5cm]{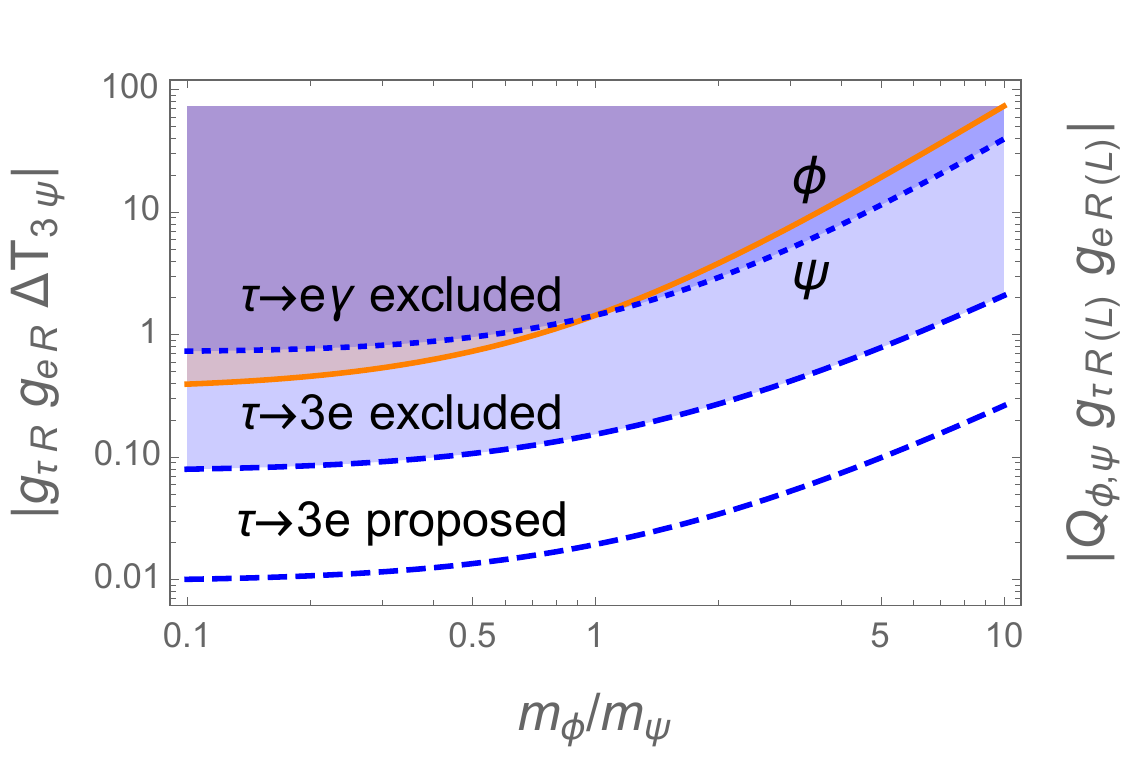}
}
\hspace{0.5cm}
\subfigure[]{
  \includegraphics[width=6.5cm]{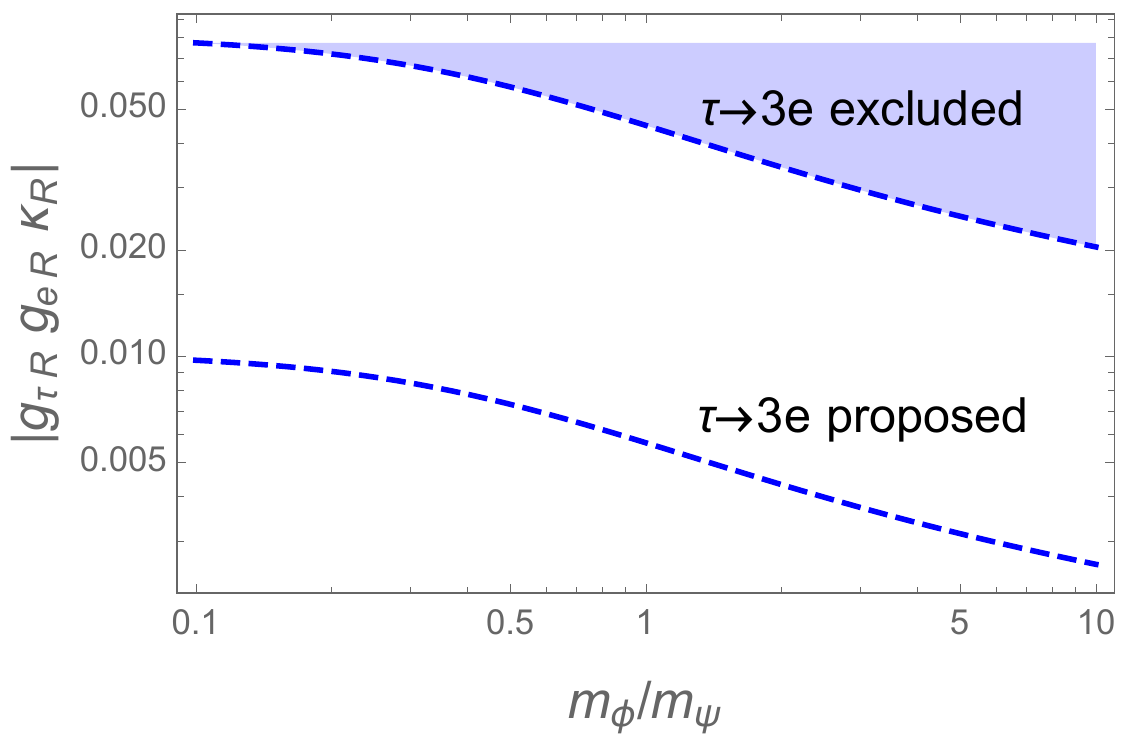}
}
\subfigure[]{
  \includegraphics[width=6.5cm]{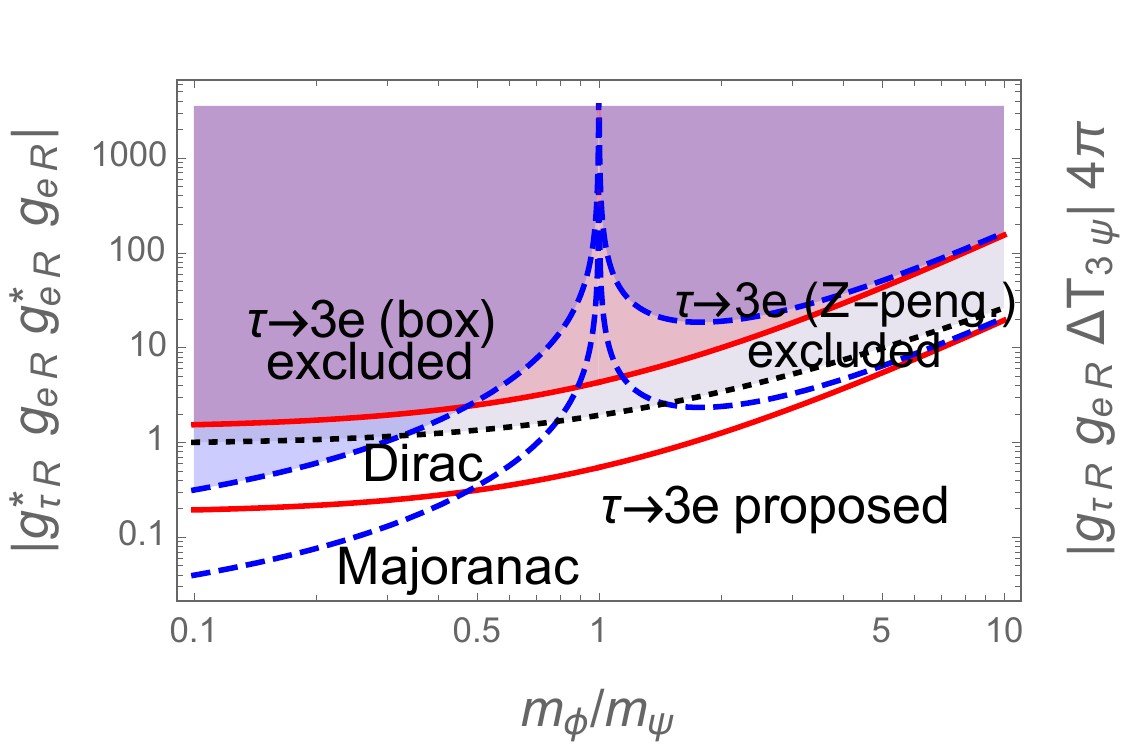}
}
\hspace{0.5cm}
\subfigure[]{
  \includegraphics[width=6.5cm]{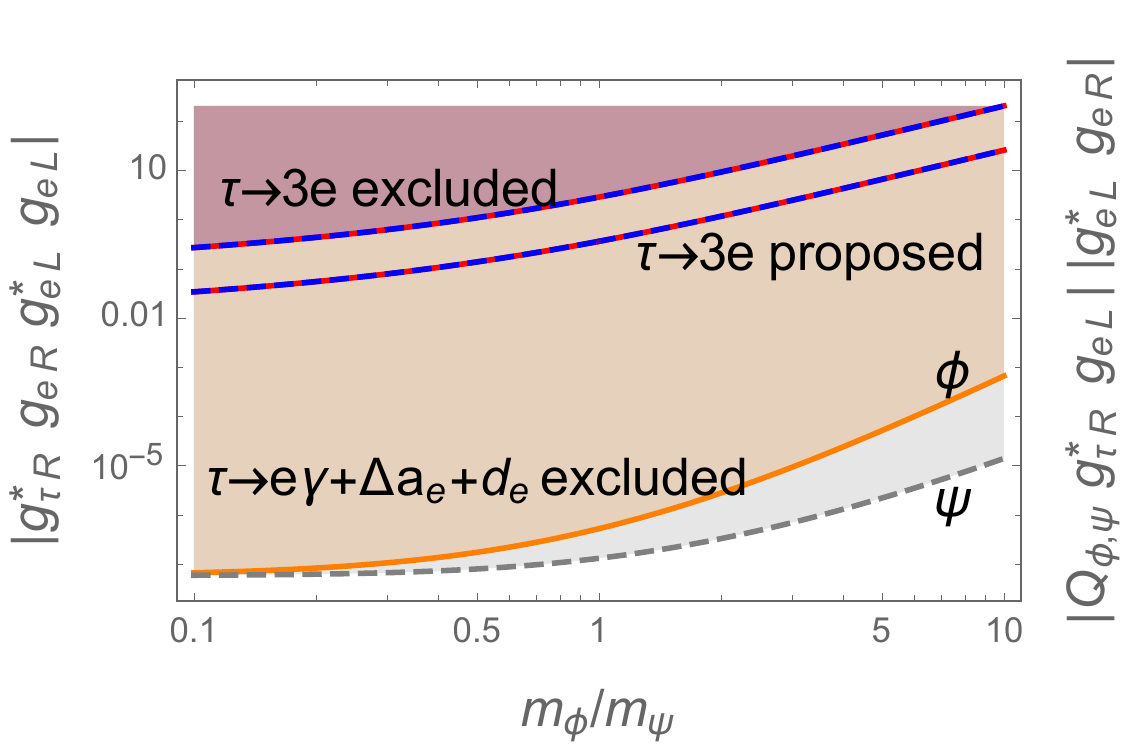}
}
\caption{
Same as Fig.~\ref{fig:mu2eLFV}, but for $\tau\to e$ transition.}
\label{fig:tau2eLFV}
\end{figure}

\begin{figure}[ht!]
\centering
\subfigure[]{
  \includegraphics[width=6.5cm]{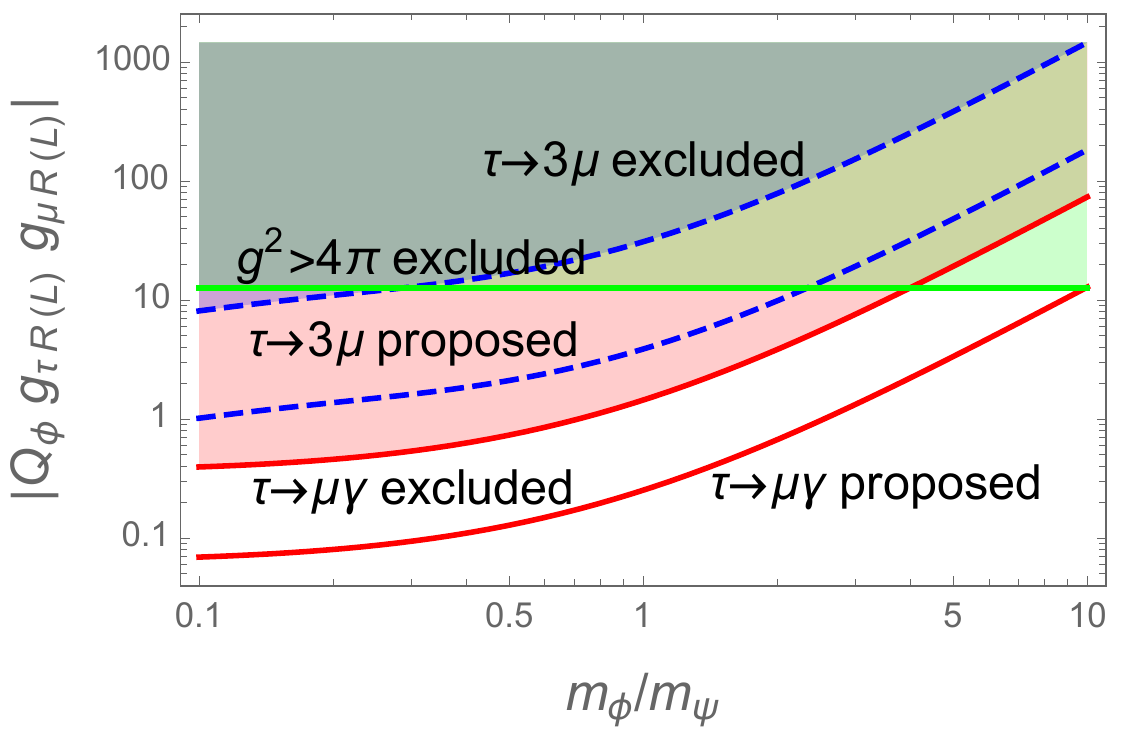}
}
\hspace{0.5cm}
\subfigure[]{
  \includegraphics[width=6.5cm]{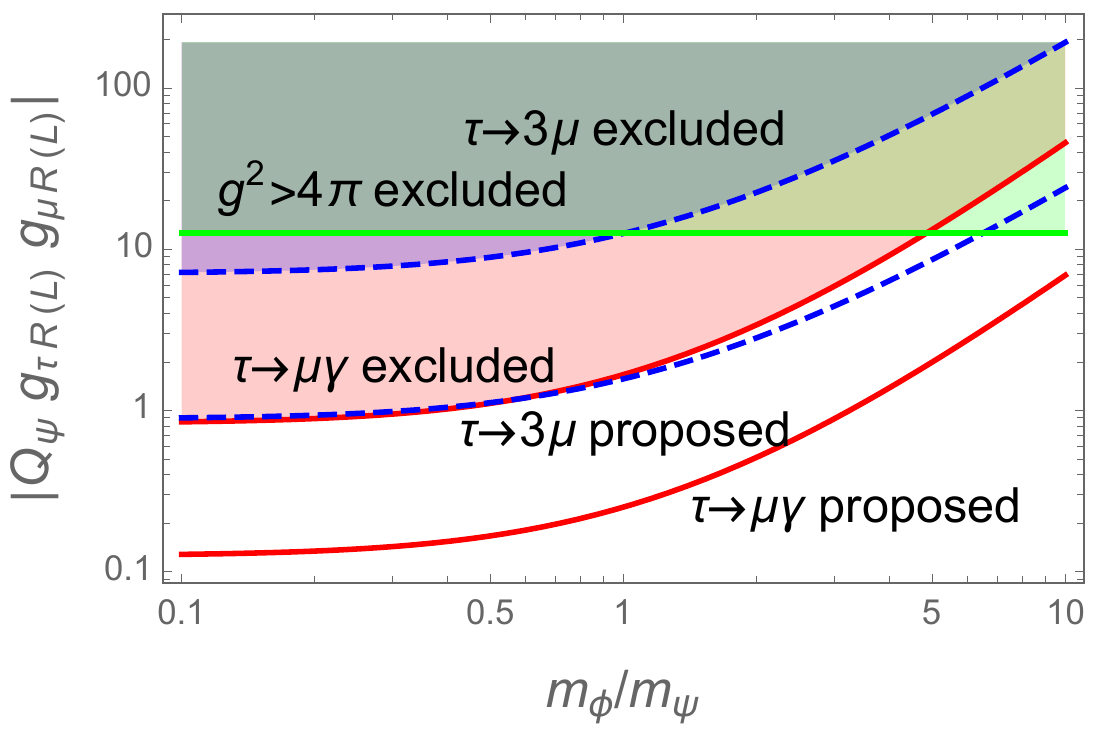}
}
\subfigure[]{
  \includegraphics[width=6.5cm]{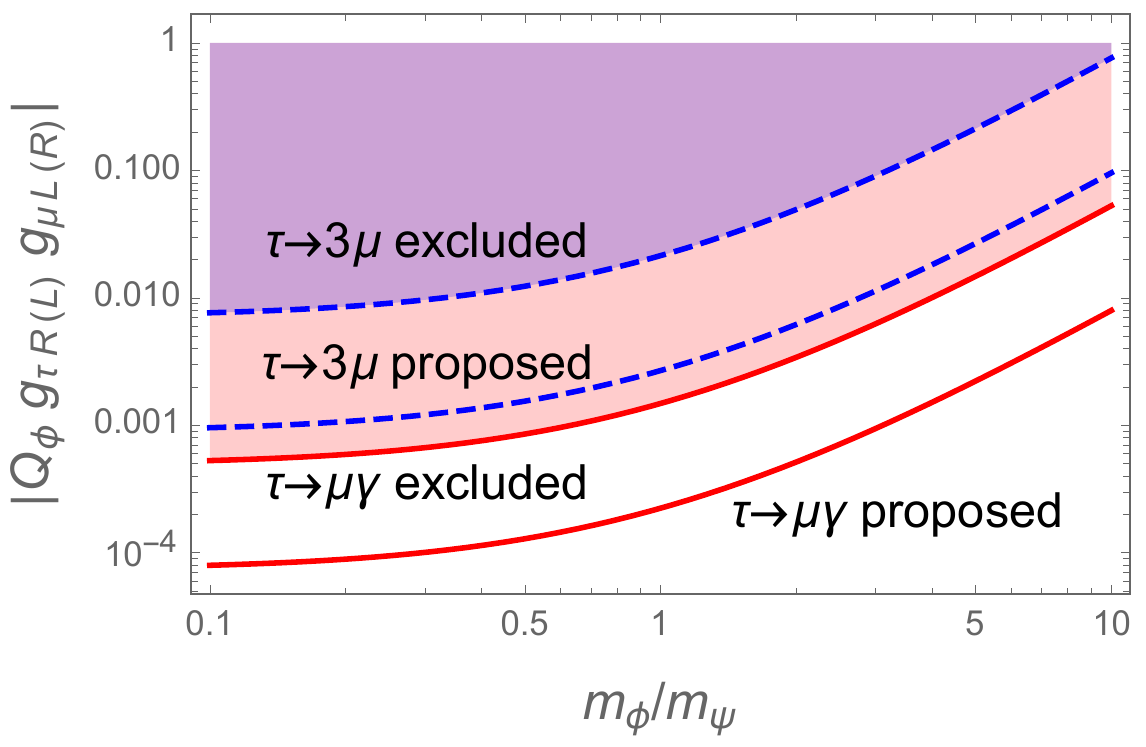}
}
\hspace{0.5cm}
\subfigure[]{
  \includegraphics[width=6.5cm]{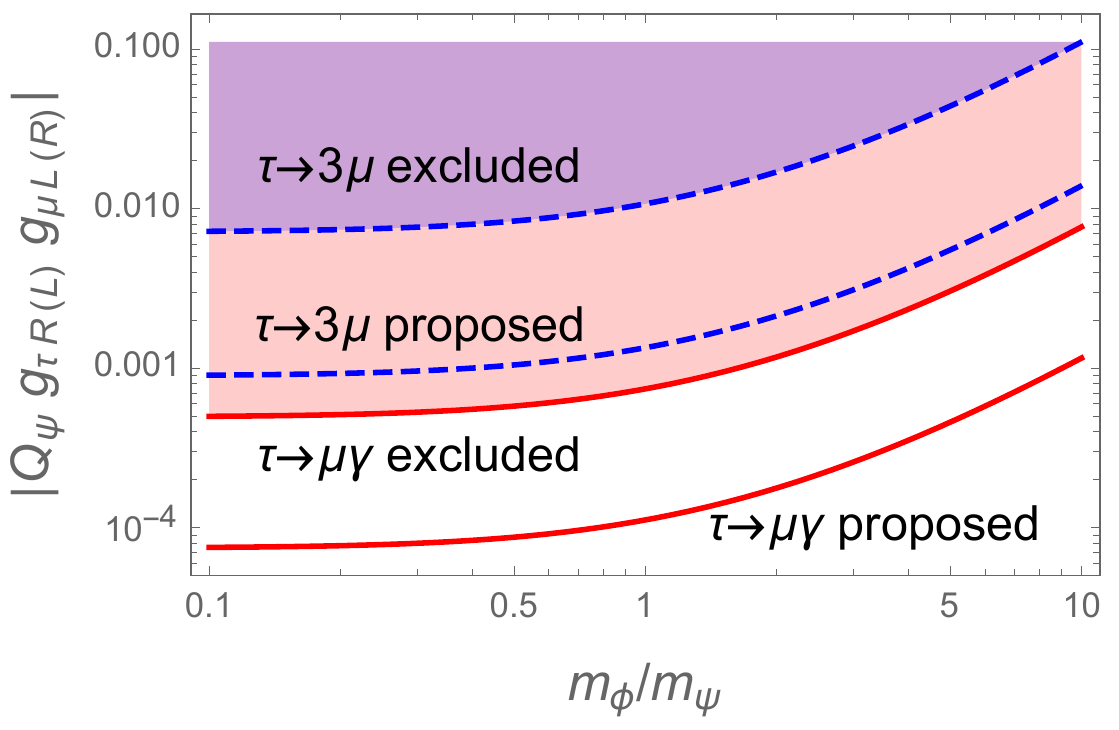}
}
\subfigure[]{
  \includegraphics[width=6.5cm]{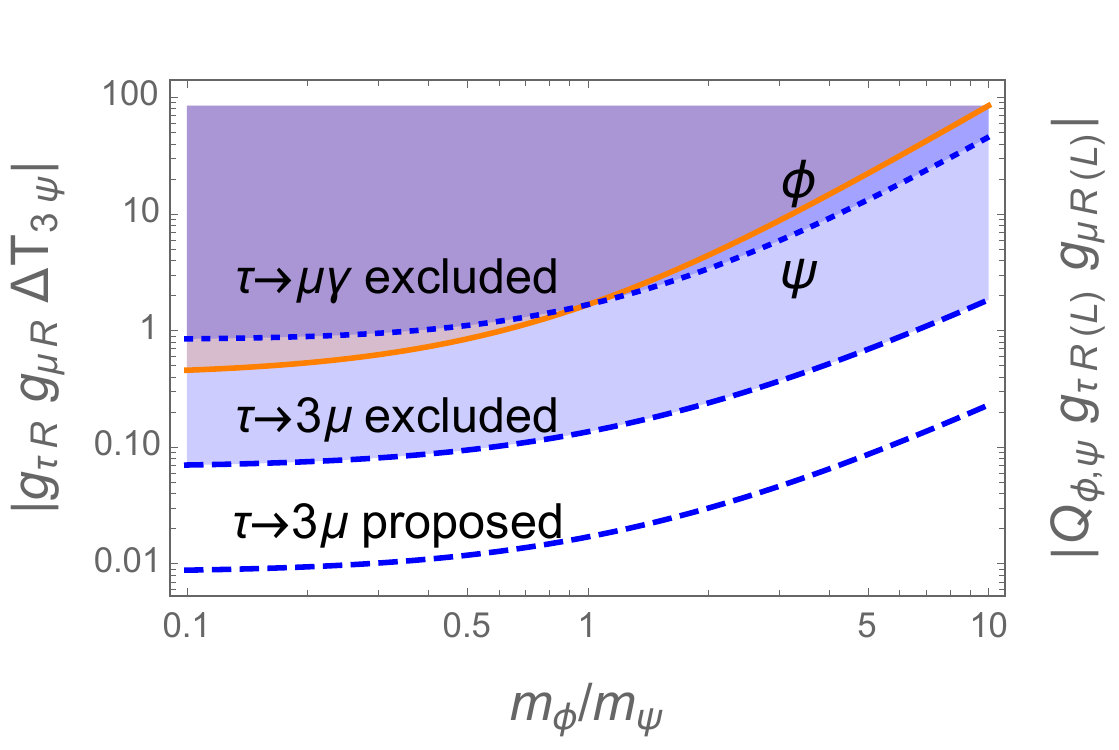}
}
\hspace{0.5cm}
\subfigure[]{
  \includegraphics[width=6.5cm]{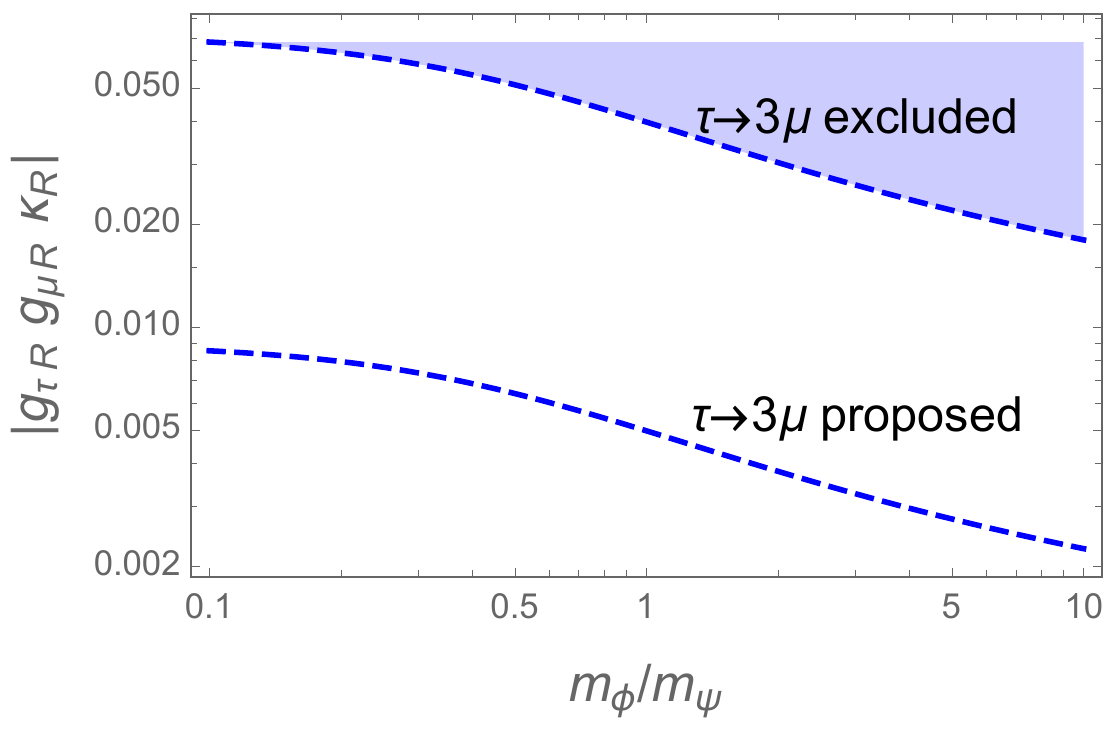}
}
\subfigure[]{
  \includegraphics[width=6.5cm]{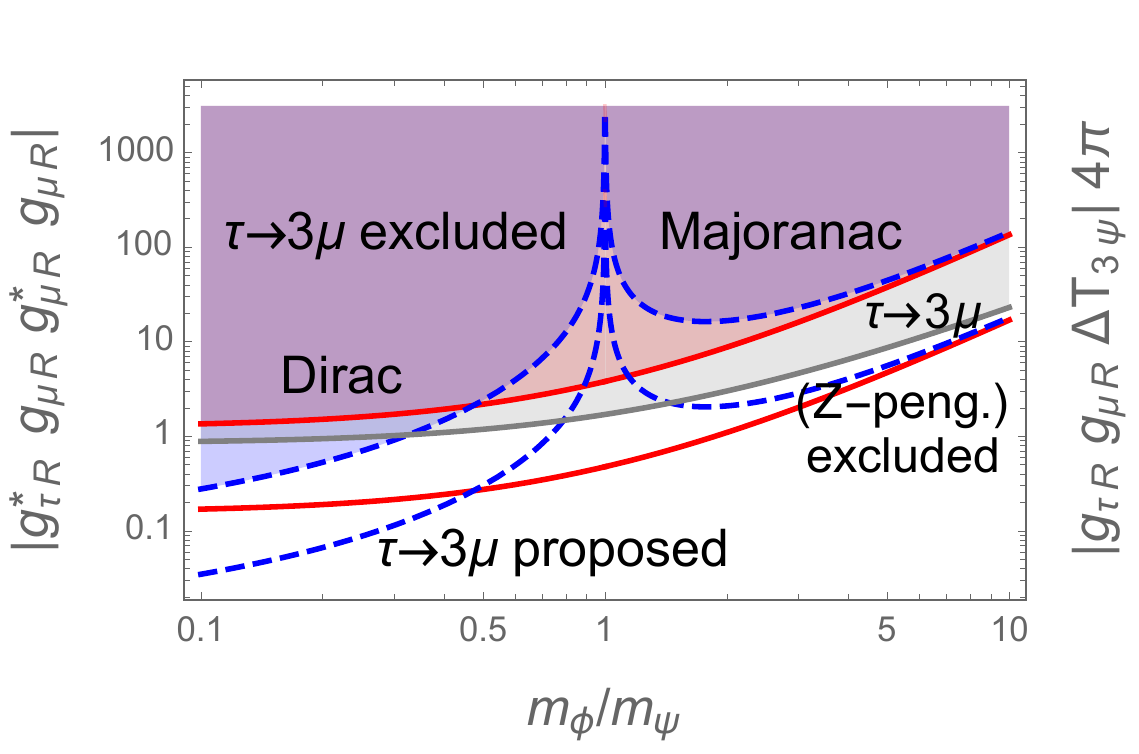}
}
\hspace{0.5cm}
\subfigure[]{
  \includegraphics[width=6.5cm]{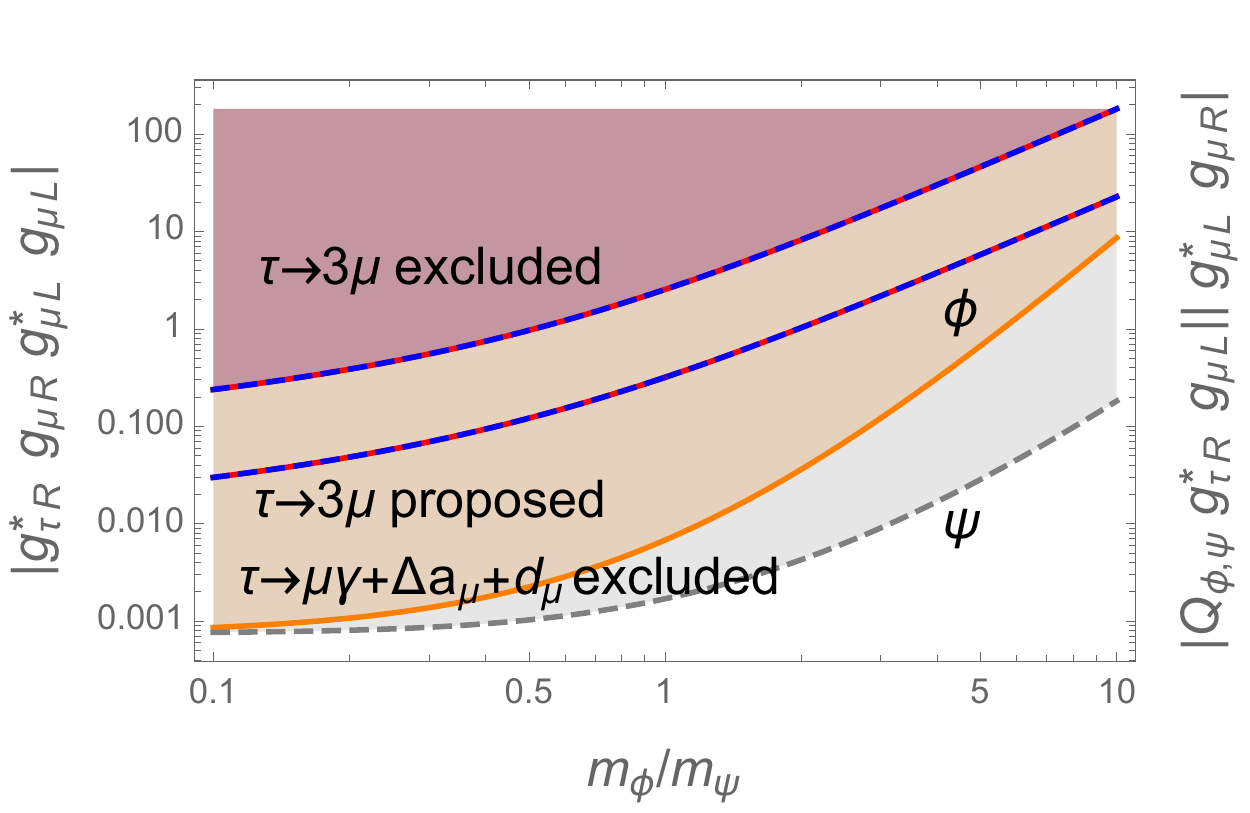}
}
\caption{Same as Fig.~\ref{fig:mu2eLFV}, but for $\tau\to \mu$ transition.}
\label{fig:tau2muLFV}
\end{figure}

\begin{figure}[ht!]
\centering
\subfigure[]{
  \includegraphics[width=6.5cm]{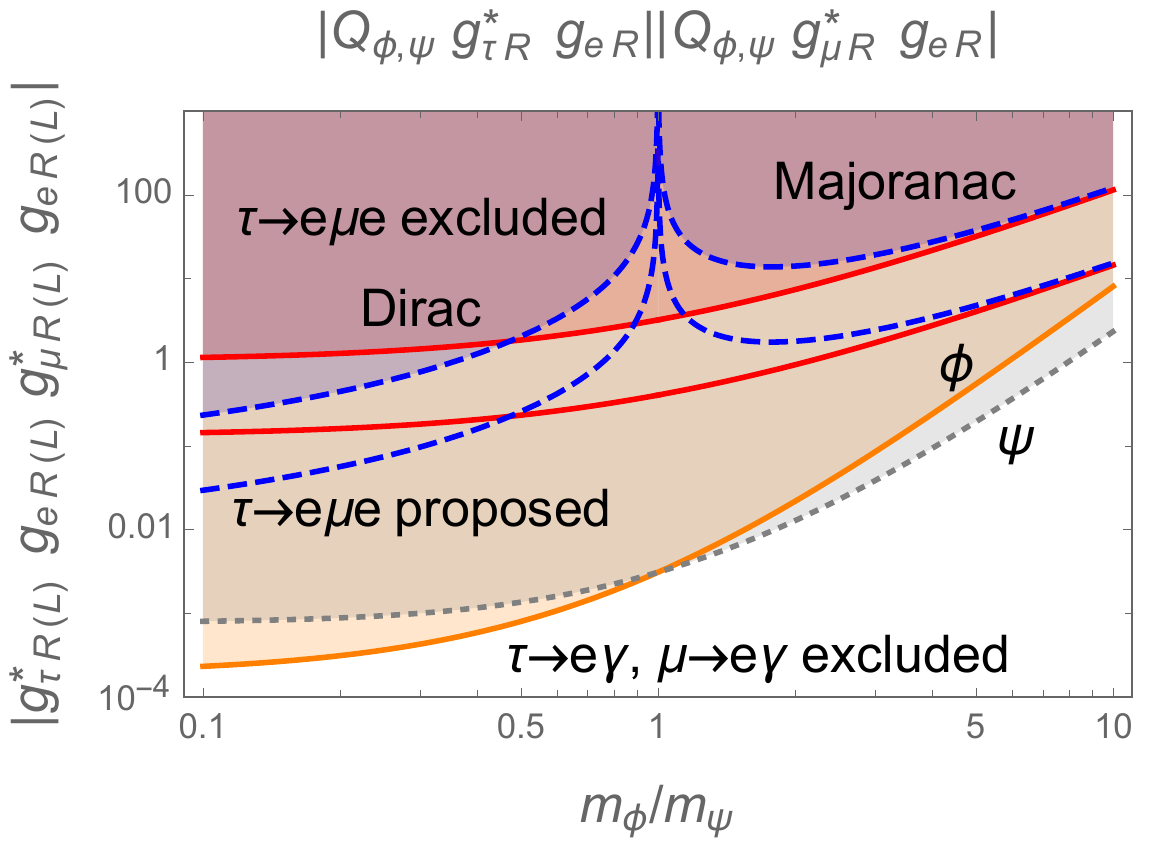}
}
\hspace{0.5cm}
\subfigure[]{
  \includegraphics[width=6.5cm]{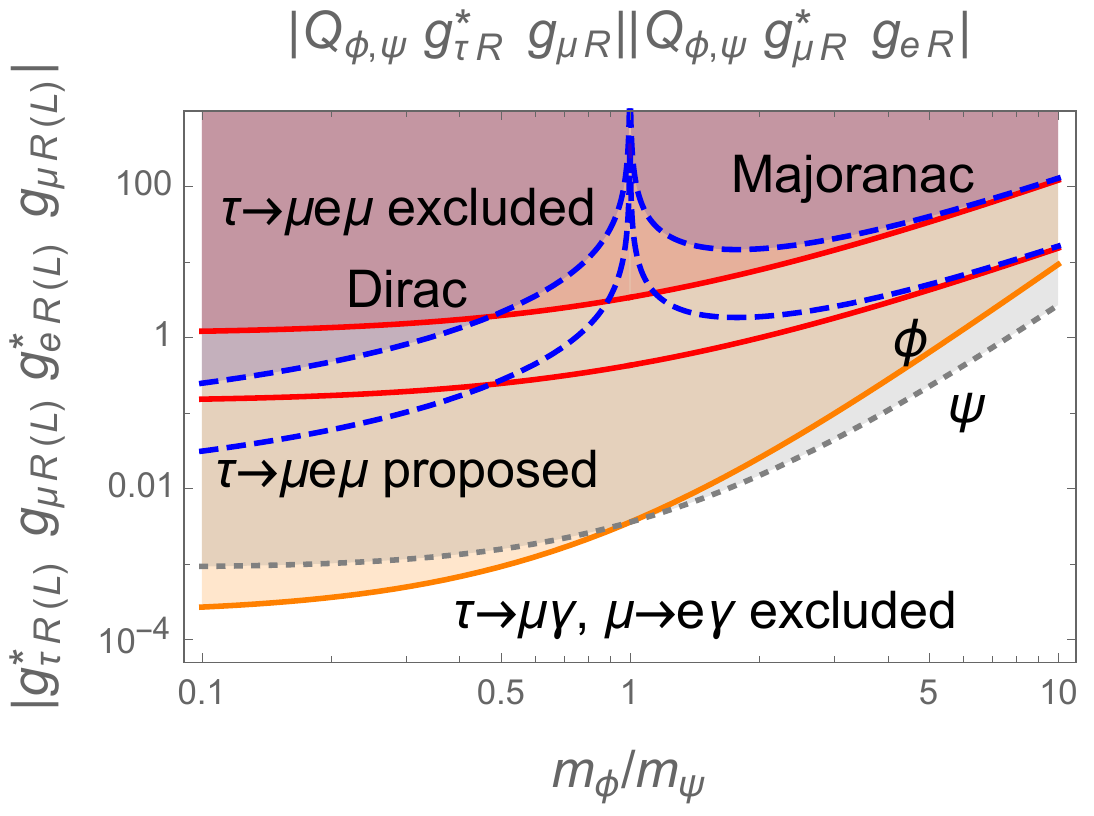}
}
\subfigure[]{
  \includegraphics[width=6.5cm]{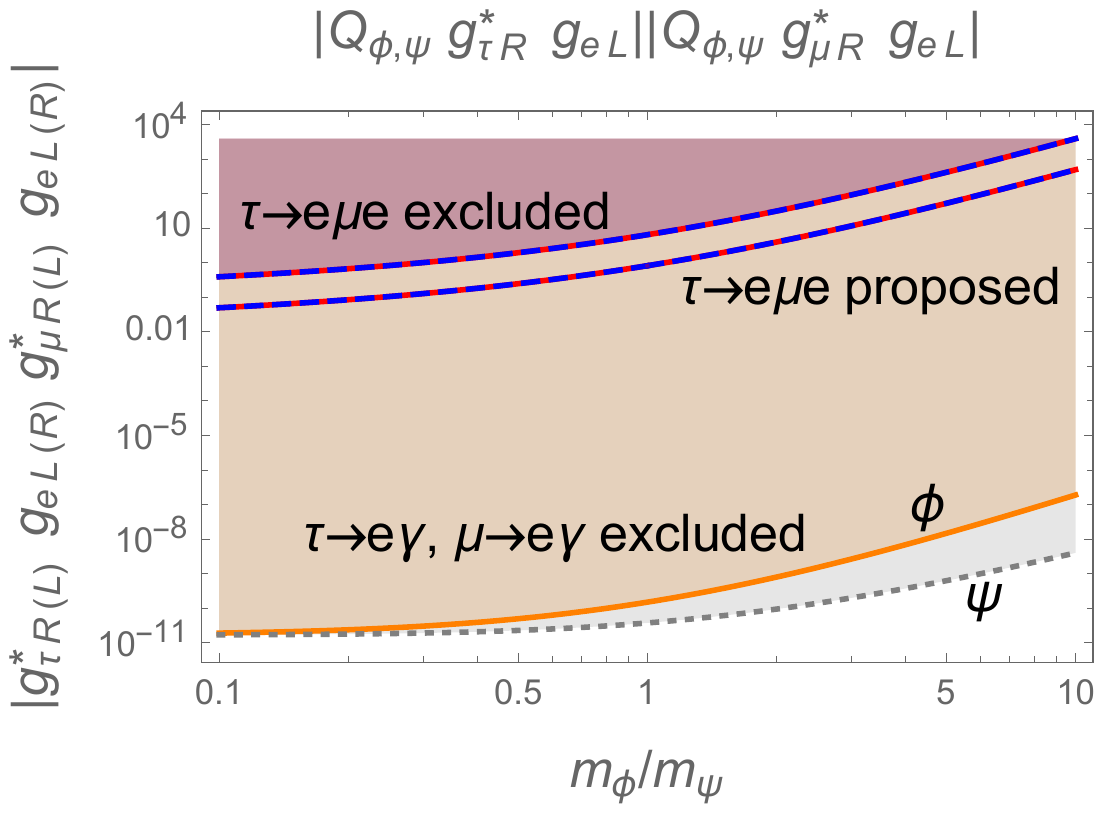}
}
\hspace{0.5cm}
\subfigure[]{
  \includegraphics[width=6.5cm]{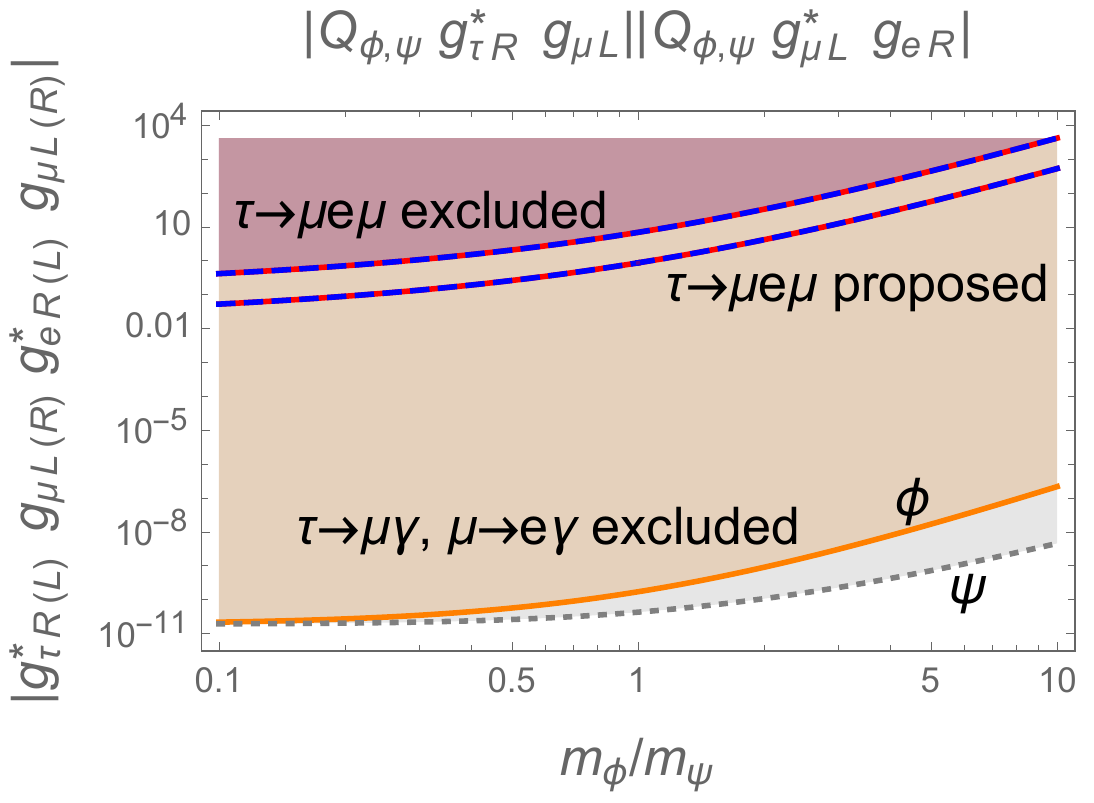}
}
\subfigure[]{
  \includegraphics[width=6.5cm]{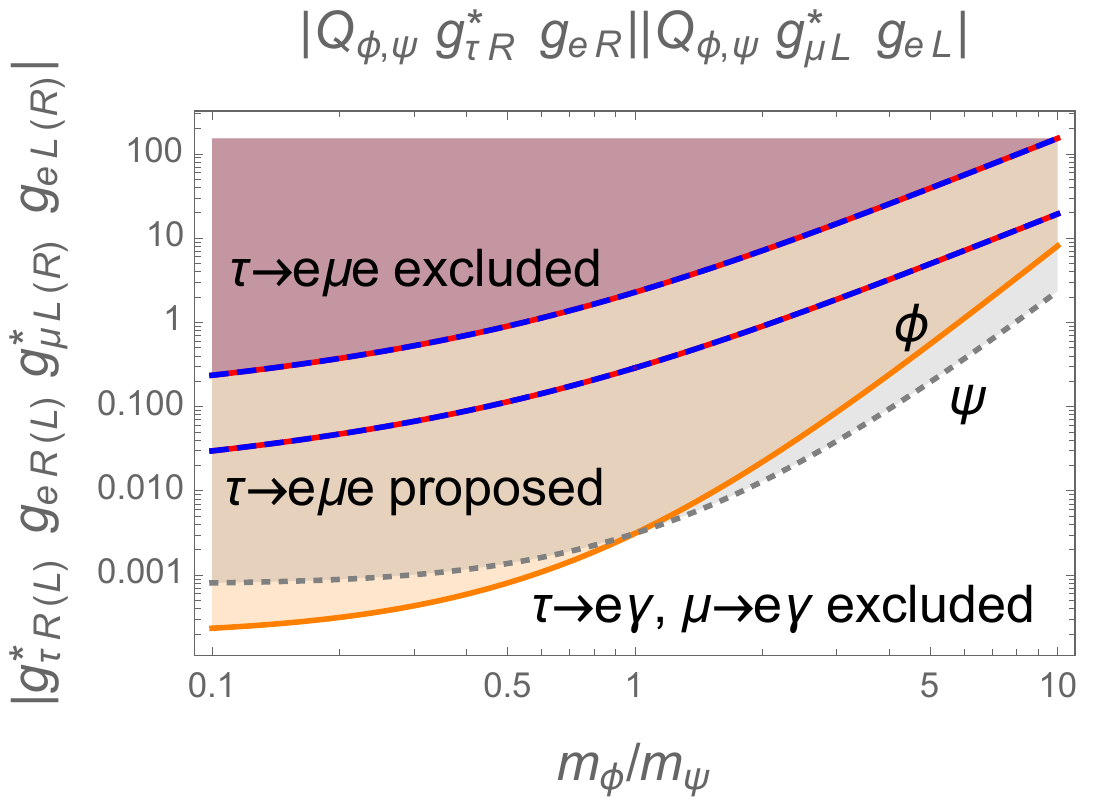}
}
\hspace{0.5cm}
\subfigure[]{
  \includegraphics[width=6.5cm]{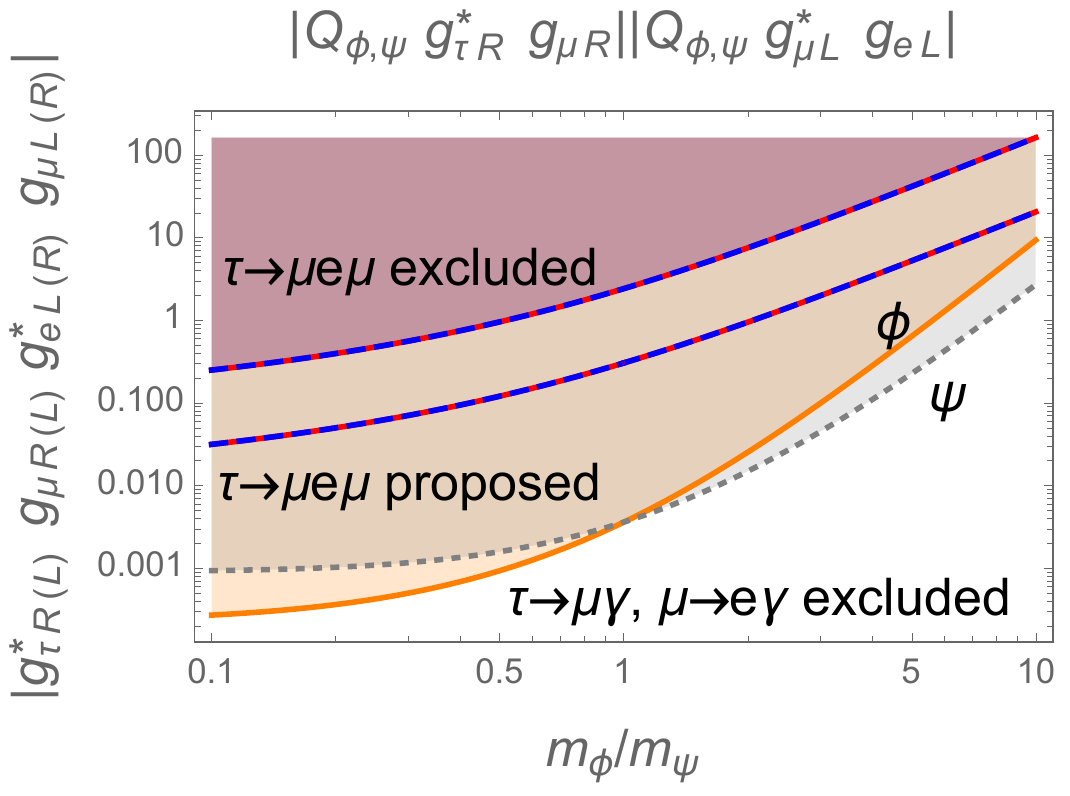}
}
\caption{Parameter space excluded or projected by various experimental bounds or expected sensitivities on $\tau^-\to e^-\mu^+ e^-, \mu^- e^+\mu^-$ processes from box contributions.} 
\label{fig:tau2emuemuemuLFV}
\end{figure}

In Table~\ref{tab: results case I x=1}, we present the constraints on parameters in case I using $x\equiv m_\phi/m_\psi=1$ and $m_\psi=500$~GeV. 
Results for other $m_\psi$ can be obtained by scaling the results with a $\frac{m_\psi}{500 {\rm GeV}}$ factor for $Q_{\phi,\psi} g^*_{l^{(\prime)} R} g_{l L}$ and $(\frac{m_\psi}{500 {\rm GeV}})^2$ for other quantities.
Results in $[...]$ are obtained by using the future experimental sensitivities. 
Both results for the cases of Dirac and Majorana fermion are given,
where results in $\{...\}$ are for the Majorana case.
Note that some of the results are unphysical. For example, the values of $Q_{\phi,\psi}|g_{e R}|^2$ and $Q_{\phi,\psi}|g_{\mu R}|^2$ required to produce large enough $\Delta a_e$ and $\Delta a_\mu$ as required by data are much larger than $4\pi$.
Perturbative calculation breaks down and the results are untrustworthy, hence, unphysical.
We na\"ively state these results simply to indicate that contributions from $Q_{\phi,\psi}|g_{e R}|^2$ and $Q_{\phi,\psi}|g_{\mu R}|^2$
cannot generate the desired results on $\Delta a_e$ and $\Delta a_\mu$.
Results for $x=0.5$ and 2 are given in Tables~\ref{tab: results case I x=0.5} and \ref{tab: results case I x=2}, respectively.

In Fig.~\ref{fig:e muon g-2 1} (a) and (b), we show the allowed parameter space for $\mp Q_{\phi,\psi} |g_{e L(R)}|^2$, $\mp Q_{\phi,\psi}{\rm  Re}(g^*_{e R} g_{e L})$ and $|Q_{\phi,\psi} {\rm Im}(g^*_{e R} g_{eL}|)$ constrained by $\Delta a_e$ and $d_e$.
In Fig.~\ref{fig:e muon g-2 1} (c) and (d), allowed parameter space for $\pm Q_{\phi,\psi} |g_{\mu L(R)}|^2$, $\pm Q_{\phi,\psi}{\rm  Re}(g^*_{\mu R} g_{\mu L})$ and $|Q_{\phi,\psi} {\rm Im}(g^*_{\mu R} g_{\mu L})|$ constrained by $\Delta a_\mu$ and $d_\mu$ are shown.
In Fig.~\ref{fig:e muon g-2 1} (e) the allowed parameter space for $|Q_{\phi,\psi} g^*_{\mu R(L)} g_{e L(R)}|$ constrained by $\mu\to e\gamma$ and the parameter space on $|Q_{\phi,\psi}| |{\rm Re}(g^*_{eR} g_{eL}) {\rm Re}(g^*_{\mu R} g_{\mu L})|^{1/2}$ to produce $\Delta a_e$ and $\Delta a_\mu$ are presented.
These results are given for $m_\psi=500$~GeV.
For other $m_\psi$, scale plots in (a) and (c) with $(500{\rm GeV}/m_\psi)^2$, and scale plots in (b), (d) and (e) with $500{\rm GeV}/m_\psi$.

In Figs.~\ref{fig:mu2eLFV},~\ref{fig:tau2eLFV}, \ref{fig:tau2muLFV}, the parameter space excluded or projected by various bounds or expected sensitivities on $\mu\to e$,  $\tau\to e$ and $\tau\to \mu$ lepton flavor violating processes are shown. They are contributed from photonic penguin, $Z$-penguin and box diagrams.
In Fig.~\ref{fig:tau2emuemuemuLFV}, the
parameter space excluded or projected by using various bounds or projected sensitivities on $\tau^-\to e^-\mu^+ e^-, \mu^- e^+\mu^-$ processes through contributions from box diagrams are shown.

From these results we can extract several messages. 
First we note that chiral interactions ($g_L\times g_R=0$) are unable to generate large enough $\Delta a_e$ and $\Delta a_\mu$ to accommodate the experimental results, Eqs. (\ref{eq: muon g-2 expt}) and (\ref{eq: e g-2 expt}). 
From Tables~\ref{tab: results case I x=1}, \ref{tab: results case I x=0.5}, \ref{tab: results case I x=2}, Fig.~\ref{fig:e muon g-2 1}(a) and (c), we see that $Q_{\phi,\psi}|g_{e R(L)}|^2$ and $Q_{\phi,\psi}|g_{\mu R(L)}|^2$ 
need to be unreasonably large to produce the experimental value of $\Delta a_e$ and $\Delta a_\mu$.
This implies the incapability of chiral interactions to generate large enough $\Delta a_e$ and $\Delta a_\mu$ to accommodate the experimental results.

Although non-chiral interactions are capable to generate $\Delta a_e$ and $\Delta a_\mu$ successfully accommodating the experimental results, they are not contributed from the same source.
From Tables~\ref{tab: results case I x=1}, \ref{tab: results case I x=0.5} and \ref{tab: results case I x=2}, 
we see that, for $x=0.5$, 1 and 2, $Q_{\phi,\psi}{\rm Re}(g^*_{e R} g_{e L})$ and $Q_{\phi,\psi}{\rm Re}(g^*_{\mu R} g_{\mu L})$  of orders $10^{-4}$ and $10^{-3}$, respectively, are able to produce the experimental values of $\Delta a_e$ and $\Delta a_\mu$.
However, the contributions cannot come from the same source, i.e. from diagrams involving the same set of $\phi$ and $\psi$. 
The reasons are as follows.
If $\Delta a_e$ and $\Delta a_\mu$ are generated from the same set of $\psi$ and $\phi$, the very same set of $\psi$ and $\phi$ will also generate $\mu\to e\gamma$ decay with rate exceeding the experimental bound.
Indeed, from Fig.~\ref{fig:e muon g-2 1}(e) we see that the $\mu \to e\gamma$ data constraints 
$|Q_{\phi,\psi} g^*_{\mu R(L)} g_{e L(R)}|$ to be less than $10^{-7}$ to $10^{-6}$, but experimental data on $\Delta a_e$ and $\Delta a_\mu$ require $|Q_{\phi,\psi}| |{\rm Re}(g^*_{eR} g_{eL}) {\rm Re}(g^*_{\mu R} g_{\mu L})|^{1/2}$ to be of the order of $10^{-3}$ to $10^{-1}$, which is larger than the constrain from $\mu\to e\gamma$ by more than 4 orders of magnitude. 
Hence, the contributions to $\Delta a_e$ and $\Delta a_\mu$ do not come from the same source.
Our finding agrees with ref.~\cite{Crivellin:2018qmi},
where a common explanation of $\Delta a_e$ and $\Delta a_\mu$ was investigated and it was found that the present $\mu\to e\gamma$ bound do not support a common explanation of the deviations. 

Presently, the upper limit in $\mu\to e\gamma$ decay gives the most severe constraints on photonic penguin contributions in $\mu\to e$ transitions, agreeing with \cite{Kuno:1999jp,Crivellin:2013hpa}, but the situation may change when we include future experimental sensitivities in the analysis.
From Tables~\ref{tab: results case I x=1}, \ref{tab: results case I x=0.5}, \ref{tab: results case I x=2} and Fig.~\ref{fig:mu2eLFV}(a) to (d), we see that the present $\mu\to e\gamma$ bound constrains the $|Q_{\phi,\psi} g^*_{\mu R} g_{e R}|$ and $|Q_{\phi,\psi} g^*_{\mu R} g_{e L}|$ much better than the present $\mu\to 3e$ and $\mu N\to e N$ upper limits.
In fact, the bounds obtained from $\mu\to e\gamma$ decay are better than those from other processes by at least one order of magnitude.
The situation is altered when considering future experimental searches. 
From the tables and the figures, we see that, on the contrary, in near future experiments the $\mu \to 3 e$ and $\mu N\to e N$ processes
may be able to probe the photonic penguin contributions from $|Q_{\phi,\psi} g^*_{\mu R} g_{e R}|$ and $|Q_{\phi,\psi} g^*_{\mu R} g_{e L}|$ better than the future experiment search on $\mu\to e\gamma$ decay.

The $Z$-penguin diagrams can constrain chiral interaction better than photonic penguin diagrams in $\mu\to e$ transitions.
From Tables~\ref{tab: results case I x=1}, \ref{tab: results case I x=0.5}, \ref{tab: results case I x=2}, Fig.~\ref{fig:mu2eLFV}(a), (b), (e) and (f) we see that the bounds on $|g^*_{\mu R} g_{e R}\Delta T_{3\psi}|$ and $|g^*_{\mu R}g_{e R}\kappa_R|$ from $Z$-penguin contributions are more severe (by two orders of magnitude) than
bounds on $|Q_{\phi,\psi} g^*_{\mu R} g_{e R}|$ from photonic penguin contributions. 
In addition, from Fig.~\ref{fig:mu2eLFV}(e) and (f) we see that $\mu N\to e N$ transitions give better constraints on 
$|g^*_{\mu R} g_{e R}\Delta T_{3\psi}|$ and $|g^*_{\mu R} g_{e R}\kappa_R|$ than the $\mu\to 3 e$ decay.

In case I, either in the Dirac or Majorana case, box contributions to $\mu\to 3e$ decay are subleading.
Furthermore, there are cancelation in box contributions in the Majorana fermionic case making the contributions even smaller.
Fig.~\ref{fig:mu2eLFV}(g) and (h) show the bounds on 
$|g^*_{\mu R} g_{e R} g^*_{eR} g_{e R}|$ 
and $|g^*_{\mu R} g_{e R} g^*_{eL} g_{e L}|$ obtained by considering box contributions to $\mu\to 3 e$ decay.
Note that the constraint on $|g^*_{\mu R} g_{e R} \Delta T_{3\psi}| |g^*_{eL} g_{e L}|$ obtained from $\mu {\rm Au}\to e {\rm Au}$ upper limit and perturbativity is much severe than the $|g^*_{\mu R} g_{e R} g^*_{eR} g_{e R}|$ bound by one to two orders of magnitude, 
while $|Q_{\phi,\psi} g^*_{\mu R} g_{e L}| |g^*_{eL} g_{e R}|$ obtained from $\mu\to e\gamma$, $\Delta a_e$ and $d_e$ experimental results is much severe than the $|g^*_{\mu R} g_{e R} g^*_{eL} g_{e L}|$ bound by more than 8 orders of magnitude.  
One can also use the values in Tables~\ref{tab: results case I x=1}, \ref{tab: results case I x=0.5}, \ref{tab: results case I x=2} to obtain similar findings.
These results imply that box contributions to $\mu\to 3 e$ decay are subleading.

From Tables~\ref{tab: results case I x=1}, \ref{tab: results case I x=0.5} and \ref{tab: results case I x=2}, 
we see that the present bounds on $\Delta a_\tau$ cannot constrain $Q_{\phi,\psi}|g_{\tau R(L)}|^2$ and $Q_{\phi,\psi}{\rm Re}(g^*_{\tau R} g_{\tau L})$ well. Even the bound on $d_\tau$ cannot give good constraints on $Q_{\phi,\psi}{\rm Im}(g^*_{\tau R} g_{\tau L})$. 
There is still a long way to go.

In $\tau\to e$ $(\mu)$ transitions, the $\tau\to e\gamma$ $(\mu \gamma)$ upper limit constrains photonic penguin contributions better than 
the $\tau\to 3 e$ $(3\mu)$ upper limit, agreeing with \cite{Crivellin:2013hpa},
and $Z$-penguin constrains chiral interaction better than photonic penguin.
From Tables~\ref{tab: results case I x=1}, \ref{tab: results case I x=0.5}, \ref{tab: results case I x=2}, Fig.~\ref{fig:tau2eLFV}(a) to (d) and Fig.~\ref{fig:tau2muLFV}(a) to (d), we see that
bounds on $|Q_{\phi,\psi} g^*_{\tau R} g_{e (\mu) R}|$ and $|Q_{\phi,\psi} g^*_{\tau R} g_{e(\mu) L}|$ are constrained by the $\tau\to e\gamma$ $(\mu\gamma)$ data more severely than by the $\tau\to 3e$ $(3\mu)$ upper limit. 
Note that the bounds of these parameters using the proposed sensitivities on $\tau\to 3 e$ and $\tau\to 3\mu$ decays by Belle II are superseded by the bounds using the present limits of $\tau\to e\gamma$ and $\tau\to \mu\gamma$ decays.
From Tables~\ref{tab: results case I x=1}, \ref{tab: results case I x=0.5}, \ref{tab: results case I x=2}, Fig.~\ref{fig:tau2eLFV}(e), (f) and Fig.~\ref{fig:tau2muLFV}(e), (f), we see that
bounds on $|g^*_{\tau R} g_{e (\mu) R}\Delta T_{3\psi}|$ and $|g^*_{\tau R}g_{e (\mu) R}\kappa_R|$ from $Z$-penguin contributions are more severe (by one order of magnitude) than
those on $|Q_{\phi,\psi} g^*_{\tau R} g_{e (\mu) R}|$ from photonic penguin contributions.
Hence, $Z$-penguin constrains chiral interaction better than photonic penguin.

\begin{table}[t!]
\caption{Current experimental upper limits, future sensitivities and bounds from consistency in case I on various muon and tau LFV 
processes. Experimental bounds are from~\cite{MEG,PDG,Mihara2019,Amhis:2016xyh,Kou:2018nap}.}
 \label{tab: expt bounds case I}
\begin{ruledtabular}
\begin{tabular}{ l c  c c }
~~~~~~
    & current limit (future sensitivity)
    & consistency bounds
    & remarks 
    \\
    \hline
${\cal B}(\mu^+\to e^+\gamma)$
    & $<4.2\times 10^{-13}$ ($6\times 10^{-14}$)
    & $<4.2\times 10^{-13}$
    & input
    \\    
${\cal B}(\mu^+\to e^+e^+e^-)$
    & $<1.0\times 10^{-12}$ ($10^{-16}$) 
    & $<1.3\times 10^{-14}$
    & from $\mu\to e\gamma$ bound
    \\ 
    & 
    & $<1.6\times 10^{-14}$
    & from $\mu {\rm Au}\to e{\rm Au}$ bound
    \\   
${\cal B}(\mu^- {\rm Ti}\to e^-{\rm Ti})$
    & $<4.3\times 10^{-12}$ ($10^{-17}$) 
    & $<9.1\times 10^{-14}$
    & from $\mu\to e\gamma$ bound
    \\
    & 
    & $<3.5\times 10^{-13}$
    & from $\mu {\rm Au}\to e{\rm Au}$ bound
    \\   
${\cal B}(\mu^- {\rm Au}\to e^-{\rm Au})$
    & $<7.0\times 10^{-13}$ ($10^{-16}$) 
    & $<1.1\times 10^{-13}$
    & from $\mu\to e\gamma$ bound
    \\ 
    & 
    & $<7.0\times 10^{-13}$
    & input
    \\
${\cal B}(\mu^- {\rm Al}\to e^-{\rm Al})$
    & $\cdots$ ($10^{-17}$) 
    & $<5.5\times 10^{-14}$
    & from $\mu\to e\gamma$ bound
    \\ 
    & 
    & $<1.7\times 10^{-13}$
    & from $\mu {\rm Au}\to e{\rm Au}$ bound
    \\     
${\cal B}(\tau^-\to e^-\gamma)$
    & $<3.3\times 10^{-8}$ ($3\times10^{-9}$) 
    & $<3.3\times 10^{-8}$
    & input
    \\
${\cal B}(\tau^-\to e^-e^+ e^-)$
    & $<2.7\times 10^{-8}$ ($4.3\times 10^{-10}$) 
    & $<1.2\times 10^{-9}$
    & from $\tau\to e\gamma$ bound
    \\       
${\cal B}(\tau^-\to \mu^-\gamma)$
    & $<4.4\times 10^{-8}$ ($1\times 10^{-9}$) 
    & $<4.4\times 10^{-8}$
    & input
    \\ 
${\cal B}(\tau^-\to \mu^-\mu^+\mu^-)$
    & $<2.1\times 10^{-8}$ ($3.3\times 10^{-10}$) 
    & $<1.2\times 10^{-9}$
    & from $\tau\to \mu\gamma$ bound
    \\        
${\cal B}(\tau^-\to \mu^- e^+ \mu^-)$
    & $<1.7\times 10^{-8}$ ($2.7\times 10^{-10}$) 
    & $\lesssim 1\times 10^{-10}$
    & from $\tau\to \mu\gamma$, $\mu\to e\gamma$ bounds
    \\  
${\cal B}(\tau^-\to e^-\mu^+ e^-)$
    & $<1.5\times 10^{-8}$ ($2.4\times 10^{-10}$) 
    & $\lesssim 7\times 10^{-11}$
    & from $\tau\to e\gamma$, $\mu\to e\gamma$ bounds
    \\                                           
\end{tabular}
\end{ruledtabular}
\end{table}

Box contributions to $\tau\to 3e$ and $\tau\to 3\mu$ decays can sometime be comparable to $Z$-penguin contributions.
In Fig.~\ref{fig:tau2eLFV}(g), (h) and Fig.~\ref{fig:tau2muLFV}(g), (h) we show the bounds on 
$|g^*_{\tau R} g_{e(\mu) R} g^*_{e(\mu) R} g_{e(\mu) R}|$ 
and $|g^*_{\tau R} g_{e(\mu) R} g^*_{e(\mu) L} g_{e(\mu) L}|$ obtained by considering box contributions to $\tau\to 3 e$ $(3\mu)$ decay.
Note that the constraint on $|g^*_{\tau R} g_{e(\mu) R} \Delta T_{3\psi}| |g^*_{e(\mu) L} g_{e(\mu) L}|$ obtained from $Z$-penguin contributions to $\tau \to 3e$ $(3\mu)$ decay and perturbativity
is much severe than the $|g^*_{\tau R} g_{e(\mu) R} g^*_{e(\mu) R} g_{e(\mu) R}|$ bound from box contributions for $x\gtrsim 0.4$, 
but it is the other way around for $x\lesssim 0.4$.
The bound on $|Q_{\phi,\psi} g^*_{\tau R} g_{e(\mu) L}| |g^*_{e(\mu) L} g_{e(\mu) R}|$ obtained using $\tau\to e\gamma $
$(\mu\gamma)$, $\Delta a_{e(\mu)}$ and $d_{e(\mu)}$ experimental results is much severe than the $|g^*_{\tau R} g_{e(\mu) R} g^*_{e(\mu) L} g_{e(\mu) L}|$ bound from box contributions by five to seven (one to three) orders of magnitude.  
One can also obtain similar results using the values in Tables~\ref{tab: results case I x=1}, \ref{tab: results case I x=0.5}, \ref{tab: results case I x=2}.
These findings imply that box contributions to $\tau\to 3 e$ $(3\mu)$ can sometime be comparable to $Z$-penguin contributions.

The $\tau^-\to e^- \mu^+ e^-$ rate is highly constrained by $\tau\to e\gamma$ and $\mu\to e\gamma$ upper limits.
From Fig.~\ref{fig:tau2emuemuemuLFV} (a), (c), (e) and Tables~\ref{tab: results case I x=1}, \ref{tab: results case I x=0.5}, \ref{tab: results case I x=2}, we see that the bounds on 
$|g^*_{\tau R} g_{e R} g^*_{\mu R} g_{e R}|$, 
$|g^*_{\tau R} g_{e L} g^*_{\mu R} g_{e L}|$ and
$|g^*_{\tau R} g_{e R} g^*_{\mu L} g_{e L}|$,
obtained from the upper limit of the $\tau^-\to e^- \mu^+ e^-$ rate,
are larger than the bounds on
$|Q_{\phi,\psi} g^*_{\tau R} g_{e R}| |Q_{\phi,\psi} g^*_{\mu R} g_{e R}|$,
$|Q_{\phi,\psi} g^*_{\tau R} g_{e L}| |Q_{\phi,\psi} g^*_{\mu R} g_{e L}|$ and
$|Q_{\phi,\psi} g^*_{\tau R} g_{e R}| |Q_{\phi,\psi} g^*_{\mu L} g_{e L}|$, obtained from the upper limits of 
$\tau\to e\gamma$ and $\mu\to e\gamma$ rates, 
by several orders of magnitude.
Note that the $\tau^-\to e^- \mu^+ e^-$ rate is constrained to be smaller than the proposed sensitivity.
Hence, the $\tau^-\to e^- \mu^+ e^-$ rate is highly constrained by the present $\tau\to e\gamma$ and $\mu\to e\gamma$ upper limits.

The $\tau^-\to \mu^- e^+ \mu^-$ rate is also highly constrained by $\tau\to \mu\gamma$ and $\mu\to e\gamma$ upper limits.
From Fig.~\ref{fig:tau2emuemuemuLFV} (b), (d), (f) and Tables.~\ref{tab: results case I x=1}, \ref{tab: results case I x=0.5}, \ref{tab: results case I x=2}, we see that the bounds on 
$|g^*_{\tau R} g_{\mu R} g^*_{e R} g_{\mu R}|$, 
$|g^*_{\tau R} g_{\mu L} g^*_{e R} g_{\mu L}|$ and
$|g^*_{\tau R} g_{\mu R} g^*_{e L} g_{\mu L}|$,
obtained from the upper limit of the $\tau^-\to \mu^- e^+ \mu^-$ rate,
are larger than the bounds on
$|Q_{\phi,\psi} g^*_{\tau R} g_{\mu R}| |Q_{\phi,\psi} g^*_{\mu R} g_{e R}|$,
$|Q_{\phi,\psi} g^*_{\tau R} g_{\mu L}| |Q_{\phi,\psi} g^*_{\mu L} g_{e R}|$ and
$|Q_{\phi,\psi} g^*_{\tau R} g_{\mu R}| |Q_{\phi,\psi} g^*_{\mu L} g_{e L}|$, obtained from the upper limits of 
$\tau\to e\gamma$ and $\mu\to e\gamma$ rates, 
by several orders of magnitude.
Hence, the $\tau^-\to \mu^- e^+ \mu^-$ rate is highly constrained by $\tau\to \mu\gamma$ and $\mu\to e\gamma$ upper limits.
In fact, the $\tau^-\to \mu^- e^+ \mu^-$ rate is constrained to be smaller than the proposed sensitivity.

In Table~\ref{tab: expt bounds case I}, we compare the current experimental upper limits, future sensitivities and bounds from consistency for case I on various muon and tau LFV processes. We see that the present $\mu\to e\gamma$ upper limit requires the bounds on $\mu\to 3 e$, $\mu\,{\rm Ti}\to e\, {\rm Ti}$ and $\mu\,{\rm Au}\to e\, {\rm Au}$ be lower by two orders of magnitude, more than one order of magnitude and almost one order of magnitude, respectively, from their present upper limits, and the $\mu\,{\rm Al}\to e\, {\rm Al}$ rate is predicted to be smaller than $6\times 10^{-14}$. These bounds can be further pushed downward by one order of magnitude if we still cannot observed $\mu\to e\gamma$ decay in MEG II.
It is interesting that the future sensitivities of $\mu\to 3 e$ and $\mu\, {\rm N}\to e \,{\rm N}$ are much lower than the above limits based on consistency, giving them good opportunity to explore these LFV processes.
We find that the situation is similar but the bounds are slightly relaxed when the $\mu\,{\rm Au}\to e\, {\rm Au}$ upper limit instead of the present $\mu\to e\gamma$ upper limit is used as an input.
Similarly, using the present $\tau\to e\gamma$ $(\mu\gamma)$ upper limit as input, the $\tau\to 3 e$ $(3\mu)$ bound is smaller than its present upper limit by one order of magnitude.
Note that the $\B(l'\to l\bar l l)/\B(l'\to l\gamma)$ ratios are close to the values shown in Eq.~(\ref{eq: 3l per lgamma ratio})~\cite{Kuno:1999jp, Crivellin:2013hpa}, but not identical to them, as the $F_1$ terms in photonic penguins also play some roles.
Finally, the $\tau^-\to \mu^- e^+ \mu^-$ and $\tau^-\to e^-\mu^+ e^-$ bounds are lower than their present upper limits by two orders of magnitude as required from the present $\tau\to \mu\gamma$, $e\gamma$ and $\mu\to e\gamma$ upper limits.
These limits are lower than the proposed future sensitivities.

\subsection{Case II}

We now turn to the second case, where we have a built-in cancelation mechanism.

\begin{table}[ht!]
\caption{Same as Table~\ref{tab: results case I x=1} ($x=1$), but for case II.}
 \label{tab: results case II x=1}
\footnotesize{
\begin{center}
\begin{tabular}{ l |c  c c c c}
\hline
Processes
    & constraints
    & constraints
    & constraints
    & constraints
    \\
    \hline
    & $ Q_{\phi}|g_{e R}|^2$
    &  $ Q_{\psi}|g_{e R}|^2$
    &  $Q_{\phi}{\rm  Re}(g^*_{e R} g_{e L}\delta^{ee}_{RL})$
    &  $Q_{\psi}{\rm  Re}(g^*_{e R} g_{e L}\delta^{ee}_{RL})$
    \\ 
$\Delta a_e$
   & $-1597\pm653$
   & $1597\mp653$
   & $(8\mp 3)\times 10^{-4}$
   & $(-8\pm 3)\times 10^{-4}$
   \\  
   \hline  
    & $Q_{\phi}|g_{\mu R}|^2$
    &  $Q_{\psi}|g_{\mu R}|^2$
    &  $Q_{\phi}{\rm  Re}(g^*_{\mu R} g_{\mu L}\delta^{\mu\mu}_{RL})$
    & $Q_{\psi}{\rm  Re}(g^*_{\mu R} g_{\mu L}\delta^{\mu\mu}_{RL})$
    \\ 
$\Delta a_\mu$
   & $115\pm31$
   & $-115\mp31$
   & $(-12\mp 3)\times 10^{-3}$
   & $(-12\pm 3)\times 10^{-3}$
   \\
   \hline        
    &  $Q_{\phi} |g_{\tau R}|^2$
    &  $Q_{\psi} |g_{\tau R}|^2$
    &  $Q_{\phi}{\rm  Re}(g^*_{\tau R} g_{\tau L}\delta^{\tau\tau}_{RL})$
    &  $Q_{\psi}{\rm  Re}(g^*_{\tau R} g_{\tau L}\delta^{\tau\tau}_{RL})$
    \\ 
$\Delta a_\tau$
   & $(-7\sim2)\times 10^6$
   & $(-2\sim7)\times 10^6$
   & $(-3\sim 13)\times 10^3$
   & $(-13\sim 3)\times 10^3$
   \\   
   \hline             
    &  $|Q_{\phi}{\rm  Im}(g^*_{e R} g_{e L}\delta^{e e}_{RL})|$
    &  $|Q_{\phi}{\rm  Im}(g^*_{e R} g_{e L})\delta^{e e}_{RL}|$
    &  \scriptsize{$|Q_{\phi}{\rm  Im}(g^*_{l R} g_{l L}\delta^{ll}_{RL})|_{l=\mu (\tau)}$}
    &  \scriptsize{$|Q_{\psi}{\rm  Im}(g^*_{l R} g_{l L}\delta^{ll}_{RL})|_{l=\mu (\tau)}$}
    \\ 
$d_e$, $d_\mu$, $d_\tau$ 
   & $5.3\times 10^{-10}$
   & $5.3\times 10^{-10}$
   & $9.1\,(76.5)$
   & $9.1\,(76.5)$
   \\  
   \hline 
    &  $|Q_{\phi} g^*_{\mu R} g_{e R}\delta^{\mu e}_{RR}|$
    &  $|Q_{\psi} g^*_{\mu R} g_{e R}\delta^{\mu e}_{RR}|$
    &  $|Q_{\phi} g^*_{\mu R} g_{e L}\delta^{\mu e}_{RL}|$
    &  $|Q_{\psi} g^*_{\mu R} g_{e L}\delta^{\mu e}_{RL}|$
    \\  
$\mu^+\to e^+ \gamma$ 
    & $0.004\,[0.0014]$
    &  $0.005\,[0.0020]$
    &  $23\,[9]\times 10^{-8}$
    &  $23\, [9]\times 10^{-8}$ 
    \\        
$\mu^+\to e^- e^+ e^-$
    & $0.077\,[0.0008]$
    &  $0.085\,[0.0008]$
    &  $448\, [4]\times 10^{-8}$
    &  $448\, [4]\times 10^{-8}$ 
    \\  
$\mu^- {\rm Au}\to e^- {\rm Au}$
    &  $0.028\,[0.0003]$
    &  $0.074\,[0.0009]$
    &  $471\,[6]\times 10^{-8}$
    &  $471\,[6]\times 10^{-8}$ 
    \\    
$\mu^- {\rm Ti}\to e^- {\rm Ti}$
    &  $0.072\,[0.0001]$
    &  $0.219\,[0.0003]$
    &  $1137\,[2]\times 10^{-8}$
    &  $1137\,[2]\times 10^{-8}$ 
    \\ 
$\mu^- {\rm Al}\to e^- {\rm Al}$
    &  $[0.0001]$
    &  $[0.0004]$
    &  $[2\times 10^{-8}]$
    &  $[2\times 10^{-8}]$ 
    \\  
    \hline             
    &  $|g^*_{\mu R} g_{e R} \Delta T_{3\psi}\delta^{\mu e}_{RR}|$
    &  $|g^*_{\mu R} g_{e R} g^*_{e R} g_{e R}\delta^{\mu e}_{RR}|$
    &  $|g^*_{\mu R} g_{e R} g^*_{e L} g_{e L}\delta^{\mu e}_{RR}|$
    \\        
$\mu^+\to e^- e^+ e^-$
    & $118\,[1]\times 10^{-5}$
    & $0.04\{0.04\}[4\{4\}\times 10^{-4}]$
    & $0.03\{0.06\}[3\{6\}\times 10^{-4}]$ 
    \\ 
$\mu^- {\rm Au}\to e^- {\rm Au}$
    &  $148\,[2] \times 10^{-6}$
    &  
    &  
    \\    
$\mu^- {\rm Ti}\to e^- {\rm Ti}$
    &  $5155\,[8]\times 10^{-7}$
    &  
    &   
    \\ 
$\mu^- {\rm Al}\to e^- {\rm Al}$
    &  $[1\times 10^{-6}]$
    &  
    &   
    \\   
    \hline                             
    & $|Q_{\phi} g_{\tau R} g^*_{e R}\delta^{e\tau}_{RR}|$
    & $|Q_{\psi} g_{\tau R} g^*_{e R}\delta^{e\tau}_{RR}|$ 
    & $|Q_{\phi} g_{\tau R} g^*_{e L}\delta^{e\tau}_{LR}|$ 
    & $|Q_{\psi} g_{\tau R} g^*_{e L}\delta^{e\tau}_{LR}|$
    \\   
$\tau^-\to e^- \gamma$ 
    &  $2.4\,[0.7]$
    &  $3.6\,[1.1]$
    &  $26\,[8]\times 10^{-4}$
    &  $26\,[8]\times 10^{-4}$
    \\               
$\tau^-\to e^- e^+ e^-$
    &  $22.2\,[2.8]$
    &  $27.3\,[3.5]$
    &  $22\,[3]\times 10^{-3}$
    &  $22\,[3]\times 10^{-3}$
    \\
    \hline
    &  $|g_{\tau R} g^*_{e R} \Delta T_{3\psi}\delta^{e\tau}_{RR}|$
    &  $|g_{\tau R} g^*_{e R} g_{e R} g^*_{e R}\delta^{e\tau}_{RR}|$
    &  $|g_{\tau R} g^*_{e R} g_{e L} g^*_{e L}\delta^{e\tau}_{RR}|$ 
    \\ 
$\tau^-\to e^- e^+ e^-$
    & $0.46\,[0.06]$
    & $17.2\,\{17.2\}[2.2\,\{2.2\}]$
    & $12.2\,\{24.3\}[1.5\,\{3.1\}]$
    \\  
    \hline         
    &  $|Q_{\phi} g_{\tau R} g^*_{\mu R}\delta^{\mu\tau}_{RR}|$
    &  $|Q_{\psi} g_{\tau R} g^*_{\mu R}\delta^{\mu\tau}_{RR}|$
    &  $|Q_{\phi} g_{\tau R} g^*_{\mu L}\delta^{\mu\tau}_{LR}|$
    &  $|Q_{\psi} g_{\tau R} g^*_{\mu R}\delta^{\mu\tau}_{LR}|$
    \\
$\tau^-\to \mu^- \gamma$ 
    &  $2.8\,[0.4]$
    &  $4.2\,[0.6]$
    &  $30\,[4]\times 10^{-4}$
    &  $30\,[4]\times 10^{-4}$
    \\   
$\tau^-\to \mu^-\mu^+\mu^-$
    & $19.5\,[2.4]$
    & $24.1\,[3.0]$
    & $20\,[2]\times 10^{-3}$
    & $20\,[2]\times 10^{-3}$
    \\  
    \hline             
    &  $|g_{\tau R} g^*_{\mu R} \Delta T_{3\psi}\delta^{\mu\tau}_{RR}|$
    &  $|g_{\tau R} g^*_{\mu R} g_{\mu R} g^*_{\mu R}\delta^{\mu\tau}_{RR}|$
    &  $|g_{\tau R} g^*_{\mu R} g_{\mu L} g^*_{\mu L}\delta^{\mu\tau}_{RR}|$
     \\ 
$\tau^-\to \mu^-\mu^+\mu^-$
    & $0.41\,[0.05]$ 
    & $15.2\,\{15.2\}\,[1.9\,\{1.9\}]$
    & $10.7\,\{21.4\}\,[1.3\,\{2.7\}]$ 
     \\     
     \hline      
    &  $|g_{\tau R} g^*_{e R} g_{\mu R} g^*_{e R}\delta^{e\tau}_{RR}\delta^{e\mu}_{RR}|$
    &  $|g_{\tau R} g^*_{e R} g_{\mu L} g^*_{e L}\delta^{e\tau}_{RR}\delta^{e\mu}_{LL}|$
    &  $|g_{\tau R} g^*_{e L} g_{\mu R} g^*_{e L}\delta^{e\tau}_{LR}\delta^{e\mu}_{LR}|$
    &  $|g_{\tau R} g^*_{e L} g_{\mu L} g^*_{e R}\delta^{e\tau}_{LR}\delta^{e\mu}_{RL}|$
    \\  
$\tau^-\to e^-\mu^+ e^-$
    & $32.0\,\{16.0\}\,[4.1\,\{2.0\}]$
    & $15.1\,\{22.7\}\,[1.9\,\{2.9\}]$
    & $21.4\,\{21.4\}\,[2.7\,\{2.7\}]$
    & $45.3\,\{22.7\}\,[5.7\,\{2.9\}]$
    \\   
    \hline    
    &  $|g_{\tau R} g^*_{\mu R} g_{e R} g^*_{\mu R}\delta^{\mu\tau}_{RR}\delta^{\mu e}_{RR}|$
    &  $|g_{\tau R} g^*_{\mu R} g_{e L} g^*_{\mu L}\delta^{\mu\tau}_{RR}\delta^{\mu e}_{LL}|$
    &  $|g_{\tau R} g^*_{\mu L} g_{e R} g^*_{\mu L}\delta^{\mu\tau}_{LR}\delta^{\mu e}_{LR}|$
    &  $|g_{\tau R} g^*_{\mu L} g_{e L} g^*_{\mu R}\delta^{\mu\tau}_{LR}\delta^{\mu e}_{RL}|$
   \\  
$\tau^-\to \mu^- e^+\mu^+$
    & $34.1\,\{17.1\}\,[4.3\,\{2.1\}]$
    & $16.1\,\{24.1\}\,[2.0\,\{3.0\}]$
    & $22.7\,\{22.7\}\,[2.9\,\{2.9\}]$
    & $48.2\,\{24.1\}\,[6.1\,\{3.0\}]$
    \\      
    \hline                  
\end{tabular}
\end{center}
}
\end{table}

\begin{table}[ht!]
\caption{Same as Table~\ref{tab: results case II x=1}, but with $x\equiv m_\phi/m_\psi=0.5$ .}
 \label{tab: results case II x=0.5}
\footnotesize{
\begin{center}
\begin{tabular}{ l |c  c c c c}
\hline
Processes
    & constraints
    & constraints
    & constraints
    & constraints
    \\
    \hline
    & $ Q_{\phi}|g_{e R}|^2$
    &  $ Q_{\psi}|g_{e R}|^2$
    &  $Q_{\phi}{\rm  Re}(g^*_{e R} g_{e L}\delta^{ee}_{RL})$
    &  $Q_{\psi}{\rm  Re}(g^*_{e R} g_{e L}\delta^{ee}_{RL})$
    \\ 
$\Delta a_e$
   & $-812\pm332$ 
   & $1059\mp433$ 
   & $(8\mp 3)\times 10^{-4}$ 
   & $(-13\pm 6)\times 10^{-4}$ 
   \\  
   \hline  
    & $Q_{\phi}|g_{\mu R}|^2$
    &  $Q_{\psi}|g_{\mu R}|^2$
    &  $Q_{\phi}{\rm  Re}(g^*_{\mu R} g_{\mu L}\delta^{\mu\mu}_{RL})$
    & $Q_{\psi}{\rm  Re}(g^*_{\mu R} g_{\mu L}\delta^{\mu\mu}_{RL})$
    \\ 
$\Delta a_\mu$
   & $58\pm16$ 
   & $-76\mp20$ 
   & $(-1.2\mp0.3)\times 10^{-2}$ 
   & $(-20\pm 5)\times 10^{-3}$ 
   \\
   \hline        
    & $Q_{\phi} |g_{\tau R}|^2$
    &  $Q_{\psi} |g_{\tau R}|^2$
    &  $Q_{\phi}{\rm  Re}(g^*_{\tau R} g_{\tau L}\delta^{\tau\tau}_{RL})$
    & $Q_{\psi}{\rm  Re}(g^*_{\tau R} g_{\tau L}\delta^{\tau\tau}_{RL})$
    \\ 
$\Delta a_\tau$
   & $(-4\sim1)\times 10^6$ 
   & $(-1\sim5)\times 10^6$ 
   & $(-3\sim 13)\times 10^3$ 
   & $(-22\sim 5)\times 10^3$ 
   \\   
   \hline    
    &  $|Q_{\phi}{\rm  Im}(g^*_{e R} g_{e L}\delta^{e e}_{RL})|$
    &  $|Q_{\phi}{\rm  Im}(g^*_{e R} g_{e L})\delta^{e e}_{RL}|$
    &  \scriptsize{$|Q_{\phi}{\rm  Im}(g^*_{l R} g_{l L}\delta^{ll}_{RL})|_{l=\mu (\tau)}$}
    &  \scriptsize{$|Q_{\psi}{\rm  Im}(g^*_{l R} g_{l L}\delta^{ll}_{RL})|_{l=\mu (\tau)}$}
    \\ 
$d_e$, $d_\mu$, $d_\tau$ 
   & $5.0\times 10^{-10}$
   & $8.7\times 10^{-10}$
   & $8.7\,(73.0)$
   & $15.0\,(126.2)$
   \\  
   \hline             
    &  $|Q_{\phi} g^*_{\mu R} g_{e R}\delta^{\mu e}_{RR}|$
    &  $|Q_{\psi} g^*_{\mu R} g_{e R}\delta^{\mu e}_{RR}|$
    &  $|Q_{\phi} g^*_{\mu R} g_{e L}\delta^{\mu e}_{RL}|$
    &  $|Q_{\psi} g^*_{\mu R} g_{e L}\delta^{\mu e}_{RL}|$
    \\  
$\mu^+\to e^+ \gamma$ 
    & $0.003\,[0.0011]$ 
    &  $0.007\,[0.0027]$ 
    &  $22\,[8]\times 10^{-8}$ 
    &  $4\, [1]\times 10^{-7}$ 
    \\        
$\mu^+\to e^- e^+ e^-$
    & $0.063\,[0.0006]$ 
    &  $0.115\,[0.0011]$ 
    &  $427\, [4]\times 10^{-8}$ 
    &  $739\, [7]\times 10^{-8}$ 
    \\  
$\mu^- {\rm Au}\to e^- {\rm Au}$
    &  $0.015\,[0.0002]$ 
    &  $0.136\,[0.0016]$ 
    &  $449\,[5]\times 10^{-8}$ 
    &  $777\,[9]\times 10^{-8}$ 
    \\    
$\mu^- {\rm Ti}\to e^- {\rm Ti}$
    &  $0.040\,[0.00006]$ 
    &  $0.416\,[0.0006]$ 
    &  $1084\,[2]\times 10^{-8}$ 
    &  $1875\,[3]\times 10^{-8}$ 
    \\ 
$\mu^- {\rm Al}\to e^- {\rm Al}$
    &  $[0.00001]$ 
    &  $[0.0008]$ 
    &  $[2\times 10^{-8}]$ 
    &  $[4\times 10^{-8}]$ 
    \\  
    \hline             
    &  $|g^*_{\mu R} g_{e R} \Delta T_{3\psi}\delta^{\mu e}_{RR}|$
    &  $|g^*_{\mu R} g_{e R} g^*_{e R} g_{e R}\delta^{\mu e}_{RR}|$
    &  $|g^*_{\mu R} g_{e R} g^*_{e L} g_{e L}\delta^{\mu e}_{RR}|$
    \\        
$\mu^+\to e^- e^+ e^-$
    & $142\,[1]\times 10^{-5}$ 
    & $0.04\{0.01\}[4\{1\}\times 10^{-4}]$ 
    & $0.01\{0.02\}[1\{2\}\times 10^{-4}]$ 
    \\ 
$\mu^- {\rm Au}\to e^- {\rm Au}$
    &  $178\,[2] \times 10^{-6}$ 
    &  
    &  
    \\    
$\mu^- {\rm Ti}\to e^- {\rm Ti}$
    &  $6226\,[9]\times 10^{-7}$ 
    &  
    &   
    \\ 
$\mu^- {\rm Al}\to e^- {\rm Al}$
    &  $[1\times 10^{-6}]$ 
    &  
    &   
    \\   
    \hline                             
    & $|Q_{\phi} g_{\tau R} g^*_{e R}\delta^{e\tau}_{RR}|$
    & $|Q_{\psi} g_{\tau R} g^*_{e R}\delta^{e\tau}_{RR}|$ 
    & $|Q_{\phi} g_{\tau R} g^*_{e L}\delta^{e\tau}_{LR}|$ 
    & $|Q_{\psi} g_{\tau R} g^*_{e L}\delta^{e\tau}_{LR}|$
    \\   
$\tau^-\to e^- \gamma$ 
    &  $1.9\,[0.6]$ 
    &  $4.7\,[1.4]$ 
    &  $24\,[7]\times 10^{-4}$ 
    &  $42\,[13]\times 10^{-4}$ 
    \\               
$\tau^-\to e^- e^+ e^-$
    &  $18.0\,[2.3]$ 
    &  $36.8\,[4.6]$ 
    &  $21\,[3]\times 10^{-3}$ 
    &  $37\,[5]\times 10^{-3}$ 
    \\
    \hline
    &  $|g_{\tau R} g^*_{e R} \Delta T_{3\psi}\delta^{e\tau}_{RR}|$
    &  $|g_{\tau R} g^*_{e R} g_{e R} g^*_{e R}\delta^{e\tau}_{RR}|$
    &  $|g_{\tau R} g^*_{e R} g_{e L} g^*_{e L}\delta^{e\tau}_{RR}|$ 
    \\ 
$\tau^-\to e^- e^+ e^-$
    & $0.56\,[0.07]$ 
    & $16.4\,\{4.5\}[2.1\,\{0.6\}]$ 
    & $5.0\,\{6.4\}[0.6\,\{0.8\}]$ 
    \\  
    \hline         
    &  $|Q_{\phi} g_{\tau R} g^*_{\mu R}\delta^{\mu\tau}_{RR}|$
    &  $|Q_{\psi} g_{\tau R} g^*_{\mu R}\delta^{\mu\tau}_{RR}|$
    &  $|Q_{\phi} g_{\tau R} g^*_{\mu L}\delta^{\mu\tau}_{LR}|$
    &  $|Q_{\psi} g_{\tau R} g^*_{\mu R}\delta^{\mu\tau}_{LR}|$
    \\
$\tau^-\to \mu^- \gamma$ 
    &  $2.2\,[0.3]$ 
    &  $5.4\,[0.8]$ 
    &  $28\,[4]\times 10^{-4}$ 
    &  $49\,[7]\times 10^{-4}$ 
    \\   
$\tau^-\to \mu^-\mu^+\mu^-$
    & $15.8\,[2.0]$ 
    & $32.4\,[4.1]$ 
    & $18\,[2]\times 10^{-3}$ 
    & $33\,[4]\times 10^{-3}$ 
    \\  
    \hline             
    &  $|g_{\tau R} g^*_{\mu R} \Delta T_{3\psi}\delta^{\mu\tau}_{RR}|$
    &  $|g_{\tau R} g^*_{\mu R} g_{\mu R} g^*_{\mu R}\delta^{\mu\tau}_{RR}|$
    &  $|g_{\tau R} g^*_{\mu R} g_{\mu L} g^*_{\mu L}\delta^{\mu\tau}_{RR}|$
     \\ 
$\tau^-\to \mu^-\mu^+\mu^-$
    & $0.49\,[0.06]$ 
    & $14.4\,\{4.0\}\,[1.8\,\{0.5\}]$ 
    & $4.4\,\{5.6\}\,[0.6\,\{0.7\}]$ 
     \\     
     \hline      
    &  $|g_{\tau R} g^*_{e R} g_{\mu R} g^*_{e R}\delta^{e\tau}_{RR}\delta^{e\mu}_{RR}|$
    &  $|g_{\tau R} g^*_{e R} g_{\mu L} g^*_{e L}\delta^{e\tau}_{RR}\delta^{e\mu}_{LL}|$
    &  $|g_{\tau R} g^*_{e L} g_{\mu R} g^*_{e L}\delta^{e\tau}_{LR}\delta^{e\mu}_{LR}|$
    &  $|g_{\tau R} g^*_{e L} g_{\mu L} g^*_{e R}\delta^{e\tau}_{LR}\delta^{e\mu}_{RL}|$
    \\  
$\tau^-\to e^-\mu^+ e^-$
    & $41.9\,\{6.1\}\,[6.3\,\{0.8\}]$ 
    & $7.5\,\{8.6\}\,[1.0\,\{1.1\}]$ 
    & $10.7\,\{10.7\}\,[1.3\,\{1.3\}]$ 
    & $59.3\,\{29.6\}\,[7.5\,\{3.7\}]$ 
    \\   
    \hline    
    &  $|g_{\tau R} g^*_{\mu R} g_{e R} g^*_{\mu R}\delta^{\mu\tau}_{RR}\delta^{\mu e}_{RR}|$
    &  $|g_{\tau R} g^*_{\mu R} g_{e L} g^*_{\mu L}\delta^{\mu\tau}_{RR}\delta^{\mu e}_{LL}|$
    &  $|g_{\tau R} g^*_{\mu L} g_{e R} g^*_{\mu L}\delta^{\mu\tau}_{LR}\delta^{\mu e}_{LR}|$
    &  $|g_{\tau R} g^*_{\mu L} g_{e L} g^*_{\mu R}\delta^{\mu\tau}_{LR}\delta^{\mu e}_{RL}|$
   \\  
$\tau^-\to \mu^- e^+\mu^+$
    & $44.6\,\{6.5\}\,[5.6\,\{0.8\}]$ 
    & $8.0\,\{9.2\}\,[1.0\,\{1.2\}]$ 
    & $11.3\,\{11.3\}\,[1.4\,\{1.4\}]$ 
    & $63.1\,\{31.5\}\,[7.9\,\{4.0\}]$ 
    \\      
    \hline                  
\end{tabular}
\end{center}
}
\end{table}

\begin{table}[ht!]
\caption{Same as Table~\ref{tab: results case II x=1}, but with $x\equiv m_\phi/m_\psi=2$ .}
 \label{tab: results case II x=2}
\footnotesize{
\begin{center}
\begin{tabular}{ l |c  c c c c}
\hline
Processes
    & constraints
    & constraints
    & constraints
    & constraints
    \\
    \hline
    & $ Q_{\phi}|g_{e R}|^2$
    &  $ Q_{\psi}|g_{e R}|^2$
    &  $Q_{\phi}{\rm  Re}(g^*_{e R} g_{e L}\delta^{ee}_{RL})$
    &  $Q_{\psi}{\rm  Re}(g^*_{e R} g_{e L}\delta^{ee}_{RL})$
    \\ 
$\Delta a_e$
   & $-4234\pm1732$ 
   & $3247\mp1328$ 
   & $(13\mp 6)\times 10^{-4}$ 
   & $(-8\pm 3)\times 10^{-4}$ 
   \\  
   \hline  
    & $Q_{\phi}|g_{\mu R}|^2$
    &  $Q_{\psi}|g_{\mu R}|^2$
    &  $Q_{\phi}{\rm  Re}(g^*_{\mu R} g_{\mu L}\delta^{\mu\mu}_{RL})$
    & $Q_{\psi}{\rm  Re}(g^*_{\mu R} g_{\mu L}\delta^{\mu\mu}_{RL})$
    \\ 
$\Delta a_\mu$
   & $305\pm82$ 
   & $-234\mp63$ 
   & $(-20\mp 5)\times 10^{-3}$ 
   & $(12\pm 3)\times 10^{-3}$ 
   \\
   \hline        
    & $Q_{\phi} |g_{\tau R}|^2$
    &  $Q_{\psi} |g_{\tau R}|^2$
    &  $Q_{\phi}{\rm  Re}(g^*_{\tau R} g_{\tau L}\delta^{\tau\tau}_{RL})$
    & $Q_{\psi}{\rm  Re}(g^*_{\tau R} g_{\tau L}\delta^{\tau\tau}_{RL})$
    \\ 
$\Delta a_\tau$
   & $(-21\sim5)\times 10^6$ 
   & $(-4\sim16)\times 10^6$ 
   & $(-6\sim 23)\times 10^3$ 
   & $(-13\sim 3)\times 10^3$ 
   \\   
   \hline   
    &  $|Q_{\phi}{\rm  Im}(g^*_{e R} g_{e L}\delta^{e e}_{RL})|$
    &  $|Q_{\phi}{\rm  Im}(g^*_{e R} g_{e L})\delta^{e e}_{RL}|$
    &  \scriptsize{$|Q_{\phi}{\rm  Im}(g^*_{l R} g_{l L}\delta^{ll}_{RL})|_{l=\mu (\tau)}$}
    &  \scriptsize{$|Q_{\psi}{\rm  Im}(g^*_{l R} g_{l L}\delta^{ll}_{RL})|_{l=\mu (\tau)}$}
    \\ 
$d_e$, $d_\mu$, $d_\tau$ 
   & $8.7\times 10^{-10}$
   & $5.0\times 10^{-10}$
   & $15.0\,(126.2)$
   & $8.7\,(73.0)$
   \\  
   \hline              
    &  $|Q_{\phi} g^*_{\mu R} g_{e R}\delta^{\mu e}_{RR}|$
    &  $|Q_{\psi} g^*_{\mu R} g_{e R}\delta^{\mu e}_{RR}|$
    &  $|Q_{\phi} g^*_{\mu R} g_{e L}\delta^{\mu e}_{RL}|$
    &  $|Q_{\psi} g^*_{\mu R} g_{e L}\delta^{\mu e}_{RL}|$
    \\  
$\mu^+\to e^+ \gamma$ 
    & $0.007\,[0.0027]$ 
    &  $0.007\,[0.0027]$ 
    &  $38\,[14]\times 10^{-8}$ 
    &  $22\, [8]\times 10^{-8}$ 
    \\        
$\mu^+\to e^- e^+ e^-$
    & $0.152\,[0.0015]$ 
    &  $0.103\,[0.0010]$ 
    &  $739\, [7]\times 10^{-8}$ 
    &  $427\, [4]\times 10^{-8}$ 
    \\  
$\mu^- {\rm Au}\to e^- {\rm Au}$
    &  $0.069\,[0.0008]$ 
    &  $0.064\,[0.0008]$ 
    &  $777\,[9]\times 10^{-8}$ 
    &  $449\,[5]\times 10^{-8}$ 
    \\    
$\mu^- {\rm Ti}\to e^- {\rm Ti}$
    &  $0.177\,[0.0003]$ 
    &  $0.183\,[0.0003]$ 
    &  $1875\,[3]\times 10^{-8}$ 
    &  $1084\,[2]\times 10^{-8}$ 
    \\ 
$\mu^- {\rm Al}\to e^- {\rm Al}$
    &  $[0.0003]$ 
    &  $[0.0004]$ 
    &  $[4\times 10^{-8}]$ 
    &  $[2\times 10^{-8}]$ 
    \\  
    \hline             
    &  $|g^*_{\mu R} g_{e R} \Delta T_{3\psi}\delta^{\mu e}_{RR}|$
    &  $|g^*_{\mu R} g_{e R} g^*_{e R} g_{e R}\delta^{\mu e}_{RR}|$
    &  $|g^*_{\mu R} g_{e R} g^*_{e L} g_{e L}\delta^{\mu e}_{RR}|$
    \\        
$\mu^+\to e^- e^+ e^-$
    & $142\,[1]\times 10^{-5}$ 
    & $0.07\{0.54\}[7\{54\}\times 10^{-4}]$ 
    & $0.12\{0.76\}[1\{8\}\times 10^{-3}]$ 
    \\ 
$\mu^- {\rm Au}\to e^- {\rm Au}$
    &  $178\,[2] \times 10^{-6}$ 
    &  
    &  
    \\    
$\mu^- {\rm Ti}\to e^- {\rm Ti}$
    &  $623\,[1]\times 10^{-6}$ 
    &  
    &   
    \\ 
$\mu^- {\rm Al}\to e^- {\rm Al}$
    &  $[1\times 10^{-6}]$ 
    &  
    &   
    \\   
    \hline                             
    & $|Q_{\phi} g_{\tau R} g^*_{e R}\delta^{e\tau}_{RR}|$
    & $|Q_{\psi} g_{\tau R} g^*_{e R}\delta^{e\tau}_{RR}|$ 
    & $|Q_{\phi} g_{\tau R} g^*_{e L}\delta^{e\tau}_{LR}|$ 
    & $|Q_{\psi} g_{\tau R} g^*_{e L}\delta^{e\tau}_{LR}|$
    \\   
$\tau^-\to e^- \gamma$ 
    &  $4.8\,[1.4]$ 
    &  $4.7\,[1.4]$ 
    &  $4.2\,[1.3]\times 10^{-3}$ 
    &  $24\,[7]\times 10^{-4}$ 
    \\               
$\tau^-\to e^- e^+ e^-$
    &  $43.7\,[5.5]$ 
    &  $33.9\,[4.3]$ 
    &  $37\,[5]\times 10^{-3}$ 
    &  $21\,[3]\times 10^{-3}$ 
    \\
    \hline
    &  $|g_{\tau R} g^*_{e R} \Delta T_{3\psi}\delta^{e\tau}_{RR}|$
    &  $|g_{\tau R} g^*_{e R} g_{e R} g^*_{e R}\delta^{e\tau}_{RR}|$
    &  $|g_{\tau R} g^*_{e R} g_{e L} g^*_{e L}\delta^{e\tau}_{RR}|$ 
    \\ 
$\tau^-\to e^- e^+ e^-$
    & $0.56\,[0.07]$ 
    & $28.3\,\{210.5\}[3.6\,\{26.6\}]$ 
    & $46.3\,\{297.7\}[5.8\,\{37.6\}]$ 
    \\  
    \hline         
    &  $|Q_{\phi} g_{\tau R} g^*_{\mu R}\delta^{\mu\tau}_{RR}|$
    &  $|Q_{\psi} g_{\tau R} g^*_{\mu R}\delta^{\mu\tau}_{RR}|$
    &  $|Q_{\phi} g_{\tau R} g^*_{\mu L}\delta^{\mu\tau}_{LR}|$
    &  $|Q_{\psi} g_{\tau R} g^*_{\mu R}\delta^{\mu\tau}_{LR}|$
    \\
$\tau^-\to \mu^- \gamma$ 
    &  $5.5\,[0.8]$ 
    &  $5.4\,[0.8]$ 
    &  $49\,[7]\times 10^{-4}$ 
    &  $28\,[4]\times 10^{-4}$ 
    \\   
$\tau^-\to \mu^-\mu^+\mu^-$
    & $38.5\,[4.8]$ 
    & $29.9\,[3.7]$ 
    & $33\,[4]\times 10^{-3}$ 
    & $19\,[2]\times 10^{-3}$ 
    \\  
    \hline             
    &  $|g_{\tau R} g^*_{\mu R} \Delta T_{3\psi}\delta^{\mu\tau}_{RR}|$
    &  $|g_{\tau R} g^*_{\mu R} g_{\mu R} g^*_{\mu R}\delta^{\mu\tau}_{RR}|$
    &  $|g_{\tau R} g^*_{\mu R} g_{\mu L} g^*_{\mu L}\delta^{\mu\tau}_{RR}|$
     \\ 
$\tau^-\to \mu^-\mu^+\mu^-$
    & $0.49\,[0.06]$ 
    & $25.0\,\{185.6\}\,[3.1\,\{23.3\}]$ 
    & $40.9\,\{262.5\}\,[5.1\,\{32.9\}]$ 
     \\     
     \hline      
    &  $|g_{\tau R} g^*_{e R} g_{\mu R} g^*_{e R}\delta^{e\tau}_{RR}\delta^{e\mu}_{RR}|$
    &  $|g_{\tau R} g^*_{e R} g_{\mu L} g^*_{e L}\delta^{e\tau}_{RR}\delta^{e\mu}_{LL}|$
    &  $|g_{\tau R} g^*_{e L} g_{\mu R} g^*_{e L}\delta^{e\tau}_{LR}\delta^{e\mu}_{LR}|$
    &  $|g_{\tau R} g^*_{e L} g_{\mu L} g^*_{e R}\delta^{e\tau}_{LR}\delta^{e\mu}_{RL}|$
    \\  
$\tau^-\to e^-\mu^+ e^-$
    & $41.9\,\{194.1\}\,[5.3\,\{24.5\}]$ 
    & $48.7\,\{274.5\}\,[6.2\,\{34.7\}]$ 
    & $68.9\,\{68.9\}\,[8.7\,\{8.7\}]$ 
    & $59.3\,\{29.6\}\,[7.5\,\{3.7\}]$ 
    \\   
    \hline    
    &  $|g_{\tau R} g^*_{\mu R} g_{e R} g^*_{\mu R}\delta^{\mu\tau}_{RR}\delta^{\mu e}_{RR}|$
    &  $|g_{\tau R} g^*_{\mu R} g_{e L} g^*_{\mu L}\delta^{\mu\tau}_{RR}\delta^{\mu e}_{LL}|$
    &  $|g_{\tau R} g^*_{\mu L} g_{e R} g^*_{\mu L}\delta^{\mu\tau}_{LR}\delta^{\mu e}_{LR}|$
    &  $|g_{\tau R} g^*_{\mu L} g_{e L} g^*_{\mu R}\delta^{\mu\tau}_{LR}\delta^{\mu e}_{RL}|$
   \\  
$\tau^-\to \mu^- e^+\mu^+$
    & $44.6\,\{206.6\}\,[5.6\,\{26.0\}]$ 
    & $51.9\,\{292.2\}\,[6.5\,\{36.8\}]$ 
    & $73.4\,\{73.4\}\,[9.2\,\{9.4\}]$ 
    & $63.1\,\{31.5\}\,[7.9\,\{4.0\}]$ 
    \\      
    \hline                  
\end{tabular}
\end{center}
}
\end{table}

In Table~\ref{tab: results case II x=1}, we show the constraints on parameters in case II using $x\equiv m_\phi/m_\psi=1$ and $m_\psi=500$~GeV. 
Constraints for other $m_\psi$ can be obtained by scaling the results in the table by a $(\frac{m_\psi}{500 {\rm GeV}})^2$ or a $\frac{m_\psi}{500 {\rm GeV}}$ factor, where the latter is for $Q_{\phi,\psi} g^*_{l^{(\prime)} R} g_{l L}$.
Results in $[...]$ are obtained by using the projected sensitivities for future experiments. 
For box contributions both results of Dirac and Majorana fermion are given, where results in $\{...\}$ are for the Majorana case. 
Results for $x=0.5$ and 2 are given in Tables~\ref{tab: results case II x=0.5} and \ref{tab: results case II x=2}, respectively.

\begin{figure}[t!]
\centering
\subfigure[]{
  \includegraphics[width=6.7cm]{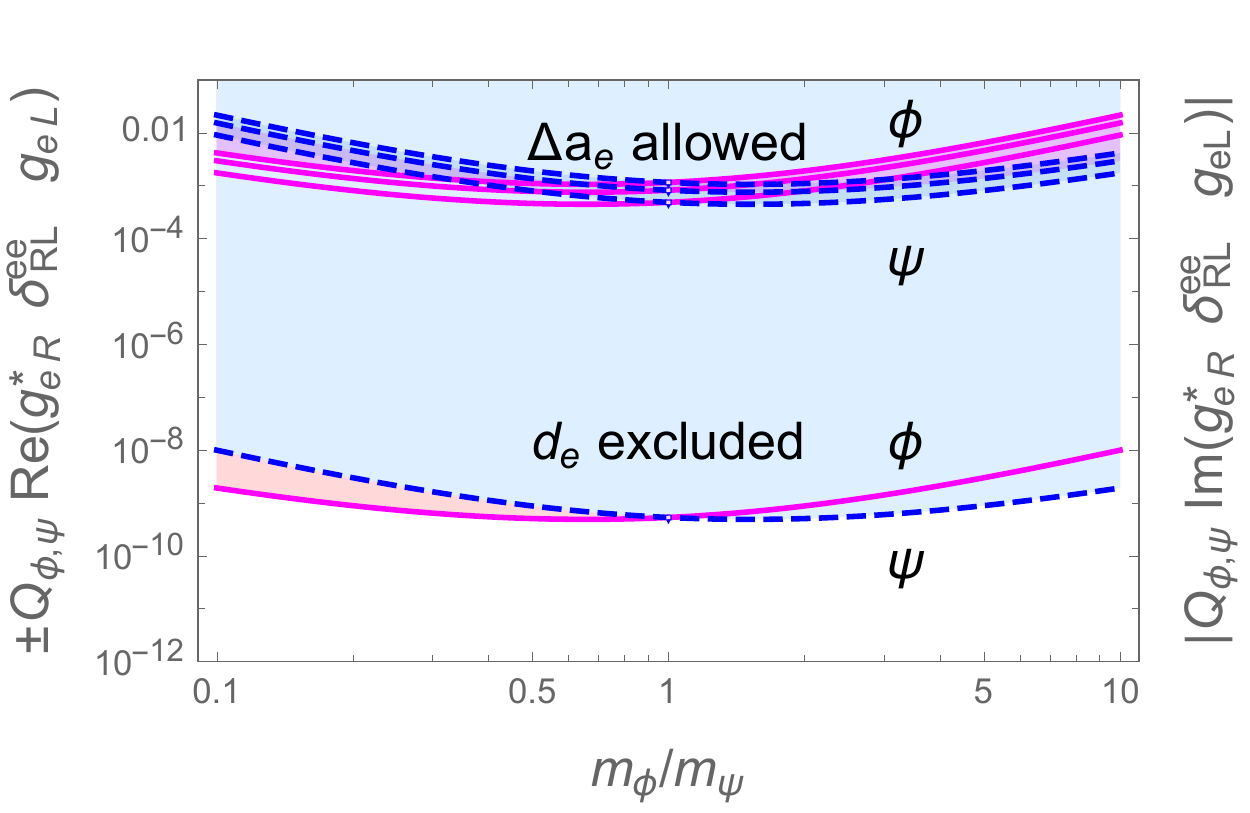}
}
\hspace{0.5cm}
\subfigure[]{
  \includegraphics[width=7cm]{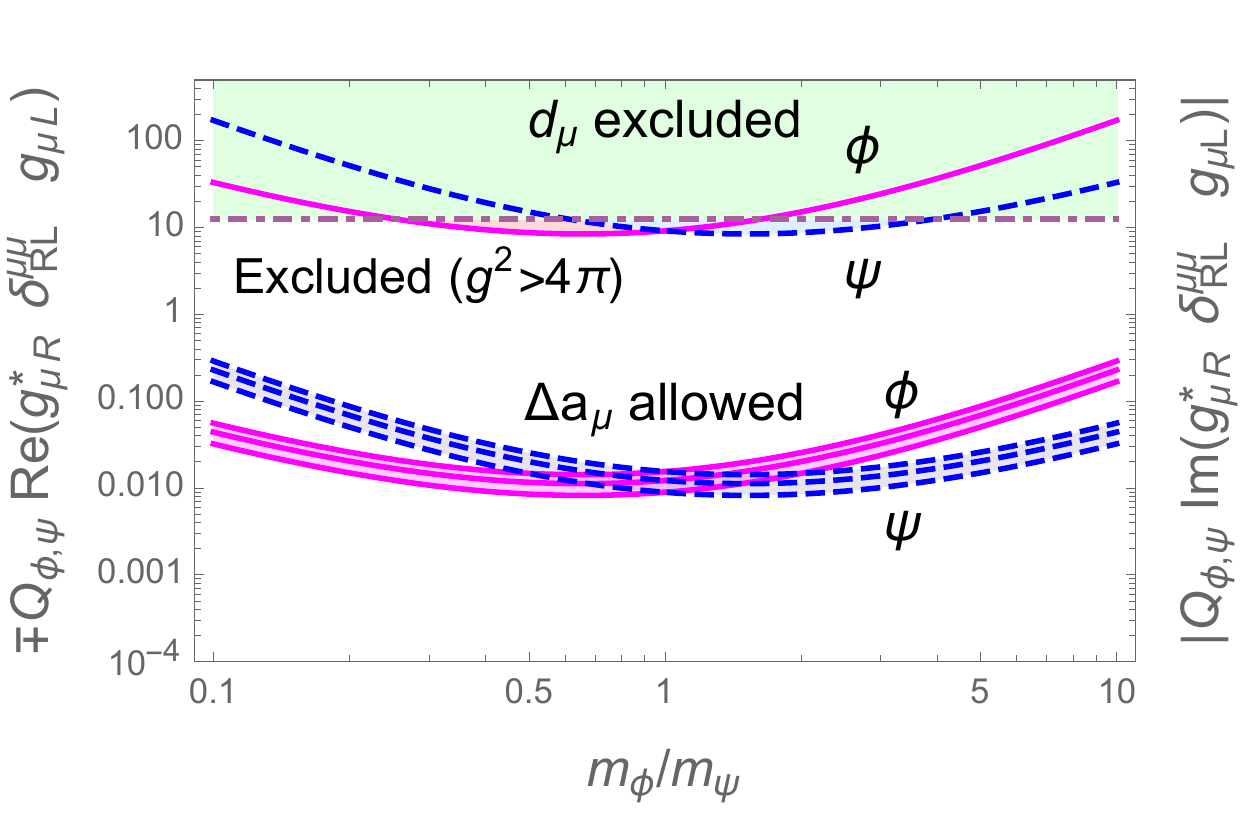}
}
\caption{
Allowed parameter space for
(a) $\pm Q_{\phi,\psi} {\rm Re}(g^*_{e R} g_{e L}\delta_{RL}^{ee})$ and 
$Q_{\phi,\psi}{\rm  Im} (g^*_{e R} g_{e L}\delta_{RL}^{ee})|$ constrained by 
$\Delta a_\mu$ and $d_e$, respectively,
and 
(b) $\mp Q_{\phi,\psi} {\rm Re}(g^*_{\mu R} g_{\mu L}\delta_{RL}^{\mu\mu})$ and 
$Q_{\phi,\psi}{\rm  Im} (g^*_{\mu R} g_{\mu L}\delta_{RL}^{\mu\mu})|$ constrained by 
$\Delta a_\mu$ and $d_\mu$, respectively.
These constrains are obtained using $m_\psi=500$~GeV, for other $m_\psi$, apply $(100{\rm GeV})/m_\psi$ to the plots.
 }
\label{fig:e muon g-2 2}
\end{figure}

In Fig.~\ref{fig:e muon g-2 2}, the allowed parameter space for
(a) $\pm Q_{\phi,\psi} {\rm Re}(g^*_{e R} g_{e L}\delta_{RL}^{ee})$ and 
$|Q_{\phi,\psi}{\rm  Im} (g^*_{e R} g_{e L}\delta_{RL}^{ee})|$ constrained by 
$\Delta a_\mu$ and $d_e$, respectively,
and 
(b) $\mp Q_{\phi,\psi} {\rm Re}(g^*_{\mu R} g_{\mu L}\delta_{RL}^{\mu\mu})$ and 
$|Q_{\phi,\psi}{\rm  Im} (g^*_{\mu R} g_{\mu L}\delta_{RL}^{\mu\mu})|$ constrained by 
$\Delta a_\mu$ and $d_\mu$, respectively, are shown.
These constrains are obtained using $m_\psi=500$~GeV. For other $m_\psi$, apply a $(500{\rm GeV})/m_\psi$ factor to the plots.

\begin{figure}[th!]
\centering
\subfigure[]{
  \includegraphics[width=6.5cm]{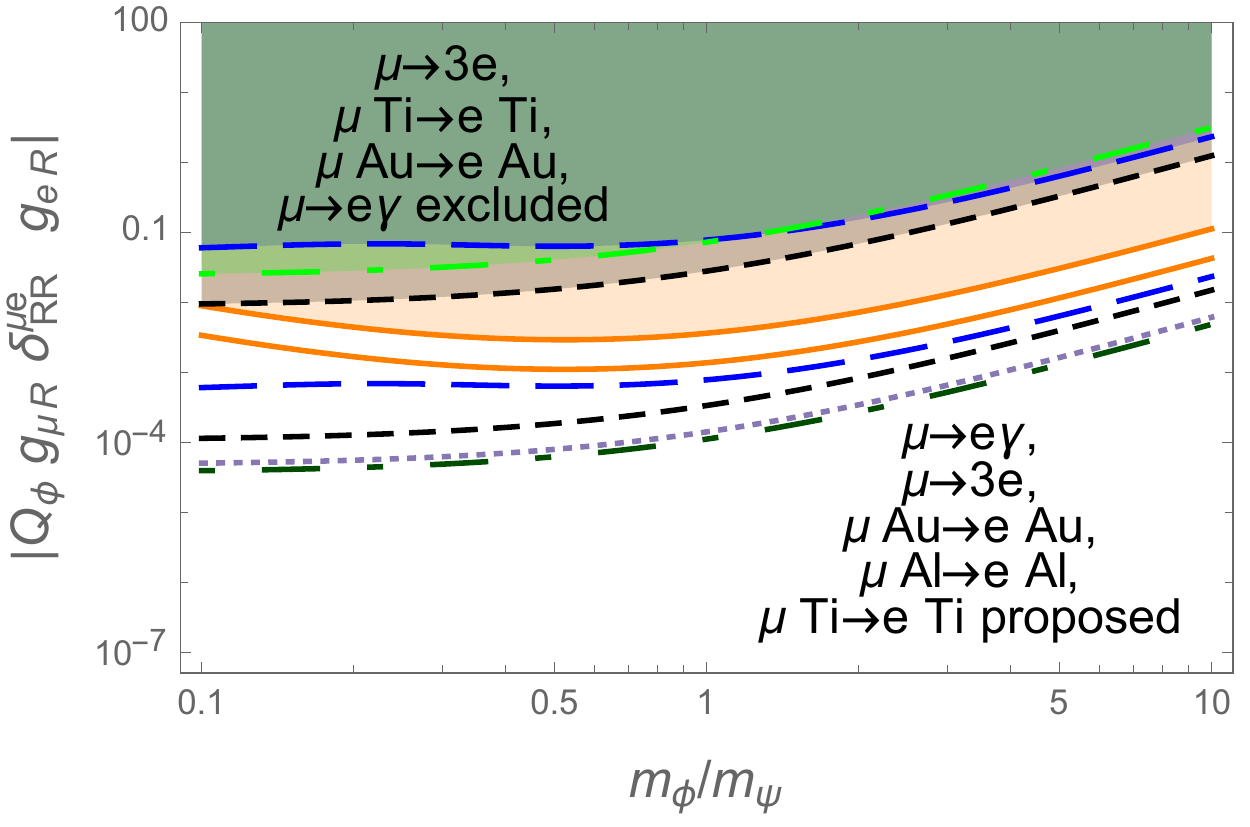}
}
\hspace{0.5cm}
\subfigure[]{
 \includegraphics[width=6.5cm]{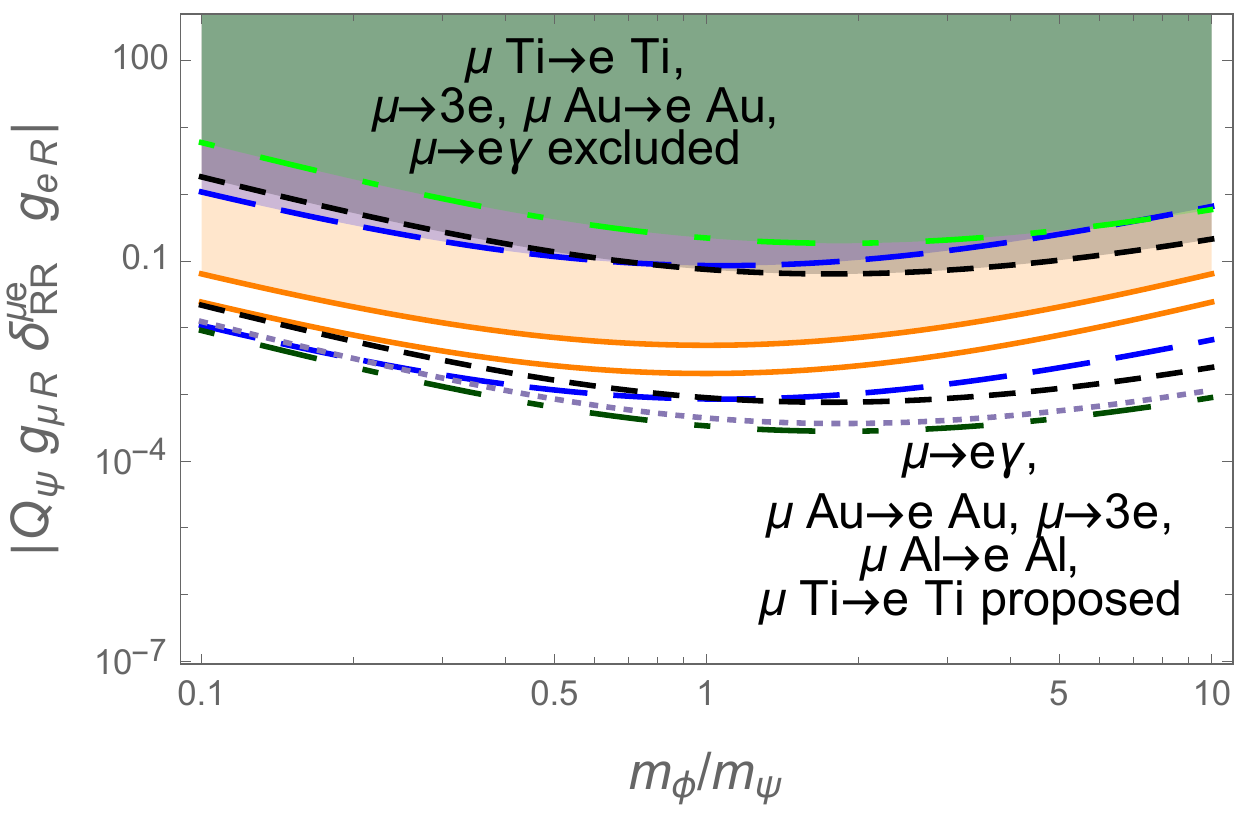}
}
\subfigure[]{
  \includegraphics[width=6.5cm]{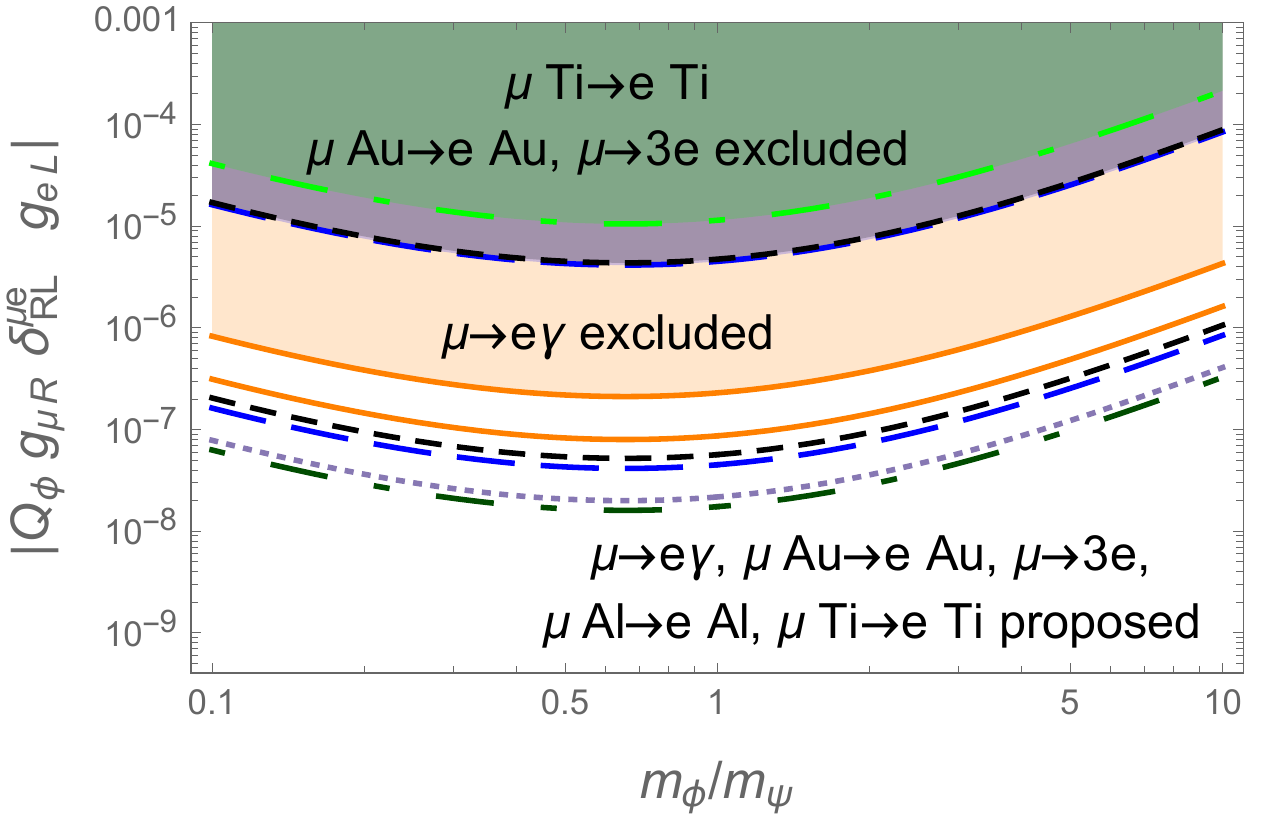}
}
\hspace{0.5cm}
\subfigure[]{
 \includegraphics[width=6.5cm]{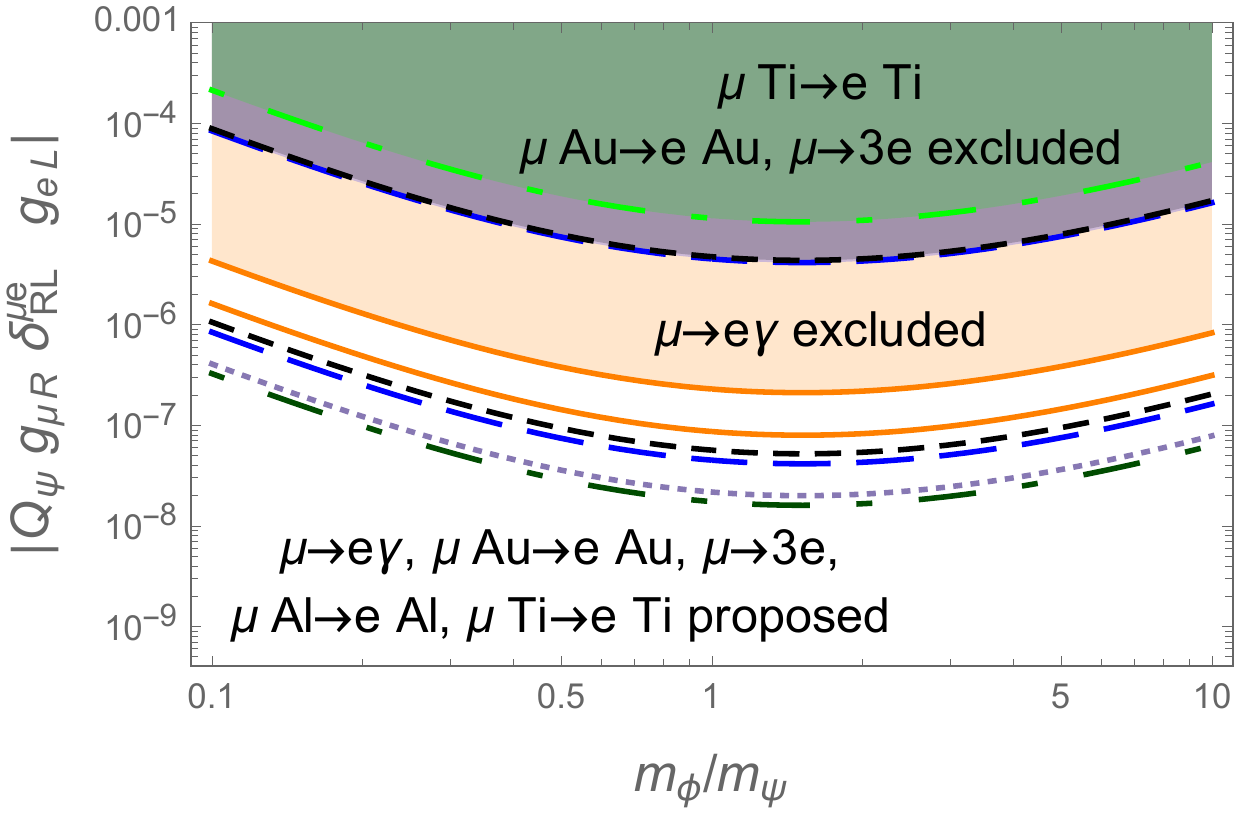}
}
\subfigure[]{
  \includegraphics[width=6.5cm]{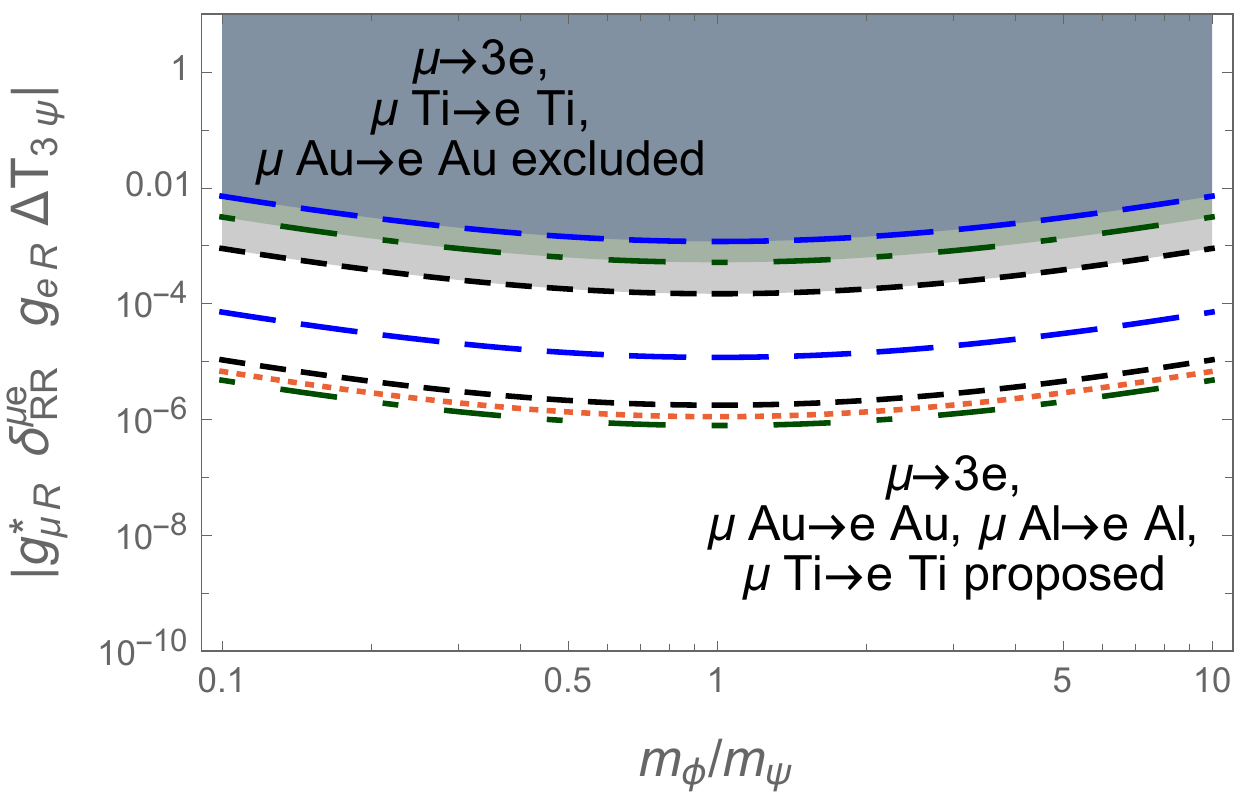}
}
\hspace{0.5cm}
\subfigure[]{
  \includegraphics[width=6.5cm]{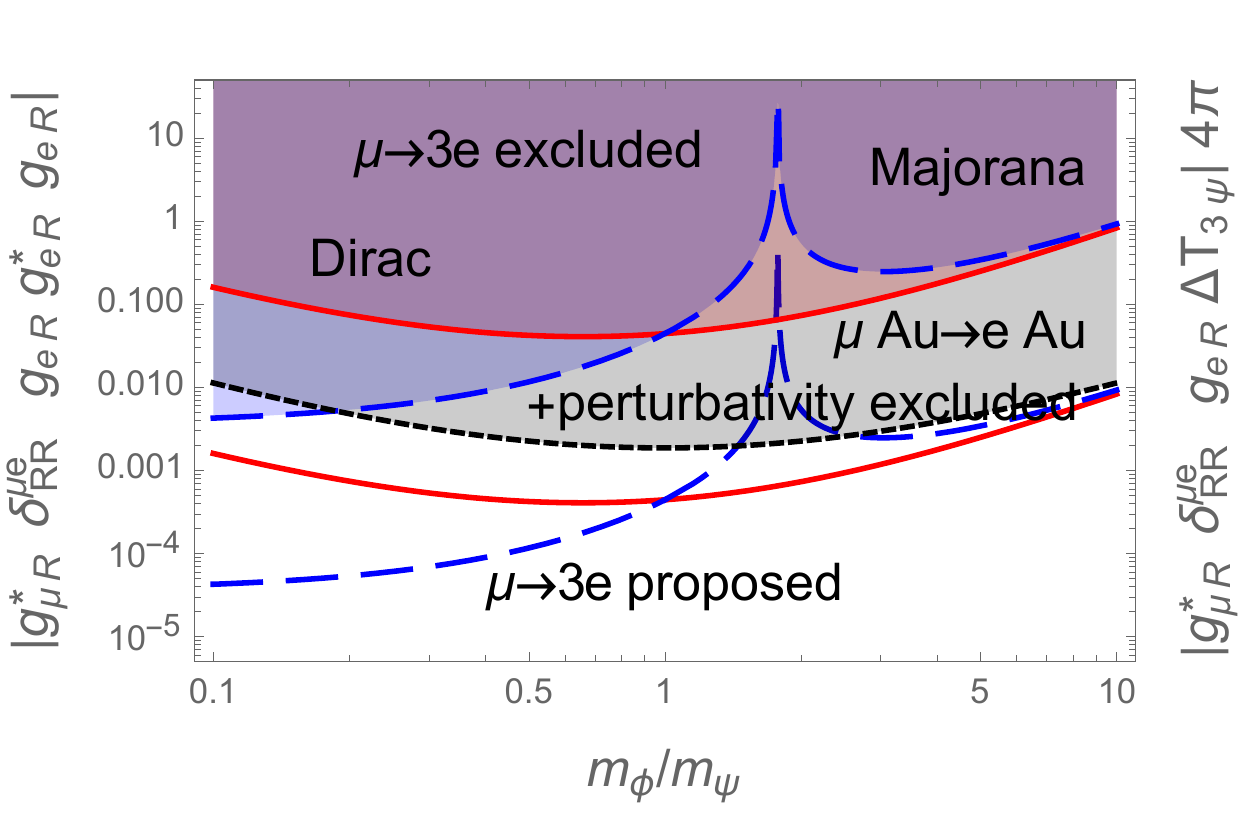}
}
\subfigure[]{
  \includegraphics[width=6.5cm]{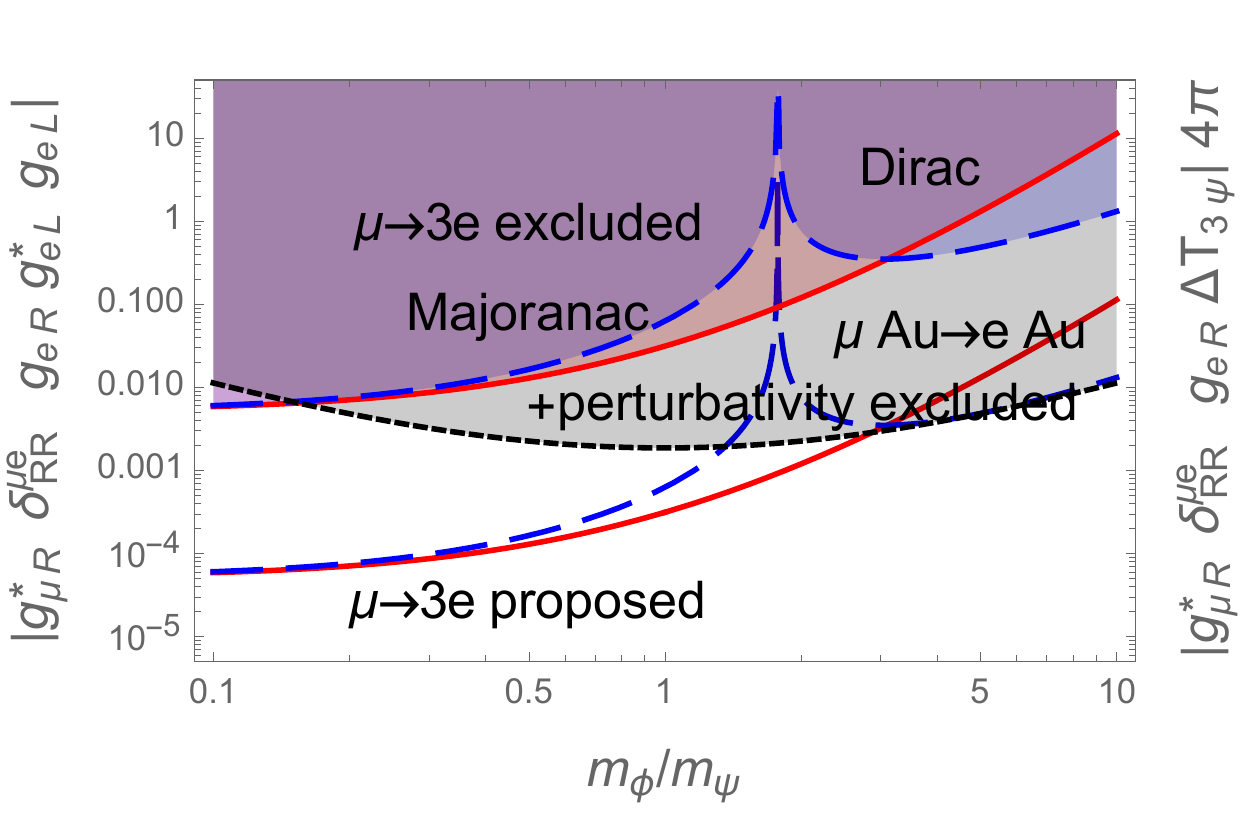}
}
\caption{Same as Fig.~\ref{fig:mu2eLFV}, but for case II.
}
\label{fig:mu2eLFV2}
\end{figure}

\begin{figure}[ht!]
\centering
\subfigure[]{
  \includegraphics[width=6.5cm]{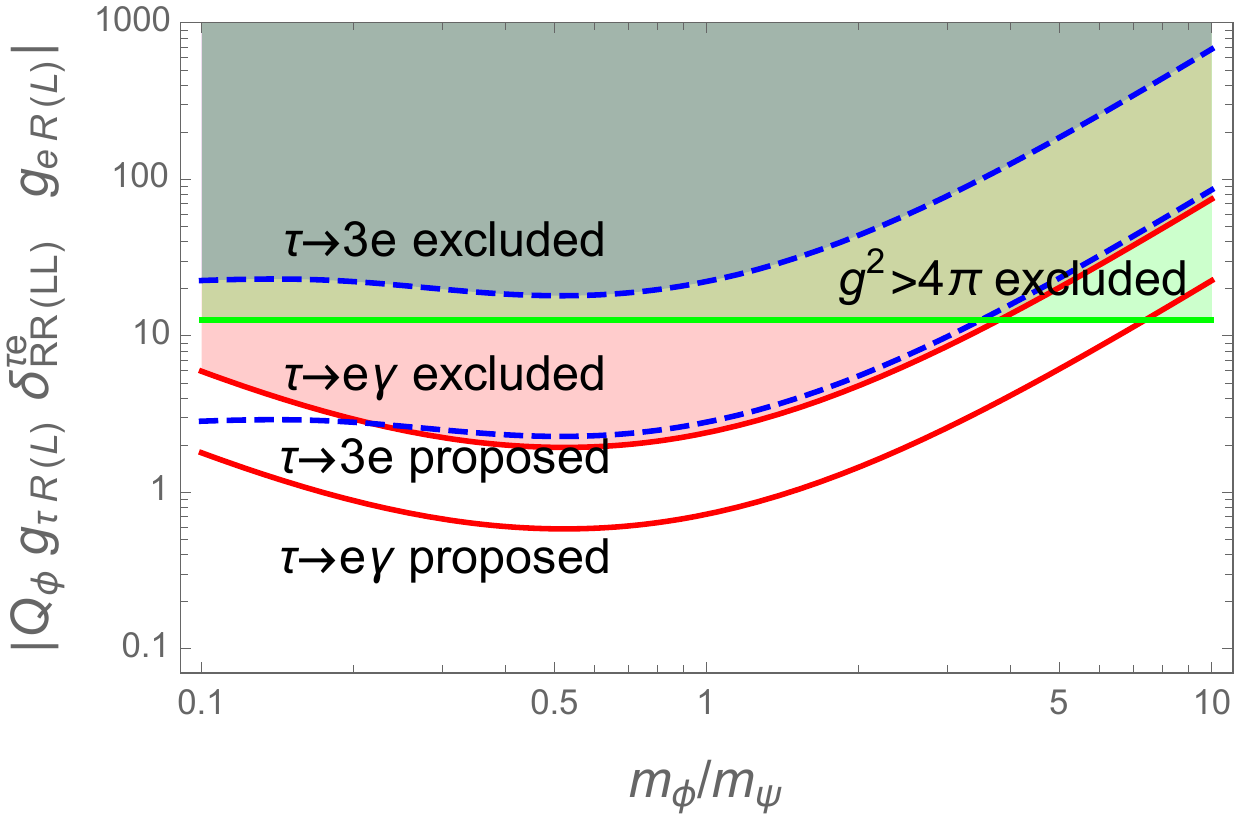}
}
\hspace{0.5cm}
\subfigure[]{
  \includegraphics[width=6.5cm]{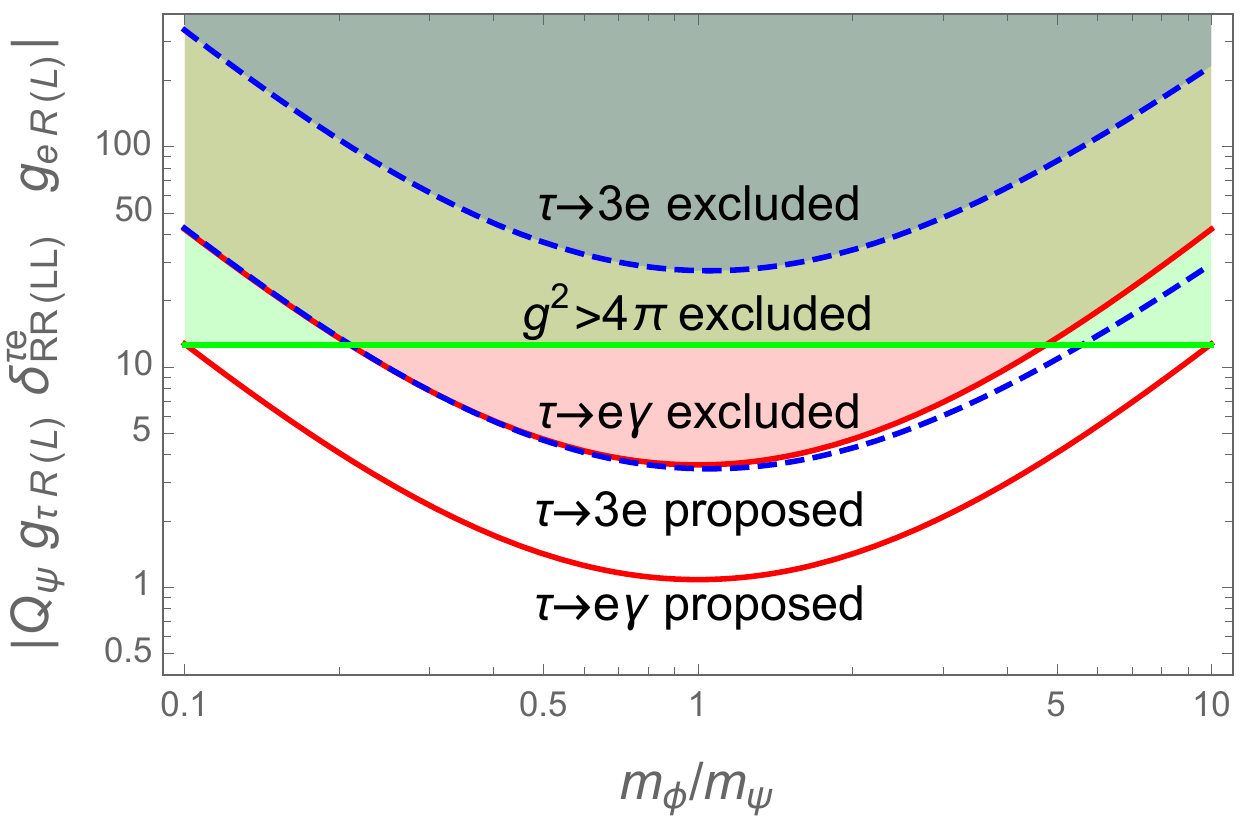}
}
\subfigure[]{
  \includegraphics[width=6.5cm]{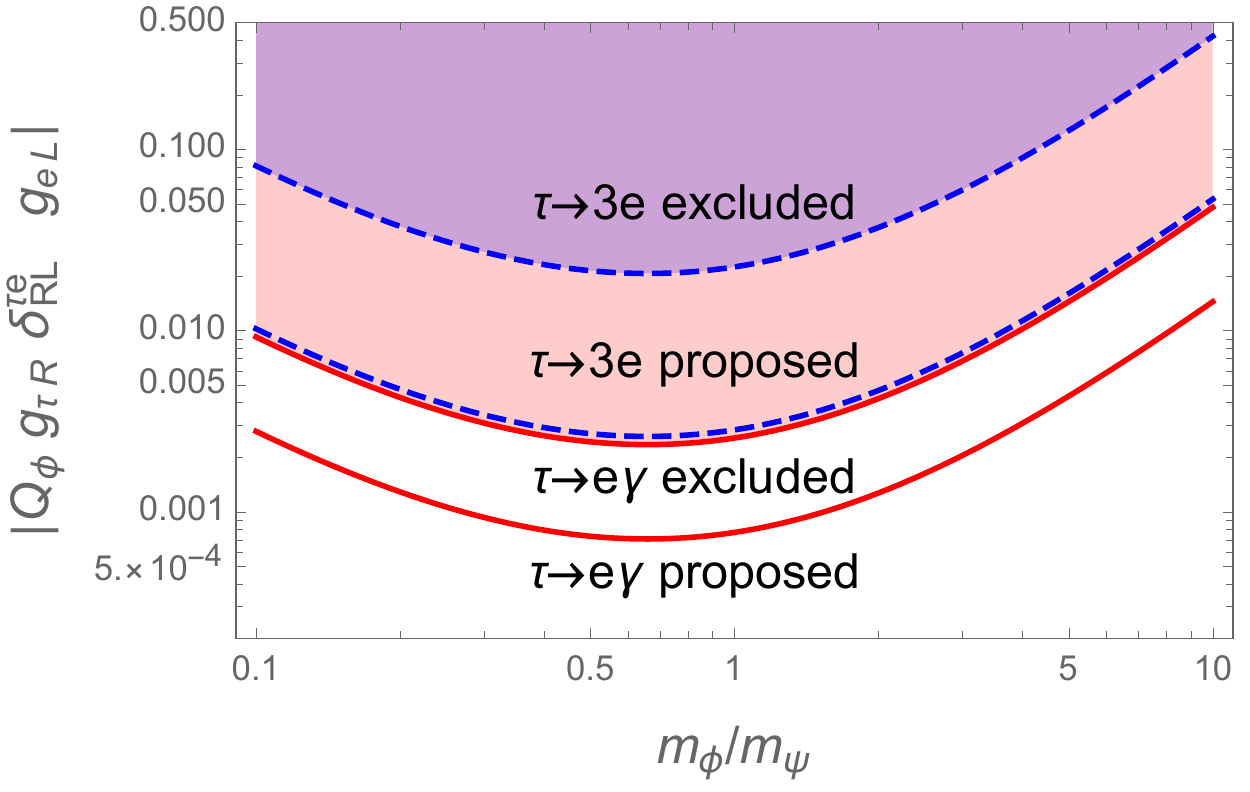}
}
\hspace{0.5cm}
\subfigure[]{
  \includegraphics[width=6.5cm]{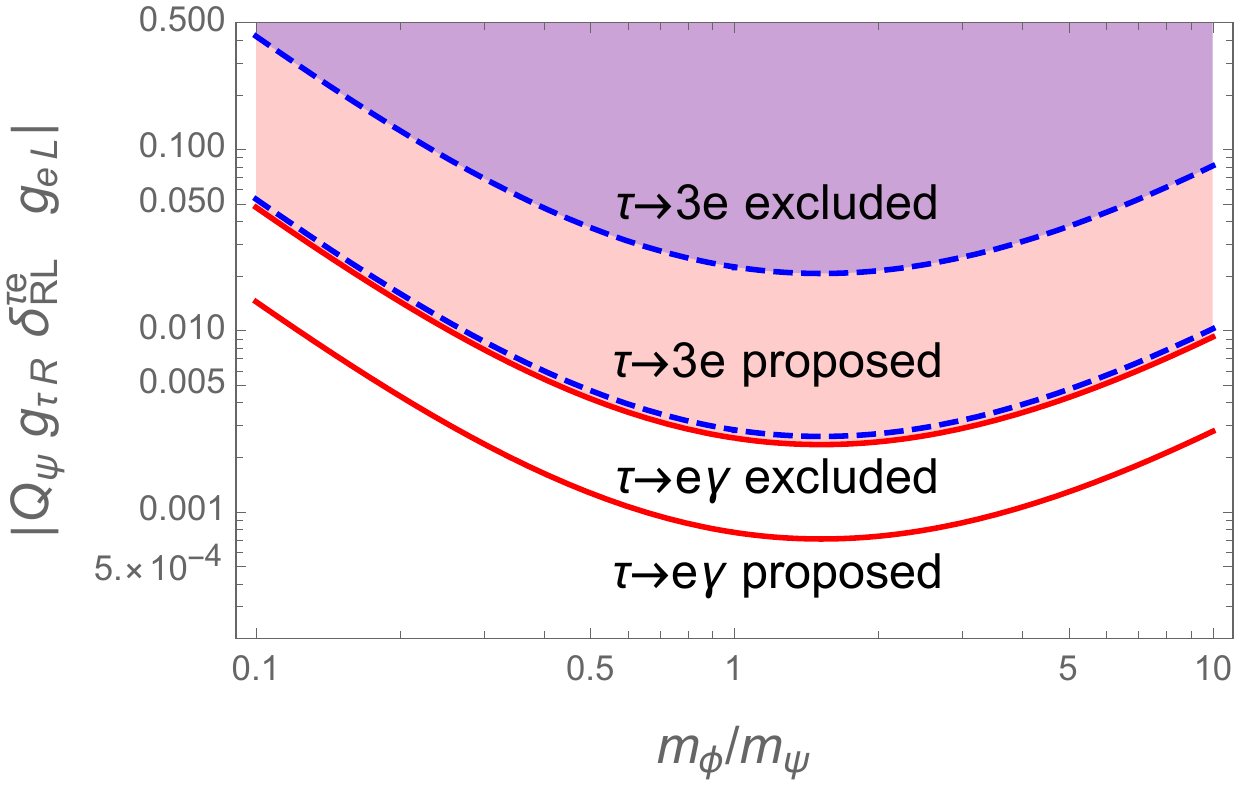}
}
\subfigure[]{
  \includegraphics[width=6.5cm]{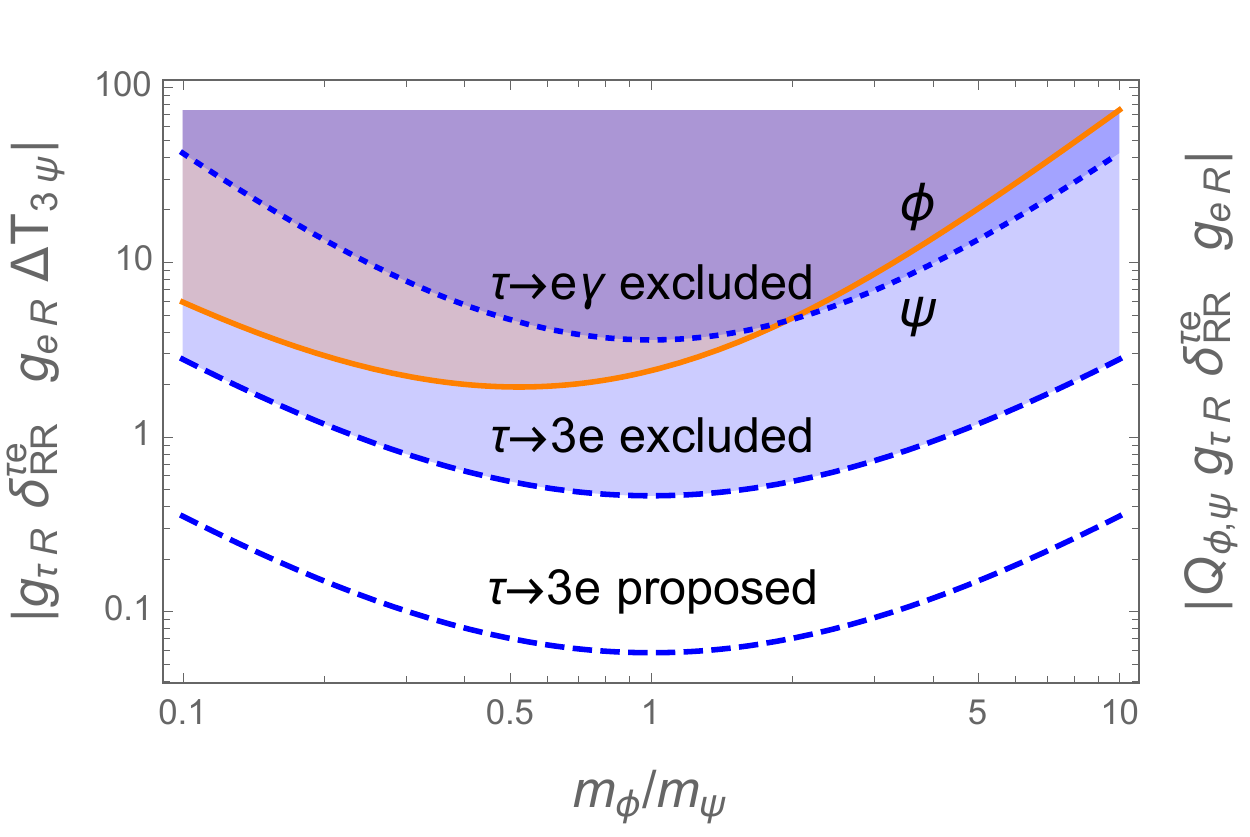}
}
\hspace{0.5cm}
\subfigure[]{
  \includegraphics[width=6.5cm]{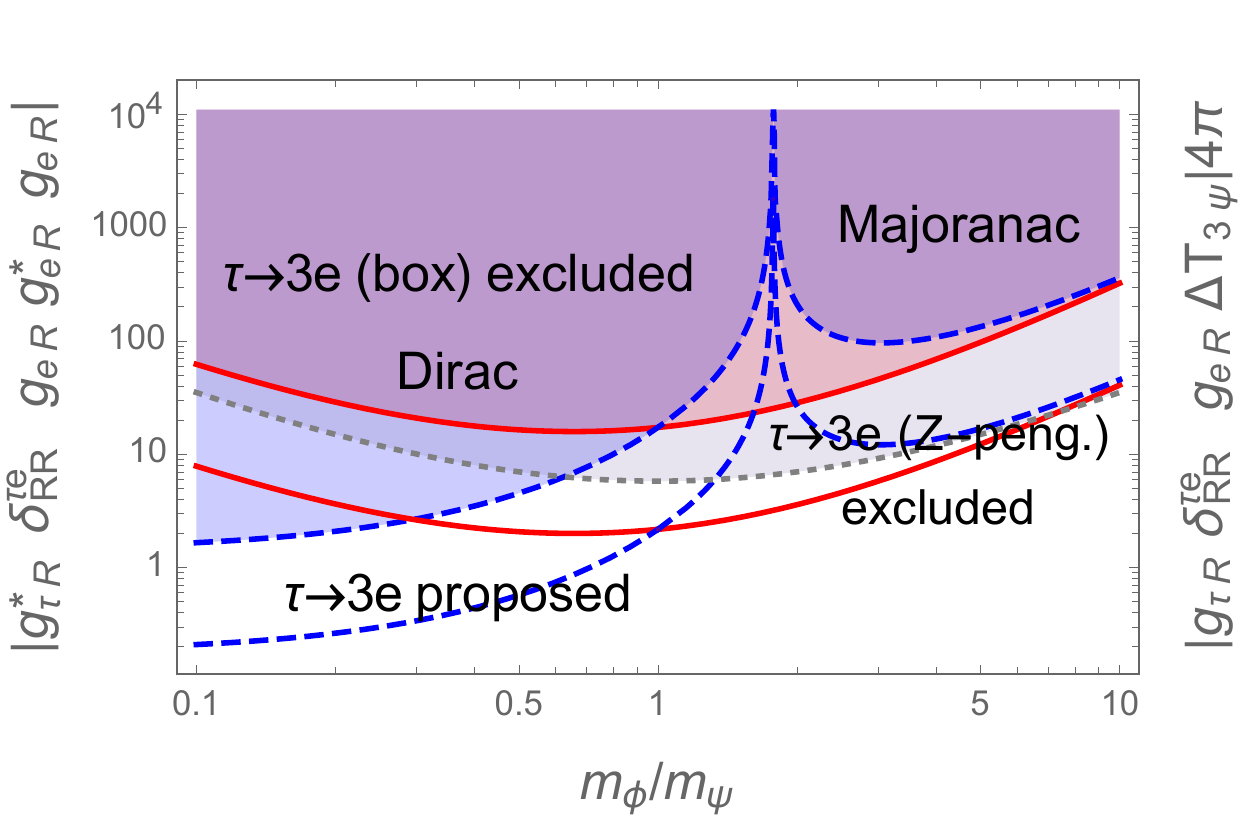}
}
\subfigure[]{
  \includegraphics[width=6.5cm]{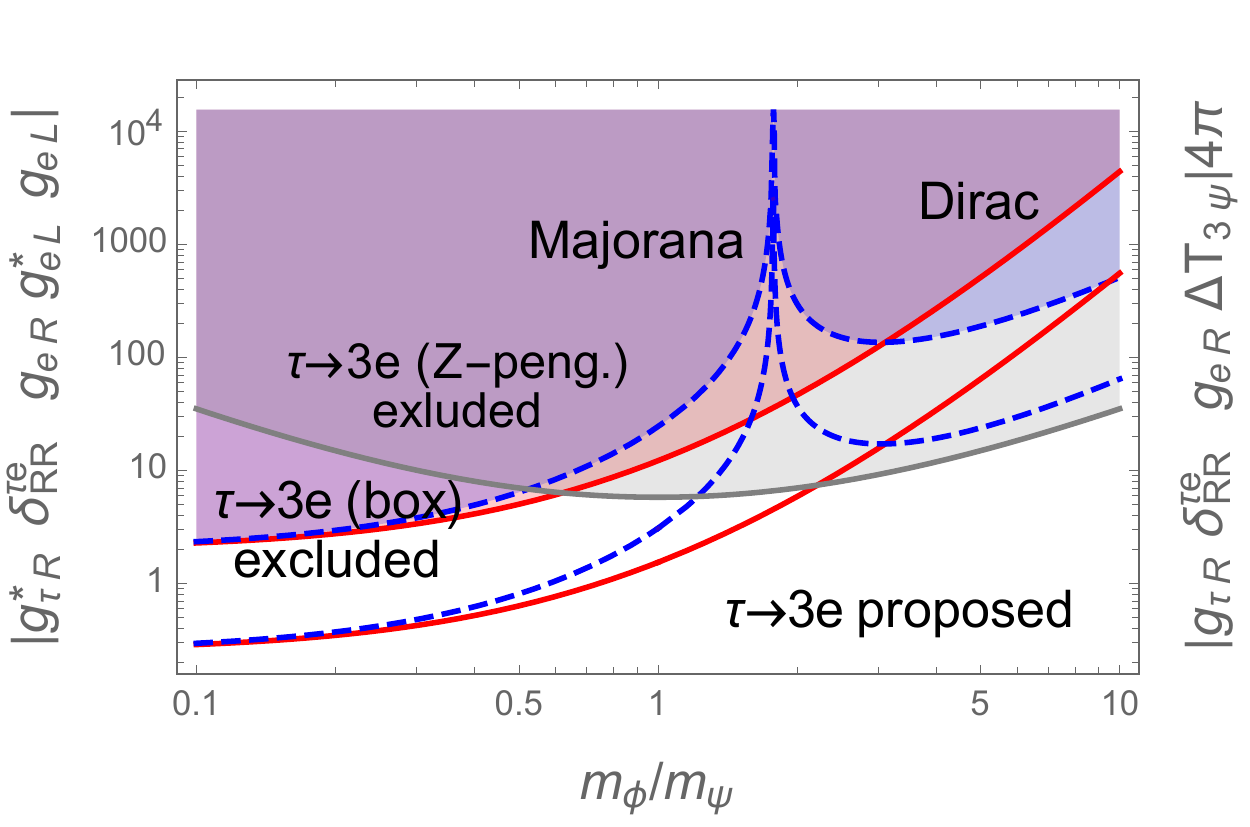}
}
\caption{Same as Fig.~\ref{fig:tau2eLFV}, but for case II.
}
\label{fig:tau2eLFV2}
\end{figure}

\begin{figure}[ht!]
\centering
\subfigure[]{
  \includegraphics[width=6.5cm]{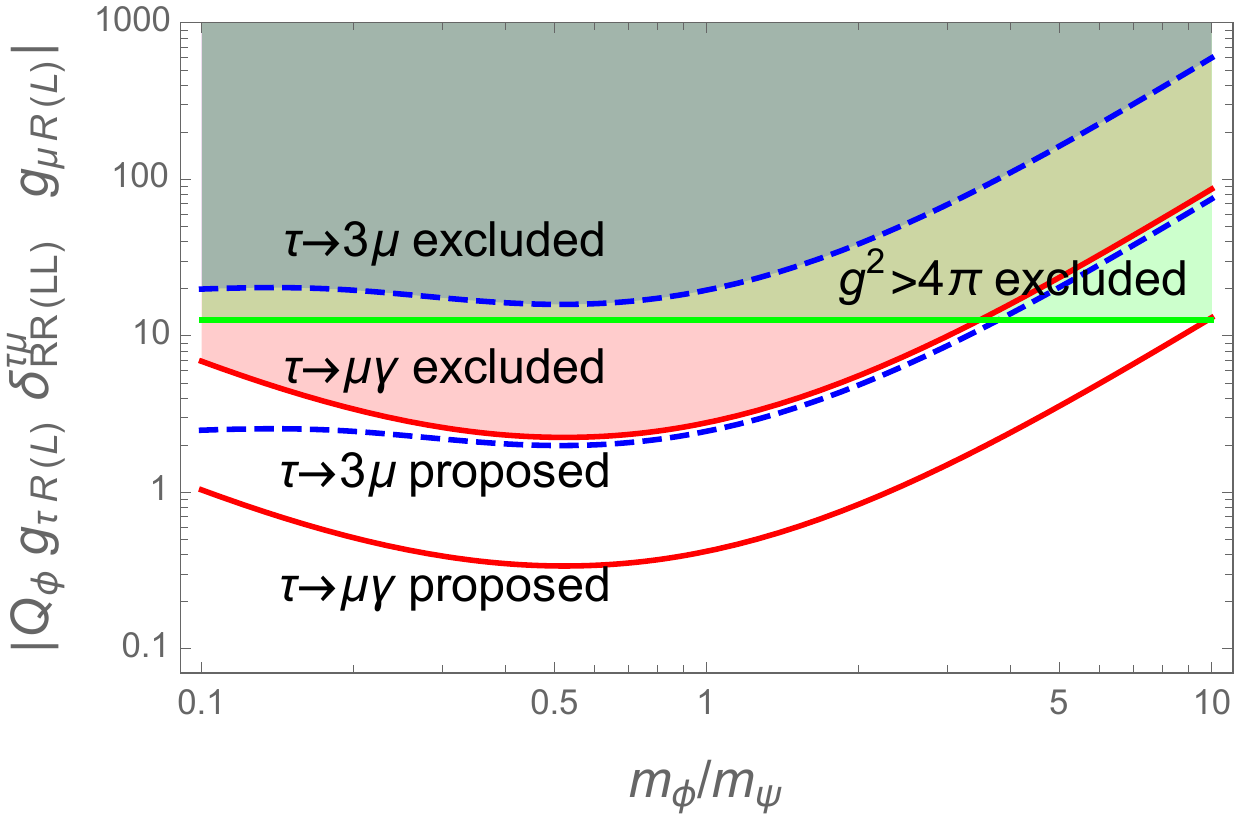}
}
\hspace{0.5cm}
\subfigure[]{
  \includegraphics[width=6.5cm]{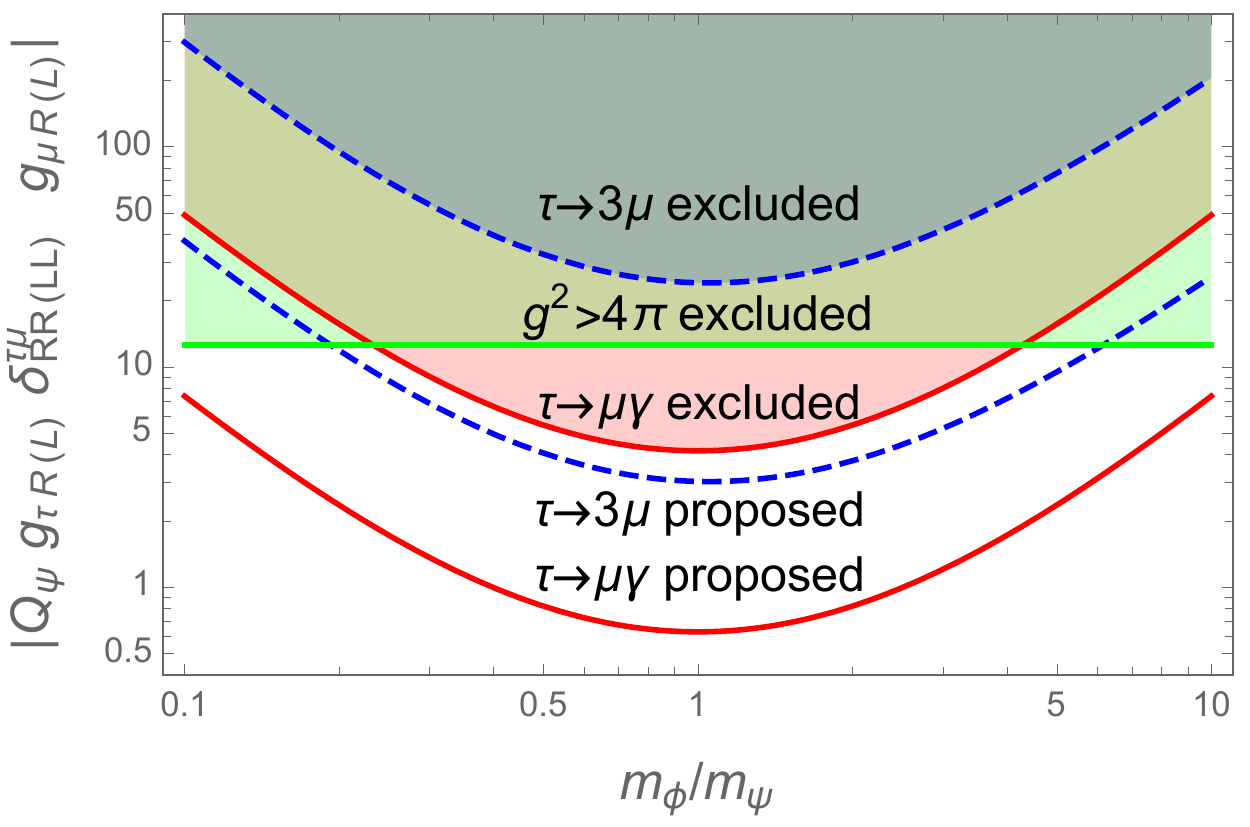}
}
\subfigure[]{
  \includegraphics[width=6.5cm]{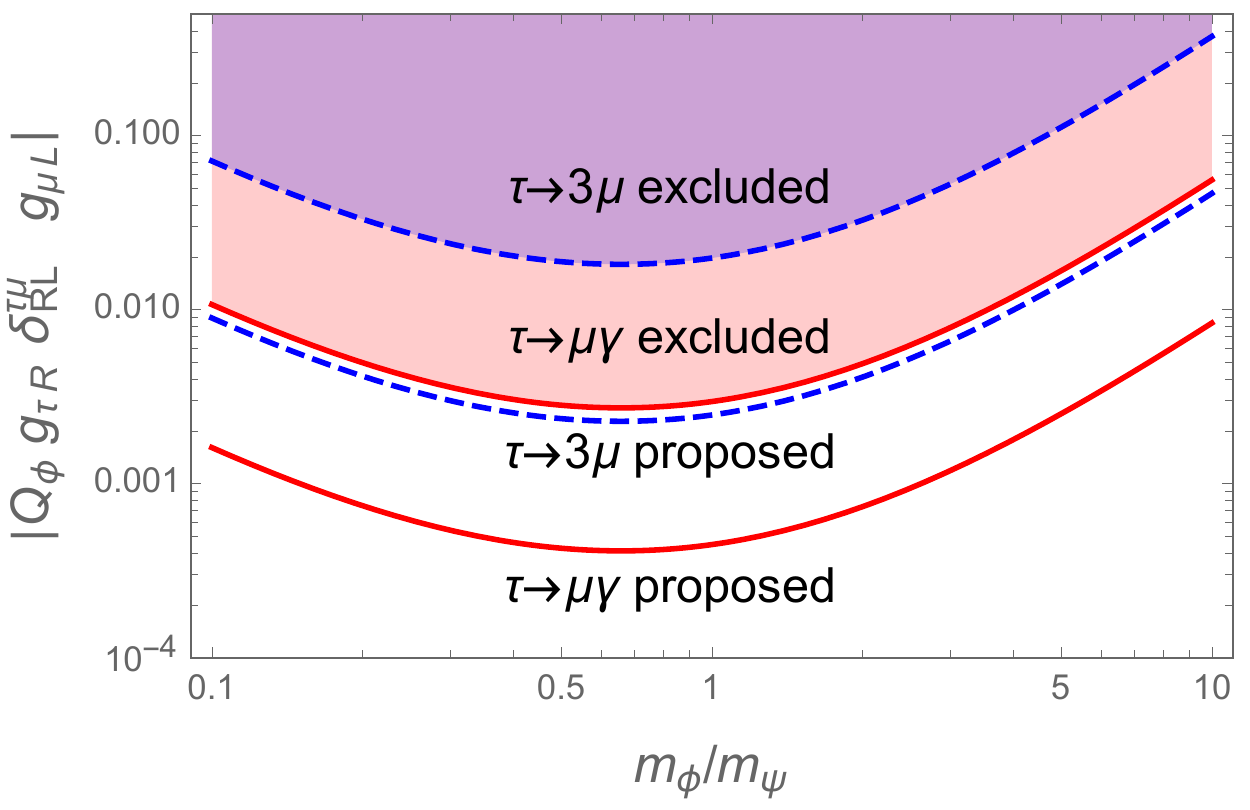}
}
\hspace{0.5cm}
\subfigure[]{
  \includegraphics[width=6.5cm]{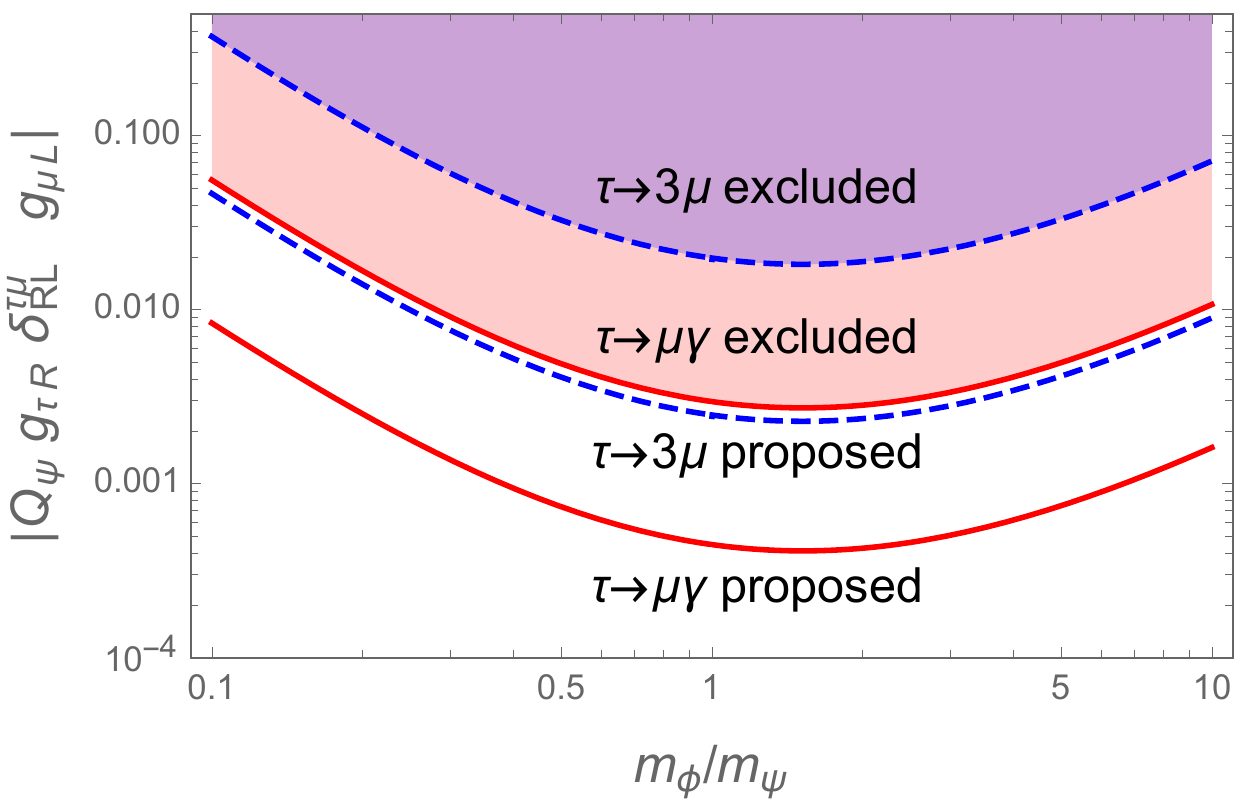}
}
\subfigure[]{
  \includegraphics[width=6.5cm]{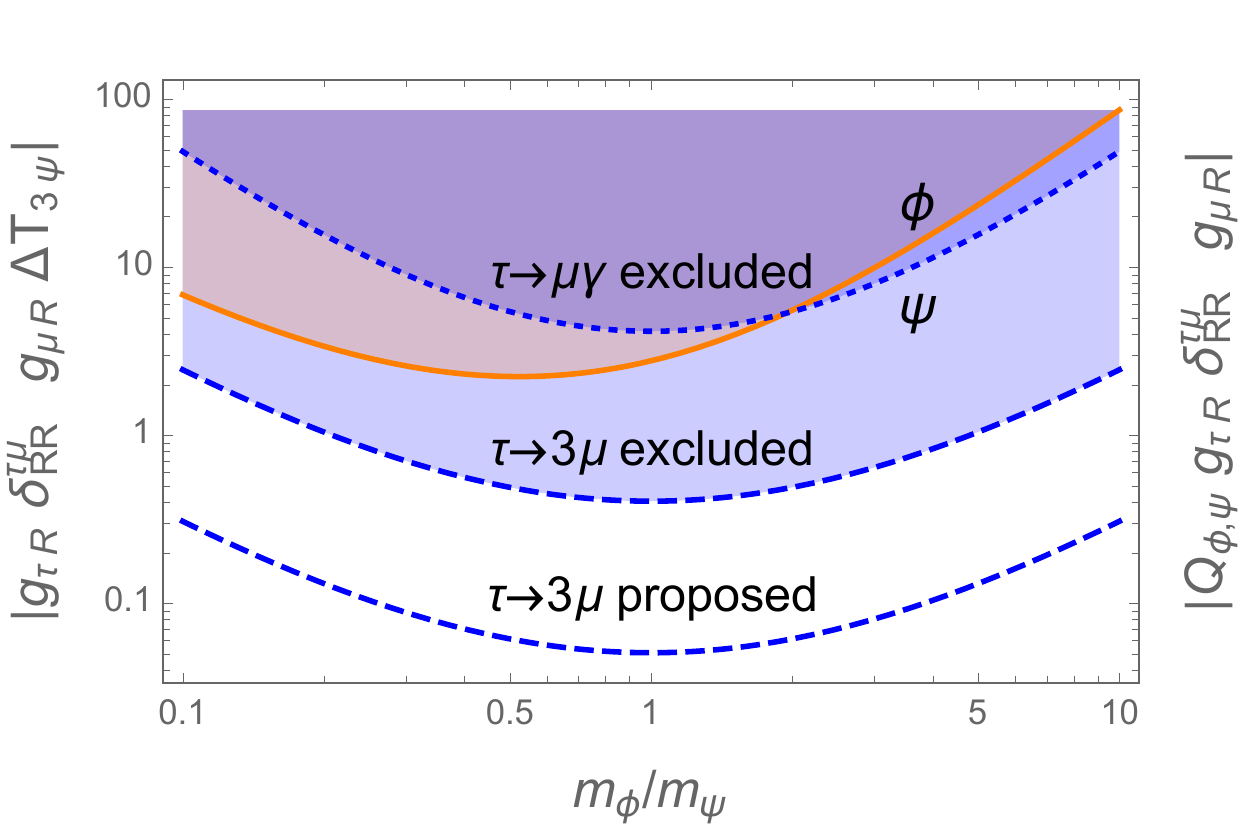}
}
\hspace{0.5cm}
\subfigure[]{
  \includegraphics[width=6.5cm]{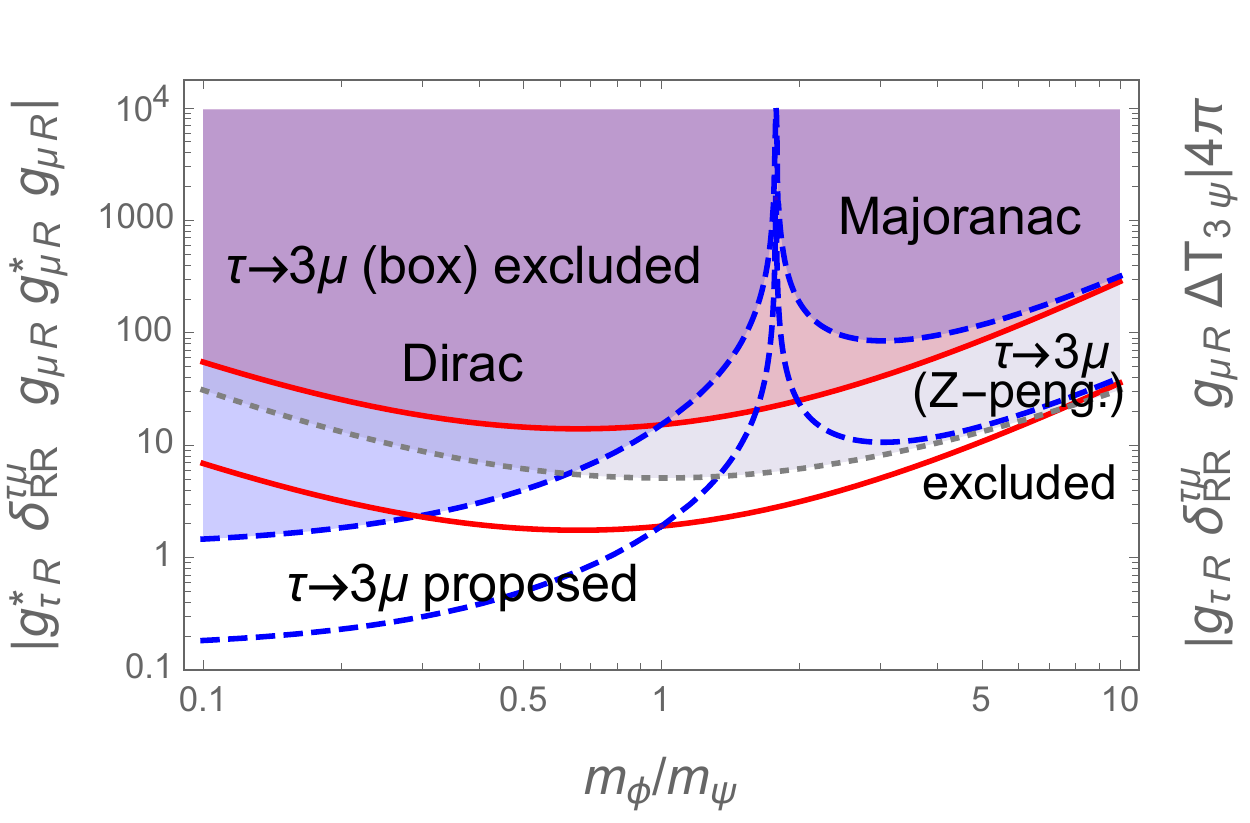}
}
\subfigure[]{
  \includegraphics[width=6.5cm]{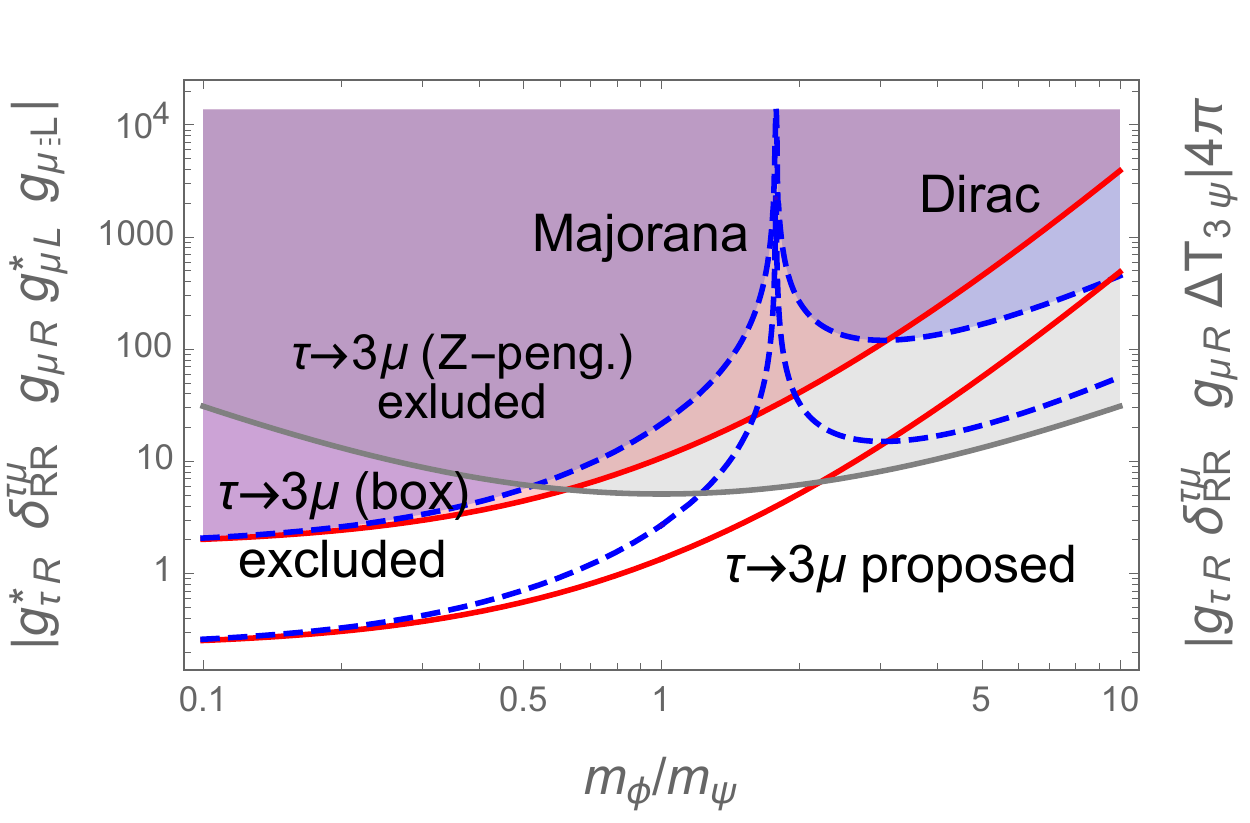}
}
\caption{Same as Fig.~\ref{fig:tau2muLFV}, but for case II.
}
\label{fig:tau2muLFV2}
\end{figure}

\begin{figure}[ht!]
\centering
\subfigure[]{
  \includegraphics[width=6.5cm]{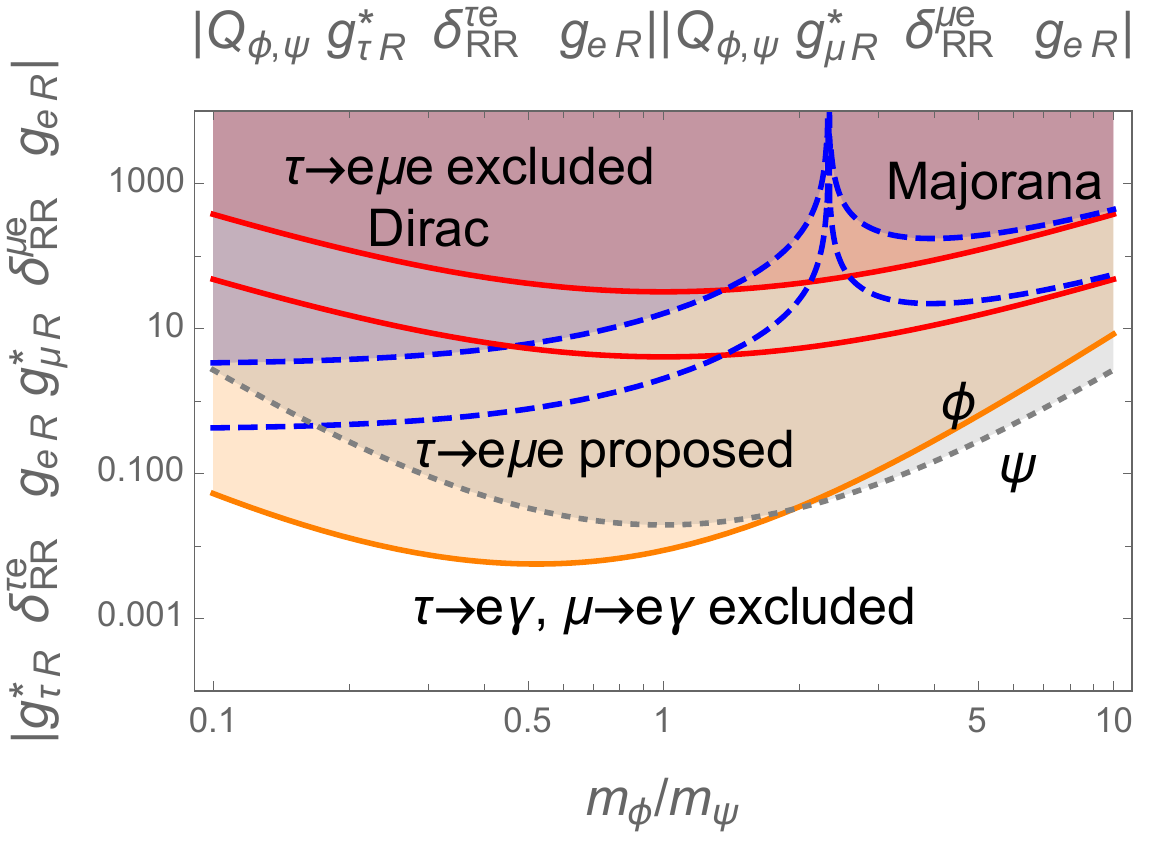}
}
\hspace{0.5cm}
\subfigure[]{
  \includegraphics[width=6.5cm]{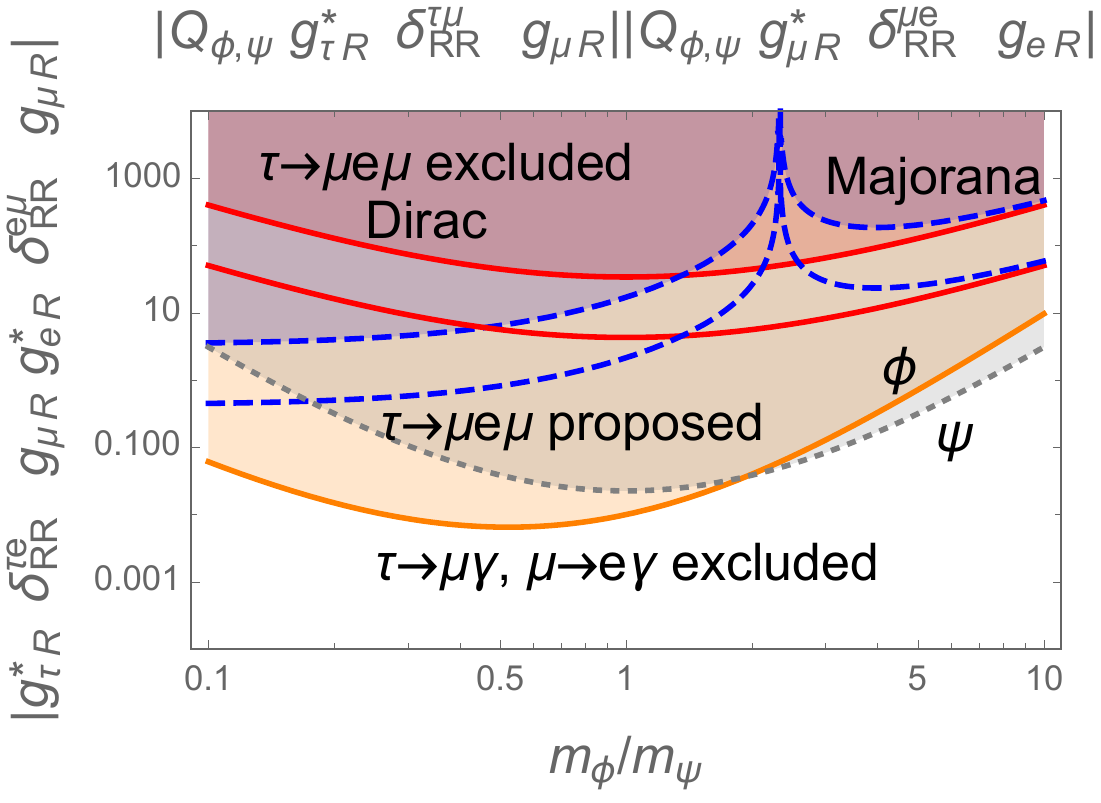}
}
\subfigure[]{
  \includegraphics[width=6.5cm]{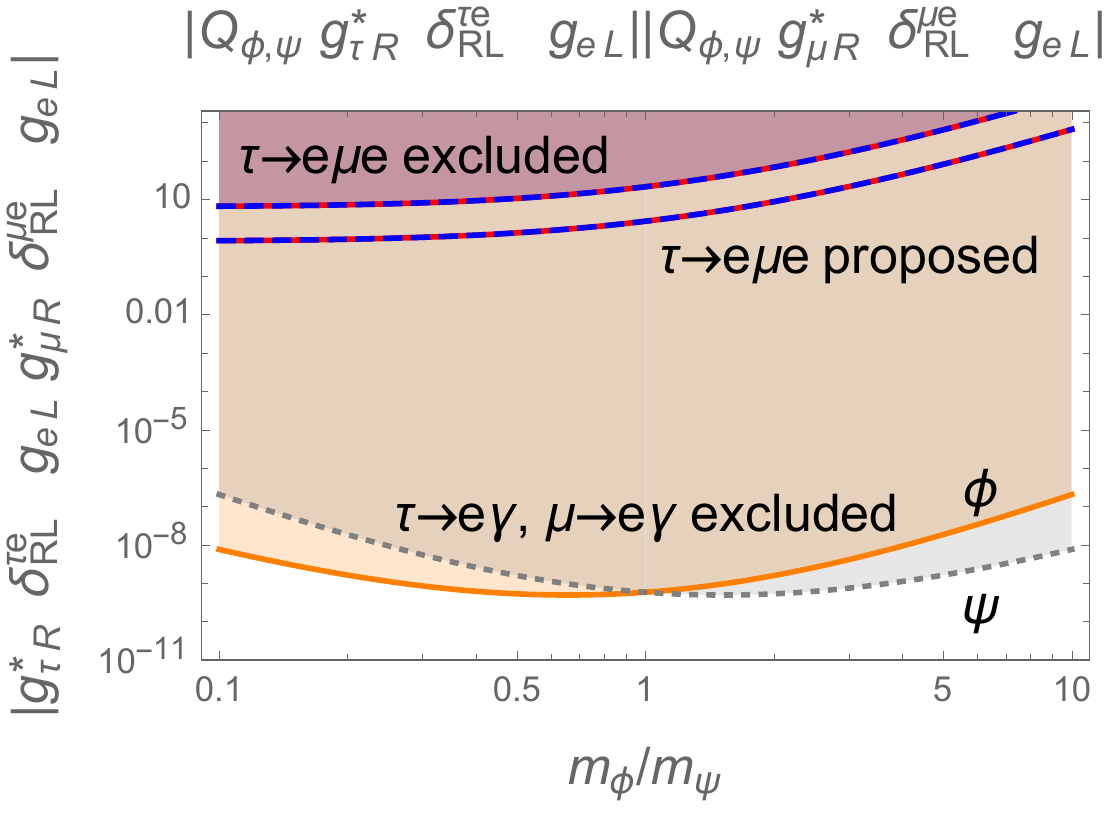}
}
\hspace{0.5cm}
\subfigure[]{
  \includegraphics[width=6.5cm]{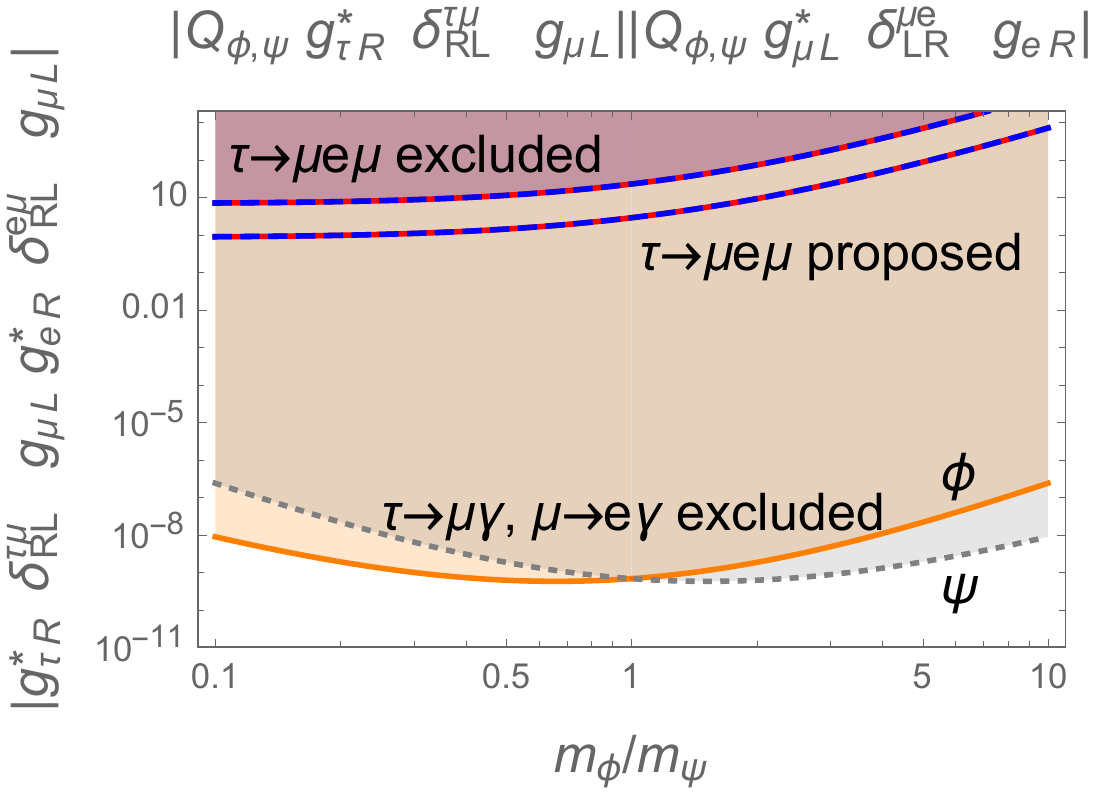}
}
\subfigure[]{
  \includegraphics[width=6.5cm]{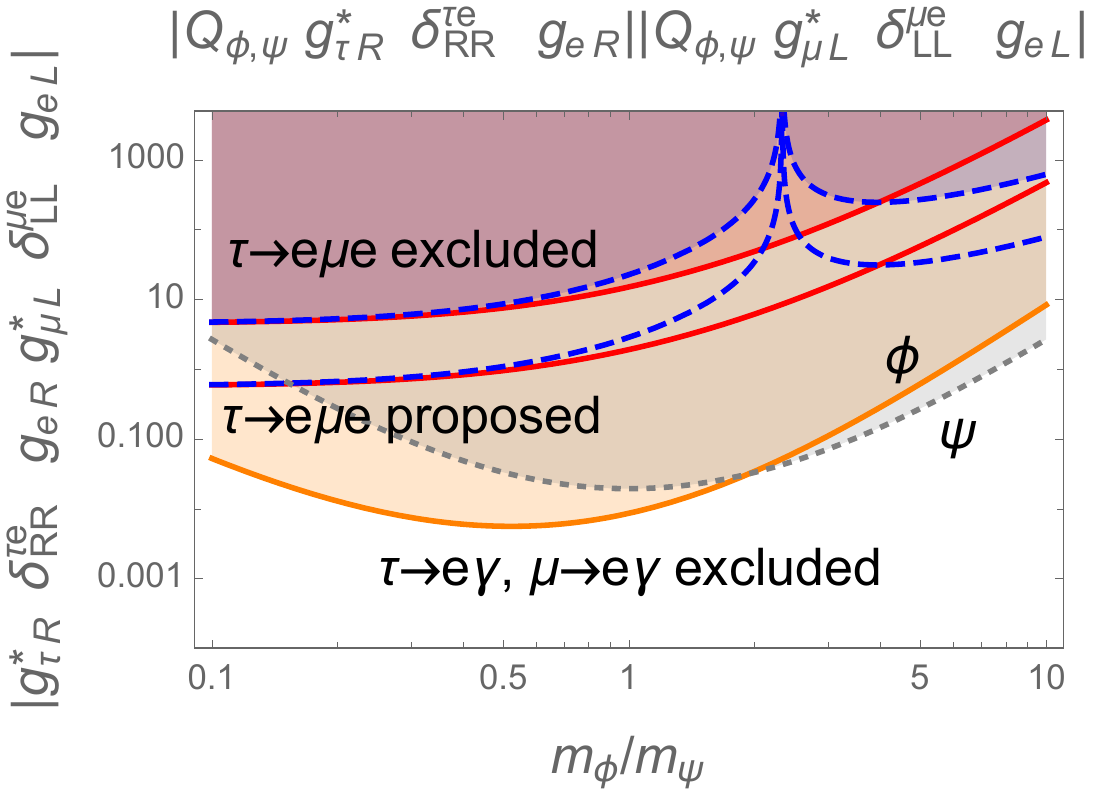}
}
\hspace{0.5cm}
\subfigure[]{
  \includegraphics[width=6.5cm]{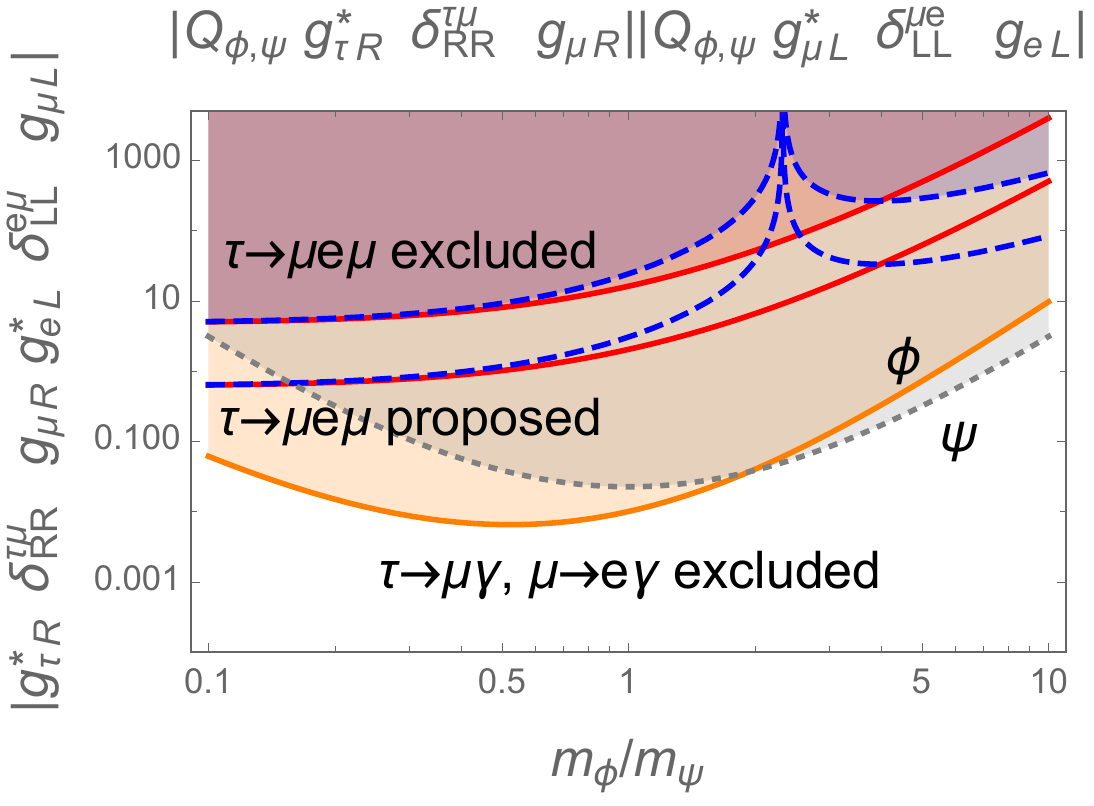}
}
\caption{Same as Fig.~\ref{fig:tau2emuemuemuLFV}, but for case II.
}
\label{fig:tau2emuemuemuLFV2}
\end{figure}

In Figs.~\ref{fig:mu2eLFV2}, ~\ref{fig:tau2eLFV2} and \ref{fig:tau2muLFV2}, we show the parameter space constrained by using various experimental bounds or expected sensitivities on $\mu\to e$, $\tau\to e$ and $\tau\to\mu$ lepton flavor violating processes.
Contributions from photonic penguin, $Z$-penguin and box diagrams are considered.
In Fig.~\ref{fig:tau2emuemuemuLFV2}, the
parameter space constrained by using various bounds or expected experimental sensitivities on $\tau^-\to e^-\mu^+ e^-, \mu^- e^+\mu^-$ processes through contributions from box contributions are shown.

There are several messages we can extracted from these results.
First we note that, comparing to case I, the built-in cancelation has more prominent effects in penguin amplitudes than in box amplitudes.
Furthermore, the cancelation affects small-$x$ ($x\equiv m_\phi/m_\psi$) region more effectively.  
We can see this clearly in the above figures by noting that the curves corresponding to penguin contributions bend upward in the small-$x$ region, hence, relaxing the constaints.

Similar to case I, we note that chiral interactions ($g_L\times g_R=0$) are unable to generate large enough contributions to $\Delta a_e$ and $\Delta a_\mu$ to accommodate the experimental results, Eqs. (\ref{eq: muon g-2 expt}) and (\ref{eq: e g-2 expt}). 
This can be seen in Tables~\ref{tab: results case I x=1}, \ref{tab: results case II x=0.5} and \ref{tab: results case II x=2}, where $Q_{\phi,\psi}|g_{e R(L)}|^2$ and $Q_{\phi,\psi}|g_{\mu R(L)}|^2$ 
need to be unreasonably and unacceptably large to produce the experimental values of $\Delta a_e$ and $\Delta a_\mu$.

Again similar to case I, we find that although non-chiral interactions are capable to generate $\Delta a_e$ and $\Delta a_\mu$ successfully accommodating the experimental results, they are contributed from different sources.
From Tables~\ref{tab: results case II x=1}, \ref{tab: results case II x=0.5}, \ref{tab: results case II x=2}, we see that $Q_{\phi,\psi}{\rm Re}(g^*_{e R} g_{e L} \delta^{ee}_{RL})$ and $Q_{\phi,\psi}{\rm Re}(g^*_{\mu R} g_{\mu L}\delta^{\mu\mu}_{RL})$ of orders $10^{-3}$ and $10^{-2}$ or larger, are able to produce the experimental values of $\Delta a_e$ and $\Delta a_\mu$.
As $\Delta a_e$ is generated from $Q_{\phi,\psi}{\rm Re}(g^*_{e R} g_{e L} \delta^{ee}_{RL})$, while $\Delta a_\mu$ is generated from $Q_{\phi,\psi}{\rm Re}(g^*_{\mu R} g_{\mu L}\delta^{\mu\mu}_{RL})$, 
the contributions are not from the same source (meaning the same $\psi$ and $\phi$).
We also note that these values are larger than $Q_{\phi,\psi}{\rm Re}(g^*_{e R} g_{e L})$ and $Q_{\phi,\psi}{\rm Re}(g^*_{\mu R} g_{\mu L})$ in case I by roughly one order of magnitude.
This is reasonable as we have cancellation in this case.
Furthermore, comparing Figs.~\ref{fig:e muon g-2 1}(b), (d) and Fig.~\ref{fig:e muon g-2 2}(a) and (b), we can clearly see the relaxation in the small-$x$ region.

The upper limit in $\mu\to e\gamma$ decay gives the most severe constraints on photonic penguin contributions in $\mu\to e$ transitions, agreeing with \cite{Kuno:1999jp,Crivellin:2013hpa},
but the constraints on parameters are relaxed, especially in the small-$x$ region,  comparing to case I.
From Tables~\ref{tab: results case II x=1}, \ref{tab: results case II x=0.5}, \ref{tab: results case II x=2} and Fig.~\ref{fig:mu2eLFV2}(a) to (d), we see that the bounds on $|Q_{\phi,\psi} g^*_{\mu R} g_{e R}\delta_{RR}^{\mu e}|$ and $|Q_{\phi,\psi} g^*_{\mu R} g_{e L} \delta_{RL}^{\mu e}|$ are severely constrained by the $\mu\to e\gamma$ upper limit. Indeed, the $\mu\to e\gamma$ bound is more severe than the $\mu\to 3e$ and $\mu N\to e N$ bounds.
The situation is altered when considering future experimental searches. 
From the tables and the figures, we see that, on the contrary, the $\mu \to 3 e$ and $\mu N\to e N$ processes
can probe the photonic penguin contributions from $|Q_{\phi,\psi} g^*_{\mu R} g_{e R}\delta_{RR}^{\mu e}|$ 
and $|Q_{\phi,\psi} g^*_{\mu R} g_{e L}\delta_{RL}^{\mu e}|$ better than the $\mu\to e\gamma$ decay in near future experiments.

Similar to case I, the $Z$-penguin diagrams can constrain chiral interaction better than photonic penguin diagrams in $\mu\to e$ transitions.
From Tables~\ref{tab: results case II x=1}, \ref{tab: results case II x=0.5}, \ref{tab: results case II x=2}, Fig.~\ref{fig:mu2eLFV2}(a), (b), and (e) we see that the bounds on $|g^*_{\mu R} g_{e R}\Delta T_{3\psi}\delta_{RR}^{\mu e}|$ from $Z$-penguin contributions are more severe (by one to two orders of magnitude) than
the bounds on $|Q_{\phi,\psi} g^*_{\mu R} g_{e R}\delta_{RR}^{\mu e}|$ from photonic penguin contributions. 
In addition, from Fig.~\ref{fig:mu2eLFV2}(e) we see that the upper limits of $\mu N\to e N$ transitions give better bounds on $|g^*_{\mu R} g_{e R}\Delta T_{3\psi}\delta_{RR}^{\mu e}|$ than the $\mu\to 3 e$ bound.

For $x$ larger than $0.2$, box contributions to $\mu\to 3e$ decay are subleading comparing to $Z$ penguin contributions,
but the former can be important for $x\lesssim 0.2$.
In Fig.~\ref{fig:mu2eLFV2}(f) and (g) we show the bounds on 
$|g^*_{\mu R} g_{e R} \delta^{\mu e}_{RR} g^*_{eR} g_{e R}|$ 
and $|g^*_{\mu R} g_{e R} \delta^{\mu e}_{RR} g^*_{eL} g_{e L}|$ obtained by considering box contributions to $\mu\to 3 e$ decay.
Note that the constraint on $|g^*_{\mu R} g_{e R} \Delta T_{3\psi}\delta_{RR}^{\mu e}| |g^*_{e R(L)} g_{e R(L)}|$ obtained from $\mu\, {\rm Au}\to e \,{\rm Au}$ upper limit and perturbativity is much severe than the $|g^*_{\mu R} \delta^{\mu e}_{RR} g_{e R} g^*_{eR(L)} g_{e R(L)}|$ bound.
However for $x$ smaller than $0.2$, box contributions can be important.
This is different from case I, as penguin contributions have larger cancellation in the small-$x$ region in the present case
and, as a result, box contributions become relatively important in this region.

From Tables~\ref{tab: results case II x=1}, \ref{tab: results case II x=0.5} and \ref{tab: results case II x=2}, 
we see that similar to case I the present bound on $\Delta a_\tau$ cannot constrain $Q_{\phi,\psi}|g_{\tau R(L)}|^2$ and $Q_{\phi,\psi}{\rm Re}(g^*_{\tau R} g_{\tau L}\delta^{\tau\tau}_{RL})$ well. Even the bound on $d_\tau$ cannot give good constraints on 
$Q_{\phi,\psi}{\rm Im}(g^*_{\tau R} g_{\tau L}\delta^{\tau\tau}_{RL})$. 

In $\tau\to e$ $(\mu)$ transitions, the $\tau\to e \gamma$ $(\mu \gamma) $ upper limit constrains photonic penguin contributions better than the $\tau\to 3 e$ $(3\mu)$ upper limit,
agreeing with \cite{Crivellin:2013hpa}, 
and the $Z$-penguin constrains chiral interaction better than the photonic penguin.
From Tables~\ref{tab: results case II x=1}, \ref{tab: results case II x=0.5}, \ref{tab: results case II x=2}, Fig.~\ref{fig:tau2eLFV2}(a) to (d) and Fig.~\ref{fig:tau2muLFV2}(a) to (d), we see that
bounds on $|Q_{\phi,\psi} g^*_{\tau R} g_{e (\mu) R}\delta^{\tau e(\tau\mu)}_{RR}|$ 
and $|Q_{\phi,\psi} g^*_{\tau R} g_{e(\mu) L} \delta^{\tau e(\tau\mu)}_{RL}|$ are constrained by the
$\tau\to e\gamma$ $(\mu\gamma)$ data more severely than by the $\tau\to 3e$ $(3\mu)$ upper limit. Note that even the bounds using the proposed sensitivities on $\tau\to 3 e$ and $\tau\to 3\mu$ decays in Belle II are superseded by the bounds using the present limits of $\tau\to e\gamma$ and $\tau\to \mu\gamma$ decays in most of the parameter space.
From Tables~\ref{tab: results case II x=1}, \ref{tab: results case II x=0.5}, \ref{tab: results case II x=2}, Fig.~\ref{fig:tau2eLFV2}(e) and Fig.~\ref{fig:tau2muLFV2}(e), we see that
bounds on $|g^*_{\tau R} g_{e (\mu) R}\Delta T_{3\psi} \delta^{\tau e(\tau\mu)}_{RR}|$ from $Z$-penguin contributions are more severe (by one order of magnitude) than
those on $|Q_{\phi,\psi} g^*_{\tau R} g_{e (\mu) R}|$ from photonic penguin contributions.
Hence, $Z$-penguin constrains chiral interaction better than photonic penguin.
These features are similar to case I, but comparing Fig.~\ref{fig:tau2eLFV}, \ref{fig:tau2muLFV}, \ref{fig:tau2eLFV2} and \ref{fig:tau2muLFV2} we can clearly see that the bounds are significant relaxed in the small-$x$ region in the present case.

Box contributions to $\tau\to 3e$ and $\tau\to 3\mu$ decays can sometime be comparable to $Z$-penguin contributions.
We show in Fig.~\ref{fig:tau2eLFV2}(g), (h) and Fig.~\ref{fig:tau2muLFV2}(g), (h) the bounds on 
$|g^*_{\tau R} g_{e(\mu) R} \delta^{\tau e(\tau\mu)}_{RR} g^*_{e(\mu) R} g_{e(\mu) R}|$ 
and $|g^*_{\tau R} g_{e(\mu) R} \delta^{\tau e(\tau\mu)}_{RR} g^*_{e(\mu) L} g_{e(\mu) L}|$ obtained by considering box contributions to $\tau\to 3 e$ $(3\mu)$ decay.
Note that the constraint on $|g^*_{\tau R} g_{e(\mu) R} \delta^{\tau e(\tau\mu)}_{RR}\Delta T_{3\psi}| |g^*_{e(\mu) L} g_{e(\mu) L}|$ obtained from $Z$-penguin contributions to $\tau \to 3e$ $(3\mu)$ decay and perturbativity
is much severe than the $|g^*_{\tau R} g_{e(\mu) R} \delta^{\tau e(\tau\mu)}_{RR} g^*_{e(\mu) R} g_{e(\mu) R}|$ bound for $x\gtrsim 0.6$, but it is the other way around for $x\lesssim 0.6$.  
One can also obtain these results using the values in Tables~\ref{tab: results case II x=1}, \ref{tab: results case II x=0.5}, \ref{tab: results case II x=2}.
These results imply that box contributions to $\tau\to 3 e, 3\mu$ can sometime be comparable to $Z$-penguin contributions.
This is similar to case I, but in different region of $x$.

The $\tau^-\to e^- \mu^+ e^-$ rate is constrained by $\tau\to e\gamma$ and $\mu\to e\gamma$ upper limits.
The bounds on 
$|g^*_{\tau R} g_{e R} \delta^{\tau e}_{RR} g^*_{\mu R} g_{e R}\delta^{\mu e}_{RR}|$, 
$|g^*_{\tau R} g_{e L} \delta^{\tau e}_{RL} g^*_{\mu R} g_{e L}\delta^{\tau e}_{RL}|$ and
$|g^*_{\tau R} g_{e R} \delta^{\tau e}_{RR} g^*_{\mu L} g_{e L}\delta^{\mu e}_{LL}|$
obtained from constraining box contributions using the upper limit of the $\tau^-\to e^- \mu^+ e^-$ rate are
shown in Fig.~\ref{fig:tau2emuemuemuLFV2} (a), (c), (e) and Tables~\ref{tab: results case II x=1}, \ref{tab: results case II x=0.5}, \ref{tab: results case II x=2}.
They
are larger than the bounds on
$|Q_{\phi,\psi} g^*_{\tau R} g_{e R}\delta^{\tau e}_{RR}| |Q_{\phi,\psi} g^*_{\mu R} g_{e R}\delta^{\mu e}_{RR}|$,
$|Q_{\phi,\psi} g^*_{\tau R} g_{e L}\delta^{\tau e}_{RL}| |Q_{\phi,\psi} g^*_{\mu R} g_{e L}\delta^{\mu e}_{RL}|$ and
$|Q_{\phi,\psi} g^*_{\tau R} g_{e R}\delta^{\tau e}_{RR}| |Q_{\phi,\psi} g^*_{\mu L} g_{e L}\delta^{\mu e}_{LL}|$ obtained by using the upper limits of 
$\tau\to e\gamma$ and $\mu\to e\gamma$ rates.
Note that for $x\gtrsim 0.2$ even the proposed sensitivity on $\tau^-\to e^- \mu^+ e^-$ rate is constrained.
Hence, the $\tau^-\to e^- \mu^+ e^-$ rate is constrained by the present $\tau\to e\gamma$ and $\mu\to e\gamma$ upper limits.
This is similar to case I, but the constraints from $\tau\to e\gamma$ and $\mu\to e\gamma$ upper limits are relatively relaxed.

Similarly the $\tau^-\to \mu^- e^+ \mu^-$ rate is constrained by $\tau\to \mu\gamma$ and $\mu\to e\gamma$ upper limits.
From Fig.~\ref{fig:tau2emuemuemuLFV2} (b), (d), (f) and Tables~\ref{tab: results case II x=1}, \ref{tab: results case II x=0.5}, \ref{tab: results case II x=2}, we see that the bounds on 
$|g^*_{\tau R} g_{\mu R} \delta^{\tau\mu}_{RR} g^*_{e R} g_{\mu R}\delta^{e\mu}_{RR}|$, 
$|g^*_{\tau R} g_{\mu L} \delta^{\tau \mu}_{RL}g^*_{e R} g_{\mu L}\delta^{e\mu}_{RL}|$ and
$|g^*_{\tau R} g_{\mu R} \delta^{\tau\mu}_{RR}g^*_{e L} g_{\mu L}\delta^{e\mu}_{RR}|$
obtained from the upper limit of the $\tau^-\to \mu^- e^+ \mu^-$ rate
are larger than the bounds on
$|Q_{\phi,\psi} g^*_{\tau R} g_{\mu R}\delta^{\tau \mu}_{RR}| |Q_{\phi,\psi} g^*_{\mu R} g_{e R}\delta^{\mu e}_{RR}|$,
$|Q_{\phi,\psi} g^*_{\tau R} g_{\mu L}\delta^{\tau \mu}_{RL}| |Q_{\phi,\psi} g^*_{\mu L} g_{e R}\delta^{\mu e}_{LR}|$ and
$|Q_{\phi,\psi} g^*_{\tau R} g_{\mu R}\delta^{\tau \mu}_{RR}| |Q_{\phi,\psi} g^*_{\mu L} g_{e L}\delta^{\mu e}_{LL}|$ obtained from the upper limits of 
$\tau\to \mu\gamma$ and $\mu\to e\gamma$ rates.
Hence, the $\tau^-\to \mu^- e^+ \mu^-$ rate is constrained by $\tau\to \mu\gamma$ and $\mu\to e\gamma$ upper limits.
Note that for $x\gtrsim0.2$ even the proposed sensitivity on $\tau^-\to \mu^- e^+ \mu^-$ rate is highly constrained.
This is similar to case I, but the constraints obtained using $\tau\to \mu\gamma$ and $\mu\to e\gamma$ upper limits are relatively relaxed.

\begin{table}[t!]
\caption{
Same as Table~\ref{tab: expt bounds case I}, but for case II.
}
 \label{tab: expt bounds case II}
\begin{ruledtabular}
\begin{tabular}{ l c  c c }
~~~~~~
    & current limit (future sensitivity)
    & consistency bounds
    & remarks 
    \\
    \hline
${\cal B}(\mu^+\to e^+\gamma)$
    & $<4.2\times 10^{-13}$ ($6\times 10^{-14}$)
    & $<4.2\times 10^{-13}$
    & input
    \\    
${\cal B}(\mu^+\to e^+e^+e^-)$
    & $<1.0\times 10^{-12}$ ($10^{-16}$) 
    & $<2.2\times 10^{-14}$ 
    & from $\mu\to e\gamma$ bound
    \\ 
    & 
    & $<1.6\times 10^{-14}$ 
    & from $\mu {\rm Au}\to e{\rm Au}$ bound
    \\   
${\cal B}(\mu^- {\rm Ti}\to e^-{\rm Ti})$
    & $<4.3\times 10^{-12}$ ($10^{-17}$) 
    & $<5.2\times 10^{-14}$ 
    & from $\mu\to e\gamma$ bound
    \\
    & 
    & $<3.5\times 10^{-13}$ 
    & from $\mu {\rm Au}\to e{\rm Au}$ bound
    \\   
${\cal B}(\mu^- {\rm Au}\to e^-{\rm Au})$
    & $<7.0\times 10^{-13}$ ($10^{-16}$) 
    & $<6.2\times 10^{-13}$ 
    & from $\mu\to e\gamma$ bound
    \\ 
    & 
    & $<7.0\times 10^{-13}$
    & input
    \\
${\cal B}(\mu^- {\rm Al}\to e^-{\rm Al})$
    & $\cdots$ ($10^{-17}$) 
    & $<3.2\times 10^{-13}$ 
    & from $\mu\to e\gamma$ bound
    \\ 
    & 
    & $<1.7\times 10^{-13}$ 
    & from $\mu {\rm Au}\to e{\rm Au}$ bound
    \\     
${\cal B}(\tau^-\to e^-\gamma)$
    & $<3.3\times 10^{-8}$ ($3\times10^{-9}$) 
    & $<3.3\times 10^{-8}$
    & input
    \\
${\cal B}(\tau^-\to e^-e^+ e^-)$
    & $<2.7\times 10^{-8}$ ($4.3\times 10^{-10}$) 
    & $<1.9\times 10^{-9}$ 
    & from $\tau\to e\gamma$ bound
    \\       
${\cal B}(\tau^-\to \mu^-\gamma)$
    & $<4.4\times 10^{-8}$ ($1\times 10^{-9}$) 
    & $<4.4\times 10^{-8}$
    & input
    \\ 
${\cal B}(\tau^-\to \mu^-\mu^+\mu^-)$
    & $<2.1\times 10^{-8}$ ($3.3\times 10^{-10}$) 
    & $<2.5\times 10^{-9}$ 
    & from $\tau\to \mu\gamma$ bound
    \\        
${\cal B}(\tau^-\to \mu^- e^+ \mu^-)$
    & $<1.7\times 10^{-8}$ ($2.7\times 10^{-10}$) 
    & $\lesssim 1.3\times 10^{-8}$ 
    & from $\tau\to \mu\gamma$, $\mu\to e\gamma$ bounds
    \\  
${\cal B}(\tau^-\to e^-\mu^+ e^-)$
    & $<1.5\times 10^{-8}$ ($2.4\times 10^{-10}$) 
    & $\lesssim 1\times 10^{-8}$ 
    & from $\tau\to e\gamma$, $\mu\to e\gamma$ bounds
    \\                                           
\end{tabular}
\end{ruledtabular}
\end{table}

In Table~\ref{tab: expt bounds case II}, we compare the current experimental upper limits, future sensitivities and bounds from consistency for case II on various muon and tau LFV processes. 
We see that the present $\mu\to e\gamma$ upper limit requires the bounds on $\mu\to 3 e$ and $\mu\,{\rm Ti}\to e \,{\rm Ti}$ be lower by more than one order of magnitude from their present upper limits, 
while the $\mu\,{\rm Au}\to e\, {\rm Au}$ bound is close to its present limit and the $\mu\,{\rm Al}\to e\, {\rm Al}$ rate is predicted to be smaller than $3\times 10^{-13}$. 
Comparing to case~I we see that the $\mu\to 3 e$,  $\mu\,{\rm Au}\to e \,{\rm Au}$ and $\mu\,{\rm Al}\to e \,{\rm Al}$ bounds are relaxed, while the $\mu\,{\rm Ti}\to e\, {\rm Ti}$ bound is tighten.
We find that the situation is similar when the present $\mu\,{\rm Au}\to e\, {\rm Au}$ upper limit instead of the present $\mu\to e\gamma$ upper limit is used as an input.
Using the present $\tau\to e\gamma$ $(\mu\gamma)$ upper limit as input, 
the $\tau\to 3 e$ $(3\mu)$ bound is smaller than its present upper limit by one order of magnitude.
These bounds are relaxed compared to those in case~I.
Note that the $\B(l'\to l\bar l l)/\B(l'\to l\gamma)$ ratios are close to the values shown in Eq.~(\ref{eq: 3l per lgamma ratio})~\cite{Kuno:1999jp, Crivellin:2013hpa}, but not identical to them, as the $F_1$ terms in photonic penguins also play some roles.
Finally, the $\tau^-\to \mu^- e^+ \mu^-$ and $\tau^-\to e^-\mu^+ e^-$ bounds are similar to their present upper limits
when the present $\tau\to \mu\gamma$, $e\gamma$ and $\mu\to e\gamma$ upper limits are used.
These limits are significant relaxed compared to those in case~I.

\section{Conclusion}

We study anomalous magnetic moments and lepton flavor violating processes of $e$, $\mu$ and $\tau$ leptons in this work. We use a data driven approach to investigate the implications of the present data on the parameters of a class of models, which has spin-0 scalar and spin-1/2 fermion fields and can contribute to $\Delta a_l$ and LFV processes. 
We compare two different cases, case I and case~II, which does not have and has a built-in cancelation mechanism, respectively. Our findings are as following.

\begin{itemize}

\item
Parameters are constrained using the present data of $\Delta a_l$, $d_l$ and lepton flavor violating processes of $e$, $\mu$ and $\tau$ leptons.

\item
The built-in cancelation has more prominent effects in penguin amplitudes than in box amplitudes.
Furthermore, the cancelation affects amplitudes in small-$x$ ($x\equiv m_\phi/m_\psi$) region more effectively.  

\item
Chiral interactions are unable to generate large enough $\Delta a_e$ and $\Delta a_\mu$ to accommodate the experimental results.

\item
Although $\Delta a_e$ and $\Delta a_\mu$ can be successfully generated to accommodate the experimental results by using non-chiral interactions, they are not contributed from the same source.
This agree with the finding in \cite{Crivellin:2018qmi}.

\item 
Presently, the upper limit in $\mu\to e\gamma$ decay gives the most severe constraints on photonic penguin contributions in $\mu\to e$ transitions, agreeing with \cite{Kuno:1999jp,Crivellin:2013hpa},
but the situation may change in considering future experimental sensitivities.
In fact, the future $\mu \to 3 e$ and $\mu N\to e N$ experiments may probe the photonic penguin contributions better than the future $\mu\to e\gamma$ experiment.

\item
The $Z$-penguin diagrams can constrain chiral interaction better than photonic penguin diagrams in $\mu\to e$ transitions.
In addition, $\mu N\to e N$ transitions constrain $Z$-penguin contributions better $\mu\to 3 e$ decay. 

\item 
In case I,  either in the Dirac or Majorana case, box contributions to $\mu\to 3e$ decay are subleading.
Furthermore, there are cancelation in box contributions in the Majorana fermionic case making the contributions even smaller.
In case II, we find that for $x\gtrsim 0.2$, box contributions to $\mu\to 3e$ decay are subleading comparing to $Z$ penguin contributions, but they can be important for $x\lesssim 0.2$.

\item
The present bounds on $\Delta a_\tau$ and $d_\tau$ are unable to give useful constraints on parameters. 

\item
In $\tau\to e$ $(\mu)$ transitions, the $\tau\to e \gamma$ $(\mu \gamma)$ upper limit constrains photonic penguin contributions better than 
the $\tau\to 3 e$ $(3\mu)$ upper limit, agreeing agrees with \cite{Crivellin:2013hpa},
and $Z$-penguin constrains chiral interaction better than photonic penguin.
Note that even the bounds using the proposed sensitivities on $\tau\to 3 e$ and $\tau\to 3\mu$ decays by Belle II are superseded by the bounds using the present limits of $\tau\to e\gamma$ and $\tau\to \mu\gamma$ decays for most of the parameter space.
Bounds are significant relaxed in small-$x$ region in case II.

\item
Box contributions to $\tau\to 3e$ and $\tau\to 3\mu$ decays can sometime be comparable to $Z$-penguin contributions.

\item
The $\tau^-\to e^- \mu^+ e^-$ rate is highly constrained by $\tau\to e\gamma$ and $\mu\to e\gamma$ upper limits.
Note that in case I even the proposed sensitivity on $\tau^-\to e^- \mu^+ e^-$ rate is highly constrained,
but in case II, for $x\lesssim 0.2$ the constraints are relaxed.

\item
The $\tau^-\to \mu^- e^+ \mu^-$ rate is also highly constrained by $\tau\to \mu\gamma$ and $\mu\to e\gamma$ upper limits.
Note that in case I even the proposed sensitivity on $\tau^-\to \mu^- e^+ \mu^-$ rate is highly constrained,
but in case II, for $x\lesssim0.2$ the constraints are relaxed.

\item
We compare the current experimental upper limits, future sensitivities and bounds from consistency on various muon and tau LFV processes:

(a)
In case I, the present $\mu\to e\gamma$ upper limit requires the bounds on $\mu\to 3 e$, $\mu\,{\rm Ti}\to e\, {\rm Ti}$ and $\mu\,{\rm Au}\to e\, {\rm Au}$ be lower by two orders of magnitude, more than one order of magnitude and almost one order of magnitude, respectively, from their present upper limits, and the $\mu\,{\rm Al}\to e\, {\rm Al}$ rate is predicted to be smaller than $6\times 10^{-14}$. 
In case II, the $\mu\to 3 e$,  $\mu\,{\rm Au}\to e \,{\rm Au}$ and $\mu\,{\rm Al}\to e\, {\rm Al}$ bounds are relaxed, while the $\mu\,{\rm Ti}\to e\, {\rm Ti}$ bound is tighten.
We agree with \cite{Crivellin:2013hpa} that  presently the $\B(\mu\to e \gamma)$ upper limit provides the most severe constrain on NP contributing to $\mu\to e$ transitions. 

(b)
We find that the situation is similar but the bounds are slightly relaxed when the $\mu\,{\rm Au}\to e\, {\rm Au}$ upper limit instead of the present $\mu\to e\gamma$ upper limit is used as an input.

(c)
Using the present $\tau\to e\gamma$  $(\mu\gamma)$ upper limit as input, the $\tau\to 3 e$ $(3\mu)$ bound is smaller than its present upper limit by one order of magnitude.

(d)
In case I, the $\tau^-\to \mu^- e^+ \mu^-$ and $\tau^-\to e^-\mu^+ e^-$ bounds are lower than their present upper limits by two orders of magnitude as required from the present $\tau\to \mu\gamma$, $e\gamma$ and $\mu\to e\gamma$ upper limits.
These limits are lower than the proposed future sensitivities.
In case II, the $\tau^-\to \mu^- e^+ \mu^-$ and $\tau^-\to e^-\mu^+ e^-$ bounds are similar to their present upper limits
when the present $\tau\to \mu\gamma$, $e\gamma$ and $\mu\to e\gamma$ upper limits are used.
These limits are significant relaxed compared to those in case I.

\end{itemize}

\vskip 1.71cm {\bf Acknowledgments}

This research was supported in part by the Ministry of Science and Technology of R.O.C. under Grant
No. 106-2112-M-033-004-MY3.

\appendix

\section{Formulas for various processes}\label{app:Formula}

Formulas in this Appendix are taken from ref.~\cite{Chua:2012rn} and are updated.
In the weak bases of $\psi_{L p}$, $\psi_{R p}$, $\phi_{L a}$ and $\phi_{R a}$, the interacting Lagrangian is given by
\be
{\cal L}_{\rm int}= 
(g'{}^{pa}_{lL} \bar\psi_{R p} l_L\phi_{L a}^*+g'{}^{pa}_{lR} \bar \psi_{L p} l_R\phi_{R a}^*)
+h. c.,
\label{eq:Lint weak}
\en
where $\phi_{L(R)}$ are scalar fields coupling to $l_{L(R)}$ and $p$, $a$ indicate weak quantum numbers. 
Fields in the weak bases can be transformed into those in the mass bases, 
\be
\phi_i=U^L_{ia} \phi_{L a}+U^R_{i a}\phi_{R a},
\quad
\psi_{nL(R)}=V^{L(R)}_{np} \psi_{L(R) p},
\en
with the help of mixing matrices, $U$ and $V$. 
It is useful to define
\be
g^{ni}_{lL(R)}\equiv g'{}^{pa}_{lL(R)} V^{R(L)}_{np} U^{L(R)}_{ia}
\en
and, consequently, the interacting Lagrangian can be expressed as in Eq. (\ref{eq:Lint}).

The effective Lagrangian for various precesses is given by
\be
{\cal L}_{\rm eff}={\cal L}_{l'l\gamma}+{\cal L}_{l'll''l}+{\cal L}_{l'lqq}
\label{eq: Leff}
\en
with $l^{(\prime,\prime\prime)}=e,\mu,\tau$ denoting leptons and $q$ denoting quarks.
For $l'\neq l$, we have
\be
\label{eq:dipole}
{\cal L}_{l'l\gamma}=\bar l'_L\sigma_{\mu\nu} l_R F^{\mu\nu} A_{L' R}
                              +\bar l'_R\sigma_{\mu\nu} l_L F^{\mu\nu} A_{R' L}
                              +h.c.,                              
\en
and
\be
A_{L R'}=A^*_{R' L},
\quad
A_{R L'}=A^*_{L' R},
\en
while for $l'=l$, the additional hermitian conjugated terms in Eq.~(\ref{eq:dipole}) are not required.
These $A$s are from the so-called photonic dipole penguin.
The relevant effective Lagrangians responsible for $\bar l'\to \bar l l'' \bar l$ decays and $l'\to l$ conversion processes are given by~\cite{review}
\be
{\cal L}_{l'll''l}
&=&g_{RLRL}(\bar {l'}_R l_L)(\bar l''_R l_L)
        +g_{LRLR}(\bar {l'}_L l_R)(\bar l''_L l_R)
\non\\ 
&&        +g_{RRRR}(\bar{l'}_R\gamma^{\mu} l_R)(\bar l''_R\gamma_\mu l _R)
        +g_{LLLL}(\bar{l'}_L\gamma^{\mu} l_L)(\bar l''_L\gamma_{\mu} l_L)
\non\\        
&&  +g_{RRLL}(\bar{l'}_R\gamma^{\mu} l_R)(\bar l''_L\gamma_{\mu} l_L)      
        +g_{LLRR}(\bar{l'}_L\gamma^{\mu} l_L)(\bar l''_R\gamma_{\mu} l_R)
        +h.c.,
\\
{\cal L}_{l'lqq}
&=&\sum_{q=u,d}
 [g_{LV}(q)\bar{l'}_L\gamma^{\mu} l_L+g_{RV}(q)\bar{l'}_R\gamma^{\mu} l_R]\bar q\gamma_{\mu} q
   +h.c., 
\label{eq: Leff1}   
\en        
where 
\be
g_{MNOP}&\equiv&e^2 Q_l g^\gamma_{M'M}\delta_{MN}\delta_{OP}\delta_{l l''}
+g^Z_{M'M} g^Z_{l_O}\delta_{MN}\delta_{OP}\delta_{l l''}+g^B_{MNOP},
\non\\
g_{M'V}(q)&\equiv&e^2 Q_q g^\gamma_{M'M}+\frac{1}{2}g^Z_{M'M}(g^Z_{q_L}+g^Z_{q_R}),
\non\\
g^Z_X&\equiv&\frac{e}{\sin\theta_W\cos\theta_W}(T_3-\sin^2\theta_W Q)_X,
\label{eq:effective} 
\en
with $M$, $N$, $O$, $P$=$L$, $R$, $g^\gamma_{M'M}$ from the non-photonic dipole penguin, $g^Z_{M'M}$ from the $Z$-penguin, $g^B_{MNOP}$ from the box diagrams and $X=l_L, l_R, q_L, q_R$ and so on.
 
Using Eq.~(\ref{eq:Lint}), the Wilson coefficients for ${\cal L}_{l'l\gamma}$ in Eq.~(\ref{eq:dipole}) can be calculated to be~\cite{Chua:2012rn}
\be
A_{M' N}&=&\frac{e}{32\pi^2}
[(m_{l'} g^{ni*}_{l'N} g^{ni}_{lN}+m_l g_{l'M}^{ni*} g^{ni}_{lM})(Q_{\phi_i} F_1(m_{\psi_n}^2,m_{\phi_i}^2)-Q_{\psi_n} F_1(m_{\phi_i}^2,m_{\psi_n}^2))
\non\\
&&+m_{\psi_n} g_{l'M}^{ni*} g^{ni}_{lN} (Q_{\phi_i} F_3(m_{\psi_n}^2,m_{\phi_i}^2)-Q_{\psi_n} F_2(m_{\phi_i}^2,m_{\psi_n}^2))],
\en
for $M$ different from $N$, 
and $F_i$ are loop functions with the explicit forms to be given below.
The Wilson coefficients for ${\cal L}_{l'll''l}$ and ${\cal L}_{l'lqq}$ in Eq.~(\ref{eq:effective}) are given by
\be
g^\gamma_{R'R}&=&\frac{1}{16\pi^2}\{g^{ni*}_{l'R} g^{ni}_{lR} 
                                             [Q_{\psi_n} G_2(m^2_{\phi_i},m^2_{\psi_n})+Q_{\phi_i} 
                                             G_1(m^2_{\psi_n},m^2_{\phi_i})]
\non\\
           &&  + m_{\psi_n}(m_{l'} g^{ni*}_{l'L} g^{ni}_{lR}+m_l g^{ni*}_{l'R} g^{ni}_{lL})
                                             [Q_{\psi_n} G_3(m^2_{\phi_i},m^2_{\psi_n})
                                             +Q_{\phi_i} G_3(m^2_{\psi_n},m^2_{\phi_i})]\}                                          
\non\\
g^Z_{R'R}&=&-\frac{e}{16\pi^2 m_Z^2\sin2\theta_W}2\kappa_{R\,ijmn}
                                             g^{mi*}_{l'R} g^{nj}_{lR}
                                             F_Z(m^2_{\psi_m},m^2_{\psi_n},m^2_{\phi_i},m^2_{\phi_j},m^2_Z)
                                             \non\\
                                             &&
-\frac{e}{16\pi^2 m_Z^2\sin2\theta_W}2 \Delta T^{RL}_{3\psi mn}
                                             g^{mi*}_{l'R} g^{ni}_{lR}
                                             G_Z(m^2_{\psi_m},m^2_{\psi_n},m^2_{\phi_i}),
\non\\
g^B_{RLRL}&=&\frac{1}{16\pi^2} F(m^2_{\psi_m},m^2_{\psi_n},m^2_{\phi_i},m^2_{\phi_j}) 
                           (g_{l' R}^{mi^*} g_{lL}^{mj} g_{l''R}^{nj*} g_{l L}^{ni} 
                           -2\eta g_{l' R}^{mi^*} g_{l''R}^{mj*} g_{l L}^{ni}  g_{lL}^{nj}), 
\non\\
g^B_{RRRR}&=&\frac{1}{16\pi^2}
                                \bigg [\frac{\eta}{2}g_{l' R}^{mi*} g_{l R}^{ni} g_{l''R}^{mj*} g_{lR}^{nj}
                                 F(m^2_{\psi_m},m^2_{\psi_n},m^2_{\phi_i},m^2_{\phi_j}) 
\non\\                                 
 &&                          -\frac{1}{4}g_{l' R}^{mi*} g_{l R}^{ni} g_{l''R}^{nj*} g_{lR}^{mj} 
                                 G(m^2_{\psi_m},m^2_{\psi_n},m^2_{\phi_i},m^2_{\phi_j})\bigg] , 
\non\\
g^B_{RRLL}&=&\frac{1}{16\pi^2}\bigg\{ 
                            -\frac{1}{4}G(m^2_{\psi_m}, m^2_{\psi_n},m^2_{\phi_i},m^2_{\phi_j})
                            (g_{l' R}^{mi^*} g_{l L}^{ni}  g_{l''L}^{nj*} g_{lR}^{mj}
                            +\eta g_{l' R}^{mi^*} g_{l L}^{ni} g_{l''L}^{mj*} g_{lR}^{nj}) 
\non\\
 &&         -\frac{1}{2}g_{l' R}^{mi*} g_{l R}^{ni}g_{l''L}^{nj*} g_{lL}^{mj}  
                                    F(m^2_{\psi_m}, m^2_{\psi_n},m^2_{\phi_i},m^2_{\phi_j}) 
\non\\          
 &&                +\frac{\eta}{4} g_{l' R}^{mi*}  g_{l R}^{ni} g_{l''L}^{mj*} g_{lL}^{nj}
                           G(m^2_{\psi_m}, m^2_{\psi_n},m^2_{\phi_i},m^2_{\phi_j}) \bigg\}, 
\label{eq:gPgB}
\en
with 
\be
\kappa_{L(R)ijmn}&\equiv& \sin2\theta_W (g_{l_{L(R)}}^Z\delta_{ij}\delta_{mn}-g^Z_{\psi_{R(L)}\,mn}\delta_{ij}-g^Z_{\phi\,ij}\delta_{mn})/2e, 
\non\\
\Delta T^{RL}_{3\psi mn}&\equiv& V^R_{mp}T_{3\psi_R p} V^{\dagger L}_{pn}-V^L_{mp}T_{3\psi_L p} V^{\dagger R}_{pn}
\equiv \Delta T_{3\psi mn}
=-\Delta T^{LR}_{3\psi mn},
\label{eq: kappa and Delta T3}
\en 
$\eta=1(0)$ for Majorana (Dirac) fermionic $\psi$
and the
loop functions $F_{(Z)}$ and $G_{(i,Z)}$ will be given shortly.
Other $g$ can be obtained by exchanging $R$ and $L$. 
Note that $\Delta T_{3\psi}$ is basically the difference of weak isospin quantum numbers of $\psi_R$ and $\psi_L$ and in the case of no mixing, $\kappa_{L,R}$ are vanishing.
Therefore, we expect $\Delta T_{3\psi}$ to be an order one quantity, while $\kappa$ to be a much smaller quantity.
Note that in case II the leading order contributions to the $Z$ penguin amplitudes are at the level of $\delta_{LR}\delta_{RL}$, which is beyond the accuracy of the this analysis and their contributions are, hence, neglected.

\begin{table}[t!]
\caption{The overlap integrate parameters and total capture rates $\omega_{\rm capt}$ taken from \cite{KKO,capt} are collected.}
 \label{tab:parameters}
\begin{ruledtabular}
\begin{tabular}{ l c  c c c}
~~~~~~
    & $D(m_\mu^{5/2})$
    & $V^{(p)}(m_\mu^{5/2})$
    & $V^{(n)}(m_\mu^{5/2})$
    & $\omega_{\rm capt}(10^{6} s^{-1})$
    \\
    \hline
${}^{27}_{13}{\rm Al}$
    & 0.0362
    &  0.0161
    &  0.0173
    & 0.7054
    \\    
${}^{48}_{22}{\rm Ti}$
    & 0.0864
    &  0.0396
    &  0.0468
    & 2.59
    \\
${}^{197}_{79}{\rm Au}$
    & 0.189
    &  0.0974
    &  0.146
    & 13.07
    \\    
${}^{205}_{81}{\rm Tl}$
    & 0.161
    &  0.0834
    &  0.128
    & 13.90
    \\
\end{tabular}
\end{ruledtabular}
\end{table}

The above loop functions are defined as~\cite{Chua:2012rn}
\be
F_{1}(a,b)&=&\frac{1}{12(a-b)^4}\left(2 a^3+3 a^2 b-6 a b^2+b^3+6 a^2 b \ln\frac{b}{a}\right),
\non\\
F_{2}(a,b)&=&\frac{1}{2(a-b)^3}\left(-3 a^2+4 a b- b^2-2 a^2  \ln\frac{b}{a}\right),
\non\\
F_{3}(a,b)&=&\frac{1}{2(a-b)^3}\left(a^2- b^2+2 a b  \ln\frac{b}{a}\right),
\non\\
G_{1}(a,b)&=&\frac{1}{36(a-b)^4}\left(-(a-b)(11 a^2-7ab+2b^2)-6a^3 \ln\frac{b}{a}\right),
\non\\
G_{2}(a,b)&=&\frac{1}{36(a-b)^4}\left(-(a-b)(16 a^2-29ab+7b^2)-6a^2(2a-3b) \ln\frac{b}{a}\right),
\non\\
G_{3}(a,b)&=&\frac{1}{36(a-b)^5}\left(-(a-b)(17 a^2+8ab-b^2)-6a^2(a+3b) \ln\frac{b}{a}\right),
\non\\
F_Z(a_1,a_2,b,b,c)&=&-\frac{a_1(2\sqrt{a_1 a_2}-a_1)}{2(a_1-a_2)(a_1-b)} \ln \frac{a_1}{c}
                                       +\frac{a_2(2\sqrt{a_1 a_2}-a_2)}{2(a_1-a_2)(a_2-b)} \ln \frac{a_2}{c}
\non\\
                               &&  -\frac{b(2\sqrt{a_1 a_2}-b)}{2(a_1-b)(a_2-b)} \ln \frac{b}{c},
\non\\
F_Z(a,a,b_1,b_2,c)&=&-\frac{3}{4}
                                      +\frac{a^2}{2(a-b_1)(a-b_2)} \ln \frac{a}{c}
                                      -\frac{b_1^2}{2(a-b_1)(b_1-b_2)} \ln \frac{b_1}{c}
\non\\
                                && +\frac{b_2^2}{2(a-b_2)(b_1-b_2)} \ln \frac{b_2}{c},
\non\\
G_Z(a_1,a_2,b)&=&\frac{a_1\sqrt{a_1 a_2}}{(a_1-a_2)(a_1-b)} \ln \frac{a_1}{b}
                                     -\frac{a_2\sqrt{a_1 a_2}}{(a_1-a_2)(a_2-b)} \ln \frac{a_2}{b},
\non\\
F(a,b,c,d)&=&\frac{b\sqrt{a b}}{(a-b)(b-c)(b-d)}\ln \frac{b}{a}
                       -\frac{c\sqrt{a b}}{(a-c)(b-c)(c-d)}\ln \frac{c}{a}
\non\\
                 &&+\frac{d\sqrt{a b}}{(a-d)(b-d)(c-d)}\ln \frac{d}{a},
\non\\
G(a,b,c,d)&=&-\frac{b^2}{(a-b)(b-c)(b-d)}\ln \frac{b}{a}
                       +\frac{c^2}{(a-c)(b-c)(c-d)}\ln \frac{c}{a}
\non\\
                 && -\frac{d^2}{(a-d)(b-d)(c-d)}\ln \frac{d}{a}.                       
\en
We do not need the generic expression of $F_Z(a_1,a_2,b_1,b_2,c)$, since only $a_1=a_2=a$ and/or $b_1=b_2=b$ are used in this work.

Comparing the generic expressions in Eq.~(\ref{eq:dipole}) to the following effective Lagrangians,
\be
{\cal L}_{g-2}=-\frac{e Q}{4 m_l} \Delta a_l\,\bar l\sigma_{\mu\nu}l F^{\mu\nu},
\quad
{\cal L}_{EDM}=-\frac{i}{2} d_l\,\bar l\sigma_{\mu\nu}\gamma_5 l F^{\mu\nu},
\en
the $\Delta a_l$ and $d_l$ can be readily obtained as
\be
\Delta a_l=-\frac{4 m_l}{e Q_l} {\rm Re}(A_{R L}),
\quad
d_l=2{\rm Im}( A_{RL}).
\en
The $\bar l'\to \bar l\gamma$ decay rate is related to the above $A_{M'N}$,
\be
\Gamma(\bar l'\to \bar l\gamma)=\frac{(m_{l'}^2-m_l^2)^3}{4\pi m_{l'}^3}
\left(|A_{L'R}|^2+|A_{R'L}|^2\right),
\en
the $\bar l'\to \bar l l'' \bar l$ decay rate is governed by the following formula,~\cite{review}
\be
\Gamma(\bar l'\to \bar l\,l''\,\bar l)
&=&\frac{m_{l'}^5}{3(8\pi)^3}\Bigg[\frac{|g_{RLRL}|^2}{8}+2|g_{RRRR}|^2+|g_{RRLL}|^2+
32\,\delta_{l l''}\left|\frac{e A_{R'L}}{m_{l'}}\right|^2\log (\frac{m^2_{l'}}{m^2_l}-\frac{11}{4})
\non\\
&&
+16\,\delta_{l l''} {\rm Re}\left(\frac{e A_{R'L} g^*_{LLLL}}{m_{l'}}\right)
+8\,\delta_{l l''}{\rm Re}\left(\frac{e A_{R'L} g^*_{LLRR}}{m_{l'}}\right)\Bigg]
+L\leftrightarrow R,
\en
while the $l' \to l $ conversion rate ratio is given by 
\be
{\cal B}_{l'N\to eN}=\frac{\omega_{\rm conv}}{\omega_{\rm capt}},
\en
with
\be
\omega_{\rm conv}&=&
\left|\frac{A^*_{R'L} D}{2 m_{l'}}+2 [2 g^*_{LV}(u)+g^*_{LV}(d)] V^{(p)}
+2 [g^*_{LV}(u)+2g^*_{LV}(d)] V^{(n)}\right|^2
+L\leftrightarrow R,
\label{eq: conv}
\en
and the numerical values of $D$, $V$ and $\omega_{\rm capt}$ are taken from \cite{KKO,capt} and 
are collected in Table~\ref{tab:parameters} for completeness.

\section{Gauge quantum numbers of $\phi$ and $\psi$}\label{app: QN}

The $\psi-\phi-l$ lagrangian,
\be
{\cal L}_{\rm int}=
g'_L (\bar\psi_{R } \phi_{L}^*)_i (L_L)_i+g'_R \bar \psi_{L} \phi_{R }^* l_R
+h. c.,
\en
where $i$ is the weak isospin index, is gauge invariant under the SM gauge transformation. 
As the lepton quantum numbers under SU(3)$\times$SU(2)$\times$U(1) are given by
\be
L_L:(1,2,-\frac{1}{2}),
\quad
l_R: (1,1,-1),
\en
the gauge invariant requirement implies that we must have the following quantum number assignments for these combinations:
\be
\bar\psi_R\phi_L^*: (1,2,\frac{1}{2}),
\quad
\bar\psi_L\phi_R^*: (1,1,1).
\en
Consequently, the gauge quantum numbers of $\psi$ and $\phi$ are related as following:
\be
&&\psi_R: (c_R, 2I_R+1, Y_R),
\quad
\phi_L: (\bar c_R, 2(I_R\pm 1/2)+1, Y_R-1/2),
\non\\
&&\psi_L: (c_L, 2I_L+1, Y_L),
\quad
\phi_R: (\bar c_L, 2I_L+1, Y_L-1).
\label{eq: QN}
\en
Some examples of the assignments of the quantum numbers of $\psi_{L,R}$ and $\phi_{L,R}$ are given in Table~\ref{tab:QN}.

\begin{table}[t]
\caption{Some examples of the assignment of the quantum numbers of $\psi_{L,R}$ and $\phi_{L,R}$.}
 \label{tab:QN}
\begin{ruledtabular}
\begin{tabular}{ cc | cc}
     $\psi_R$
    & $\phi_L$
    &$\psi_L$
    & $\phi_R$
    \\
    \hline
     $(1,1,Y_R)$
    & $(1,2,Y_R-\frac{1}{2})$
    &  $(1,1,Y_L)$
    & $(1,1,Y_L-1)$
     \\    
    $(1,2,Y_R)$
    &  $(1,1,Y_R-\frac{1}{2})$
    & $(1,2,Y_L)$
    & $(1,2,Y_L-1)$
    \\
    $(3 (\bar 3),1,Y_R)$
    & $(\bar 3 (3),2,Y_R-\frac{1}{2})$
    & $(3 (\bar 3),1,Y_L)$
    & $(\bar 3 (3),1,Y_L-1)$
    \\    
    $(3 (\bar 3),2,Y_R)$
    &  $(\bar 3 (3),1,Y_R-\frac{1}{2})$
    &  $(3 (\bar 3),2,Y_L)$
    &  $(\bar 3 (3),2,Y_L-1)$
    \\
\end{tabular}
\end{ruledtabular}
\end{table}

\begin{table}[t]
\caption{Some examples of the assignment of the quantum numbers of $\psi_{L,R}$ and $\phi_{L,R}$ that can generate chiral enhancement in photonic dipole penguins.}
 \label{tab:QN non chiral}
\begin{ruledtabular}
\begin{tabular}{ c c c  cc}
 case   
    & $\psi_R$
    & $\phi_L$
    &$\psi_L$
    & $\phi_R$
    \\
    \hline
 (A)   
    & $(1,1,Y)$
    & $(1,2,Y-\frac{1}{2})$
    &  $(1,1,Y)$
    & $(1,1,Y-1)$
     \\ 
  (A)        
    & $(1,2,Y)$
    &  $(1,1,Y-\frac{1}{2})$
    & $(1,2,Y)$
    & $(1,2,Y-1)$
    \\
  (A)    
    & $(3 (\bar 3),1,Y)$
    & $(\bar 3 (3),2,Y-\frac{1}{2})$
    & $(3 (\bar 3),1,Y)$
    & $(\bar 3 (3),1,Y-1)$
    \\ 
  (A)       
    & $(3 (\bar 3),2,Y)$
    &  $(\bar 3 (3),1,Y-\frac{1}{2})$
    &  $(3 (\bar 3),2,Y)$
    &  $(\bar 3 (3),2,Y-1)$
    \\
    \hline
 (C)   
    & $(1,1,Y-\frac{1}{2})$
    & $(1,2,Y-1)$
    & $(1,2,Y)$
    & $(1,2,Y-1)$
     \\ 
  (C)        
    & $(1,2,Y-\frac{1}{2})$
    &  $(1,1,Y-1)$
    &  $(1,1,Y)$
    & $(1,1,Y-1)$
    \\
  (C)    
    & $(3 (\bar 3),1,Y-\frac{1}{2})$
    & $(\bar 3 (3),2,Y-1)$
    & $(3 (\bar 3),2,Y)$
    & $(\bar 3 (3),2,Y-1)$
    \\ 
  (C)       
    & $(3 (\bar 3),2,Y-\frac{1}{2})$
    &  $(\bar 3 (3),1,Y-1)$
    &  $(3 (\bar 3),1,Y)$
    &  $(\bar 3 (3),1,Y-1)$
    \\    
\end{tabular}
\end{ruledtabular}
\end{table}

As discussed in the main text chiral enhancement in photonic dipole penguins is an important ingredient to general sizable $\Delta a_\mu$ and $\Delta a_e$.
To have chiral enhancement one needs to connect $\phi_L$ and $\phi_R$ by Higgs VEV with $\psi_R$ and $\psi_L$ having identical quantum numbers or the other way around, see Fig.~\ref{fig:chiralenhancement} and the related discussion.
There are four possibilities on the quantum numbers of the $\bar\psi_L\psi_R$ and $\phi^*_L\phi_R$ combinations to achieve that: 
\be
&&(A):
\quad
\bar\psi_L\psi_R: (1,1,0),
\quad
\phi^*_L\phi_R: (1,2,-1/2),
\non\\
&&(B):
\quad
\bar\psi_L\psi_R: (1,1,0),
\quad
\phi^*_L\phi_R: (1,2,+1/2),
\non\\
&&(C):
\quad
\bar\psi_L\psi_R: (1,2,-1/2),
\quad
\phi^*_L\phi_R: (1,1,0),
\non\\
&&(D):
\quad
\bar\psi_L\psi_R: (1,2,+1/2),
\quad
\phi^*_L\phi_R: (1,1,0).
\en
The above equation imposes additional constraints on the quantum numbers of the new fields:
\be
&&(A):
\quad
c_R=c_L,
\quad
I_R=I_L,
\quad
Y_R=Y_L,
\quad
-Y_R+\frac{1}{2}+Y_L-1=-\frac{1}{2},
\non\\
&&(B):
\quad
c_R=c_L,
\quad
I_R=I_L,
\quad
Y_R=Y_L,
\quad
-Y_R+\frac{1}{2}+Y_L-1=+\frac{1}{2},
\non\\
&&(C):
\quad
c_R=c_L,
\quad
I_R\pm\frac{1}{2}=I_L,
\quad
-Y_L+Y_R=-\frac{1}{2},
\quad
-Y_R+\frac{1}{2}+Y_L-1=0,
\non\\
&&(D):
\quad
c_R=c_L,
\quad
I_R\pm\frac{1}{2}=I_L,
\quad
-Y_L+Y_R=+\frac{1}{2},
\quad
-Y_R+\frac{1}{2}+Y_L-1=0,
\en
where use of Eq.~(\ref{eq: QN}) has been made.

One can easily see that cases (B) and (D) are invalid as there are no solutions satisfying their conditions, and we are left with cases (A) and (C).
In case (A), $\psi_L$ and $\psi_R$ have identical quantum numbers, while $\phi_L$ and $\phi_R$ are mixed via the Higgs VEV. By contrast, in case (C), $\phi_L$ and $\phi_R$ have identical quantum numbers, while $\psi_L$ and $\psi_R$ are mixed via the Higgs VEV. 
To generate chiral enhancement in photonic penguins, case (A) is in general more preferable as the mass of $\psi$ is not limited by the Higgs VEV and the Yukawa coupling.

In Table~\ref{tab:QN non chiral}, we give some samples of the assignment of the quantum numbers of the new fields that can generate chiral enhancement in photonic dipole penguins.

\end{document}